\documentclass[epj,nopacs]{svjour}
\usepackage[T1]{fontenc}
\usepackage[latin9]{inputenc}
\usepackage{xspace}
\usepackage{amsmath}
\usepackage{amssymb}
\usepackage{epsfig} 
\usepackage{graphicx}
\usepackage{psfrag}

\DeclareMathAlphabet   {\mathsc}{OT1}{cmr}{m}{sc} 
 
\def\[{\left [} 
\def\]{\right ]} 
\def\({\left (} 
\def\){\right )}

\newcommand{\gappeq}{\mathrel{\rlap {\raise.5ex\hbox{$>$}} 
{\lower.5ex\hbox{$\sim$}}}} 
\newcommand{\lappeq}{\mathrel{\rlap{\raise.5ex\hbox{$<$}} 
{\lower.5ex\hbox{$\sim$}}}}

\renewcommand{\thepage}{\arabic{page}} 
\begin{document}
\onecolumn

\rightline{LPT--Orsay 05/47}
\rightline{DESY 05-128}
\rightline{ULB-TH/05-17}
\rightline{hep-ph/0507263}

\vspace{1.cm}

\begin{center}

{\Large {\bf  Dark matter and collider searches in the MSSM}}
\vspace{0.3cm}\\
{\large Yann Mambrini $^{1,2}$, Emmanuel Nezri $^{3}$}
\vspace{0.3cm}\\

$^1$ Laboratoire de Physique Th\'eorique des Hautes Energies\\
Universit\'e Paris-Sud, F-91405 Orsay, France
\vspace{0.2cm}\\

$^2$ Deutsches Elektronen-Synchrotron DESY,
Notkestrasse 85, 22607 Hamburg, Germany
\vspace{0.2cm}\\

$^3$ Service de Physique Th\'eorique

Universit\'e Libre de Bruxelles B-1050 Brussels, Belgium 
\vspace{0.2cm}\\
 
\end{center}

\vskip 1.5cm



We study the complementarity between dark matter experiments (direct detection and 
indirect detections) and accelerator
facilities (the CERN LHC and a $\sqrt{s}= 1$ TeV $e^+e^-$ Linear Collider)  in
the framework of the constrained Minimal Supersymmetric Standard Model (MSSM).
We show how non--universality in the scalar and gaugino sectors can affect the 
experimental prospects to discover the supersymmetric particles. The future
experiments will cover a large part of the parameter space of the
MSSM favored by WMAP constraint on the relic density,  but there still exist 
some regions beyond reach for some extreme
(fine tuned) values of the supersymmetric parameters. Whereas the Focus Point
region characterized by heavy scalars will be easily probed by experiments
searching for dark matter,  the regions with heavy gauginos and light sfermions
will be accessible more easily by collider experiments. More
informations on both supersymmetry and astrophysics parameters can be thus
obtained by correlating the different signals.



\vskip 2cm

\newpage

\tableofcontents

\newpage
\section{Introduction}

Several astrophysical and cosmological independent measurements point towards
the fact that the matter in our universe is dominated by a not yet identified
dark  component (see e.g. 
Refs.~\cite{Bertone:2004pz,Jungman:1995df,Olive1,carlosreview} for reviews). The
solution of this problem is very crucial for the understanding of our universe,
as it concerns different scales of astrophysics such as galaxy through rotation
curves, clusters through X-ray emission  and the cosmological scale through CMB
anisotropy measurements. The latter point provides the most stringent constraint
and gives the total fraction of dark matter in the universe with the best
accuracy.   Indeed, the recent WMAP \cite{WMAP} results lead to a flat 
concordance model universe with  a relic density of cold dark matter of
\begin{equation} 
\Omega_{CDM} h^2=0.1126^{\ +0.0161}_{\ -0.0181} \ {\rm at \ 95 \% \ CL.} 
\end{equation} 
\noindent 
The accuracy of the measurement is expected to increase with future data from  
the PLANCK satellite \cite{Planckexp} and a precision $\Delta \Omega_{CDM} h^2
\sim 2\%$ should be obtained. 

An interesting possibility for such a cold dark matter candidate is a bath of
long lived or stable Weakly--Interacting Massive Particles (WIMPs) which are
left over from the Big Bang in sufficient number to account for a significant
fraction of the relic density. Since additional constraints, especially from 
light element cosmonucleosynthesis, strongly disfavor the possibility that dark
matter is composed solely of baryons \cite{Freese:1999sh}, some form of
``non--standard'' matter is required.

The Standard Model (SM) of high--energy physics, despite of 
its success in explaining the data available today, requires an extension to
explain the stability of the hierarchy between  the weak and the Planck scales,
the unification of gauge couplings and the origin  of electroweak symmetry
breaking.  The most plebiscited extension of the model is the Minimal
Supersymmetric Standard Model (MSSM)
\cite{Fayet:1976cr,Haber:1984rc,Barbieri:1987xf,Martin:1997ns}. It predicts the
existence of several new particles, the superpartners of SM ones.  The  lightest
supersymmetric particle (LSP) is in most of the MSSM parameter space, a stable,
massive, neutral and weakly interacting particle : the lightest neutralino,
which is thus an interesting and well motivated dark matter candidate. On the
other hand, at future colliders such as the Large Hadron Collider (LHC) and the
planned International Linear $e^+ e^-$ Collider (ILC), supersymmetric  particles
are expected to be produced and observed if low energy Supersymmetry (SUSY) is
present in nature. However, even if part of the supersymmetric spectrum is
unveiled at the LHC for example, the properties of the particles which play a
dominant role in the relic density will not be measured directly or precisely. Both
types of data (from astroparticle and accelerator physics) are thus needed to
extract more complete properties of the underlying supersymmetric model
\cite{Allanach:2004xn}.

In constrained MSSM, such as the minimal supergravity model (mSUGRA), the 
minimization of the one-loop scalar potential leads to the
well--known relation between the squares of the superpotential Higgs mass term 
and the soft--SUSY breaking scalar Higgs masses $m_{H_u}, m_{H_d}$ as well
as the ratio of the vacuum expectation values of the two Higgs fields
$\tan\beta=v_d/v_u$ and the $Z$ boson mass $M_Z$, 
\begin{equation}
\mu^{2}=\frac{\(m_{H_d}^{2}+\delta m_{H_d}^{2}\) - 
  \(m_{H_u}^{2}+\delta m_{H_u}^{2}\) \tan^2{\beta}}{\tan^{2}{\beta}-1} 
-\frac{1}{2} M_{Z}^{2}
\label{mu}
\end{equation}
\noindent
imposed at the SUSY breaking scale defined by the quadratic average of the two 
top  squark
masses, $M_{SUSY}=\sqrt{m_{\tilde t_1} m_{\tilde t_2}}$. This condition
determines the absolute value of the term $\mu$, leaving its sign as a free parameter of the
theory. 

The four neutralinos ($\chi^0_1\equiv \chi,\chi^0_2,\chi^0_3,\chi^0_4,$)     are
superpositions of the neutral fermionic partners of the electroweak gauge bosons
$\tilde B^0$ and $\tilde W_3^0$ (respectively the B--ino and  W--ino fields) and
the superpartners of the neutral Higgs bosons $\tilde H_u^0$,  $\tilde H^0_d$
(respectively up and down Higgsinos fields).  In the
($\tilde{B},\tilde{W}^3,\tilde{H}^0_d,\tilde{H}^0_u$) basis,  the neutralino
mass matrix is given by 
\begin{equation}
\arraycolsep=0.01in
{\cal M}_N=\left( \begin{array}{cccc}
M_1 & 0 & -m_Z\cos \beta \sin \theta_W^{} & m_Z\sin \beta \sin \theta_W^{}
\\
0 & M_2 & m_Z\cos \beta \cos \theta_W^{} & -m_Z\sin \beta \cos \theta_W^{}
\\
-m_Z\cos \beta \sin \theta_W^{} & m_Z\cos \beta \cos \theta_W^{} & 0 & -\mu
\\
m_Z\sin \beta \sin \theta_W^{} & -m_Z\sin \beta \cos \theta_W^{} & -\mu & 0
\end{array} \right)\;.
\label{eq:matchi}
\end{equation}

\noindent where $M_1$, $M_2$ are the bino and wino
mass  parameters, respectively. This matrix can be diagonalized by a single
orthogonal matrix $z$  and we can express the LSP $\chi$ (often referred
in the following as {\it the neutralino}) as
\begin{equation}
\chi = z_{1 1} \tilde B+  z_{1 2}\tilde W  
+ z_{1 3}\tilde H_d  +z_{1 4} \tilde H_u.
\end{equation}
\noindent 
This combination determines the nature, the couplings and the phenomenology of
the neutralino. The neutralino is usually called ``gaugino--like'' if
$P \equiv |z_{11}|^2 + |z_{12}|^2> 0.9$, ``Higgsino--like'' if $P < 0.1$, and
``mixed'' otherwise.

Depending on the nature of the neutralino, the WMAP constraint can be fulfilled
essentially by bino-$\chi\tilde{\tau}$ coannihilation processes if $m_{\chi}
\sim m_{\tilde \tau_1}$, $\chi\chi\xrightarrow{A}b\bar{b}$ annihilation for
large tan$\beta$ values or a light pseudoscalar $A$ boson, and
$\chi\chi \rightarrow t\bar{t}$  for a sufficiently Higgsino--like
neutralino.  In the same time, a non negligible wino component can enhance the
annihilation process $\chi\chi \to W^+W^-$  and the $\chi \chi^\pm $
and $\chi^+ \chi^-$ coannihilation ones.

In the present work, we will consider neutralino dark matter searches in  direct
or indirect detection experiments and the prospects of superparticle  production
at future colliders like LHC or ILC. We will focus on the framework of general
supergravity scenarios but with non--universal scalar and gaugino soft--SUSY
breaking mass terms. 

The outline of the paper is as follows. We first summarize, in section 2, the
phenomenology of the different kinds of dark matter searches. Section 3 is
dedicated to the prospects for producing and detecting SUSY particles and MSSM
Higgs bosons at the LHC and at a high energy $e^+ e^-$ collider.   In section 4,
we present  a complementary analysis of each type of signal and the impact of
non--universality on the  detection potential of all types of experiments.
For our computation, we use
an interface of the latest released version of the
codes SUSPECT \cite{Suspect} for the MSSM particle spectrum, 
MICROMEGAS \cite{Micromegas} for the neutralino relic density, 
and DARKSUSY \cite{Darksusy} for the dark matter detection rates.
During the writting of this paper, the authors of \cite{Baer:2005bu} and
\cite{Baer:2005zc} have made similar analyses and reached the same conclusions 
as those presented here. Related work in a variety of frameworks and dealing
with cosmological relic density aspects, present accelerators constraints 
and/or dark matter searches and/or SUSY searches at future colliders can be found in Refs  \cite{ellissug1} - \cite{Baer:2004qq}.  


\section{Dark matter searches}

\subsection{Dark matter distribution}

The dark matter distribution in the galaxy is a crucial ingredient for all 
kinds of detection techniques. From N-body simulations, this distribution is 
commonly parameterized as :
\begin{equation}
 \rho(r)=\frac{\rho_0 [1+(R_0/a)^{\alpha}]^{(\beta-\gamma)/\alpha}}{(r/R_0)^{\gamma}  [1+(r/a)^{\alpha}]^{(\beta-\gamma)/\alpha}}
\label{profile} 
\end{equation}

\noindent
where $r$ is the galacto-centric coordinate, $\rho_0$ is the local  (sun
neighborhood) halo density, $R_0$ the solar distance to the  galactic center and
$a$ a characteristic length.  If there is an agreement concerning the behavior
at large radii ($\beta \sim 3$), the shape of the possible cusp in the 
innermost region of the galaxy  is  not well determined if we consider the
 discrepancies
between simulation results of various groups ($1 \lesssim \gamma \lesssim 1.5$).
Furthermore, the studies of systems like low surface brightness galaxies seem to
favor flat cores. Moreover, the small radius region  behavior can differ
strongly depending on the physical assumptions such as 
baryonic effects on the central dark matter density, supermassive black hole 
induced spikes, dark matter particle scattering on stars, {\it etc}... (for
discussions, see {\it e.g.} Refs.
\cite{Blumenthal:1985qy,Edsjo:2004pf,Prada:2004pi,Gnedin:2004cx,silkgondo,merrit04}).
Finally possible inhomogeneities and substructures could be present, leading to
a possible clumpyness of the halo.

In contrast, there is a general agreement on the local density  $\rho_0$ which
can be determined for each density profile assuming compatibility with the
measurements of rotational curves and the total mass of the galaxy; $\rho_0$
should range from  $0.2$ to $0.8\ {\rm GeV}.{\rm cm}^{-3}$  (see
Ref.~\cite{Jungman:1995df} for a discussion). For definiteness, our results are
presented for  $\rho_0 = 0.3 ~ {\rm GeV}.{\rm cm}^{-3}$ for all the density
profiles used in the present analysis. A more controversial topic is the
possible link between the dark matter distribution and the total relic
abundance. One can rescale the density $\rho(r)$ when the calculated value of
$\Omega_{\chi} h^2$ is smaller than the WMAP lower bound, by assuming that the
neutralino could form only a fraction of the total amount of cold dark matter.
In this study, however, we will not use this procedure  as we will mainly focus
on the dependence of the detection rates on the SUSY parameter space for a
$given$ astrophysical framework. Since the local density enters as a 
scaling factor in the signal fluxes, the effect of varying $\rho_0$ or applying
this rescaling can be taken into account in a straightforward manner.

\subsection{Direct detection}

\begin{figure}[t]
\begin{center}
\begin{tabular}{ccc}
\includegraphics[width=0.15\textwidth]{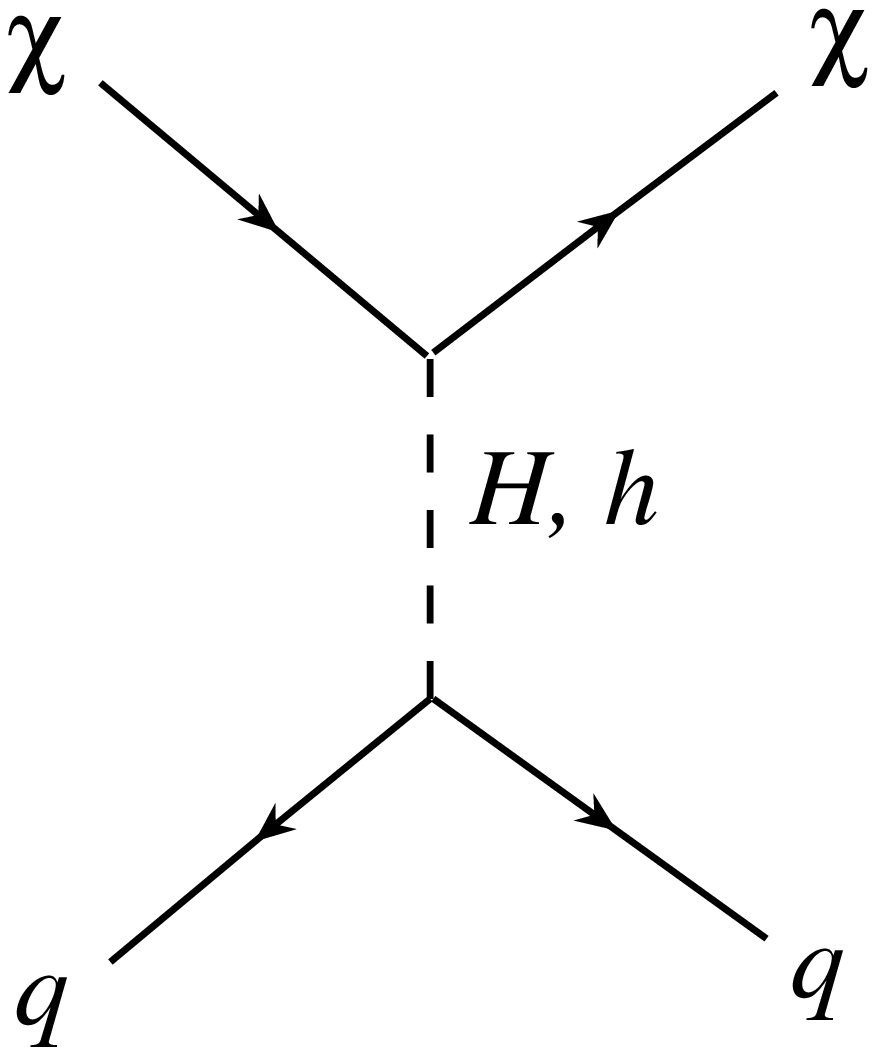}&
\includegraphics[width=0.28\textwidth]{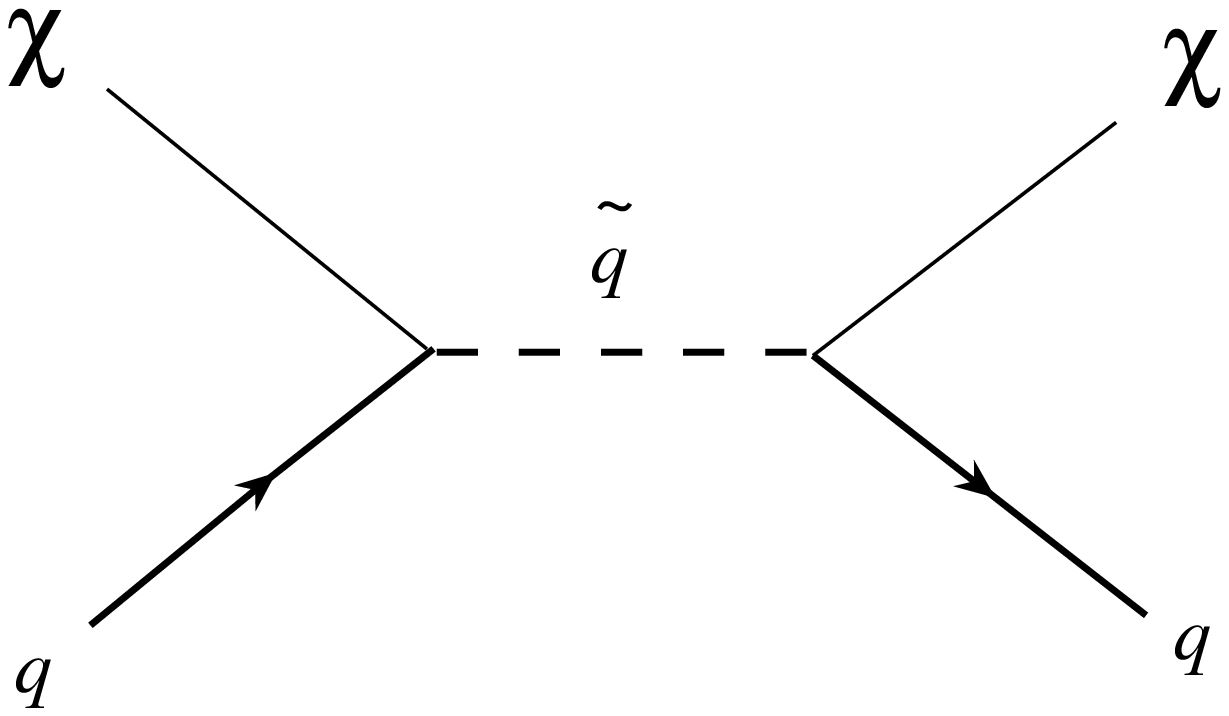}&
\includegraphics[width=0.15\textwidth]{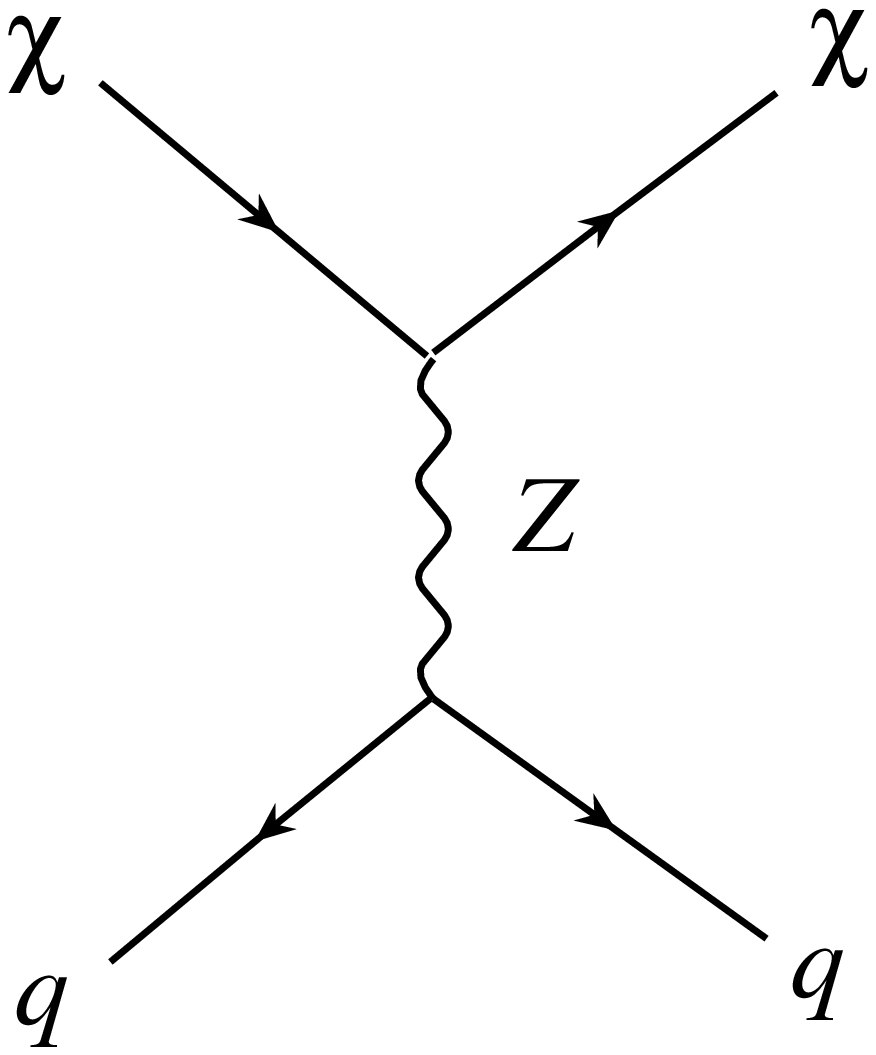}\\
a) & b) & c)
\end{tabular}
\caption{\small Feynman diagrams of the processes occuring in direct detection
of the lightest neutralinos:  a) and b) spin independent processes 
($\sigma^{scal}_{\chi-p}$); b) and c) spin dependent processes 
($\sigma^{spin}_{\chi-p}$).}
\label{diagsigma-chi-p}
\end{center}
\end{figure}

Many underground experiments have been carried out around the world in order to
detect WIMP candidates by observing their elastic  scattering on target nuclei
through nuclear   recoil \cite{Goodman}.  As pointed out before, the
astrophysical dependence on this type of detection technique is weak. Namely,
the  translation of the detection rates/sensitivities into scattering  cross
section $\sigma_{\chi -p}$  relies only on the knowledge of the local dark
matter density $\rho_0$. Depending on the spin of the target nuclei, the
detection rate is given by  the spin dependent ($\sigma^{spin}_{\chi-p}$) or the
spin independent  ($\sigma^{scal}_{\chi-p}$) neutralino--nucleon elastic  cross
section. The main contributing diagrams are shown in  Fig.
\ref{diagsigma-chi-p}. 

The squark (mainly the first generation $\tilde u$, $\tilde d$ squarks) exchange
contributions are usually suppressed by the squark masses.  The
spin--independent cross section  $\sigma^{scal}_{\chi-p}$ is then driven by
neutral CP--even Higgs boson ($h$, $H$) exchanges  ($\chi q \xrightarrow{h,H}
\chi q \varpropto z_{11(2)}z_{13(4)}$) and   the spin--dependent  cross section 
$\sigma^{spin}_{\chi-p}$ by $Z$ boson  exchange  ($\chi q \xrightarrow{Z,} \chi
q \varpropto z^2_{13(4)}$). Direct detection is thus more favored for a mixed
gaugino-Higgsino neutralino and models where the scalar Higgs boson $H$ is
sufficiently light. 


In the usual  mSUGRA scenario, where the soft terms of the  MSSM are assumed to
be universal at the unification scale, the spin--independent cross section turns
out to be constrained by $\sigma^{scal}_{\chi-p} \lesssim 3 \times 10^{-8}$ pb
\cite{carlosreview}. However, it has been shown that if the assumption of
universality in the scalar and/or gaugino sectors is relaxed, the cross section
can be increased significantly with respect to the universal scenario and  could
values of the order of reach $\sigma^{scal}_{\chi-p} \sim 10^{-6}$ pb 
\cite{Birkedal-Hansen:2001is,Mynonuniv,BirkedalnonU,Cerdeno:2003yt}. QCD
corrections to the neutralino-nucleon scattering cross sections can also be
relevant \cite{abdeldrees}.

Current experiments such as  EDELWEISS  \cite{Sanglard:2005we} and CDMS
\cite{Cdms} are sensitive to WIMP--proton cross sections larger than
approximately  $10^{-6}$ pb, excluding the DAMA region \cite{Dama}.
These sensitivities are slightly too small to probe 
minimal SUSY models if  we impose the accelerator constraints and 
the bound on the relic density from WMAP.  
Several new or upgraded direct WIMP detection experiments will soon
reach a significantly improved sensitivity  (GENIUS, EDELWEISS II
\cite{EdelweissII}, ZEPLIN(s)  \cite{Zeplin}, CDMS II and superCDMS
\cite{Brink:2005ej}). The next generation of experiments ({\it e.g } EDELWEISS
II and CDMS II) will lead to a minimum of the valley sensitivity around
$10^{-8}$ pb for a neutralino mass of $m_{\chi} = {\cal O}$(100 GeV). Though
challenging from the experimental point of view, a ton-size detector (ZEPLIN,
SuperCDMS) should be able to reach  $\sigma^{scal}_{\chi-p} \gtrsim 10^{-10}$ pb
which would be conclusive to probe WIMP dark matter models.  In our study, we
will take the neutralino mass dependent projected experiment sensitivities of
EDELWEISSII \cite{EdelweissII} and ZEPLIN \cite{Zeplin}.  

\subsection{Gamma Indirect detection}

\begin{figure}[t]
\begin{center}
 \begin{tabular}{ccccc}
\includegraphics[width=0.145\textwidth]{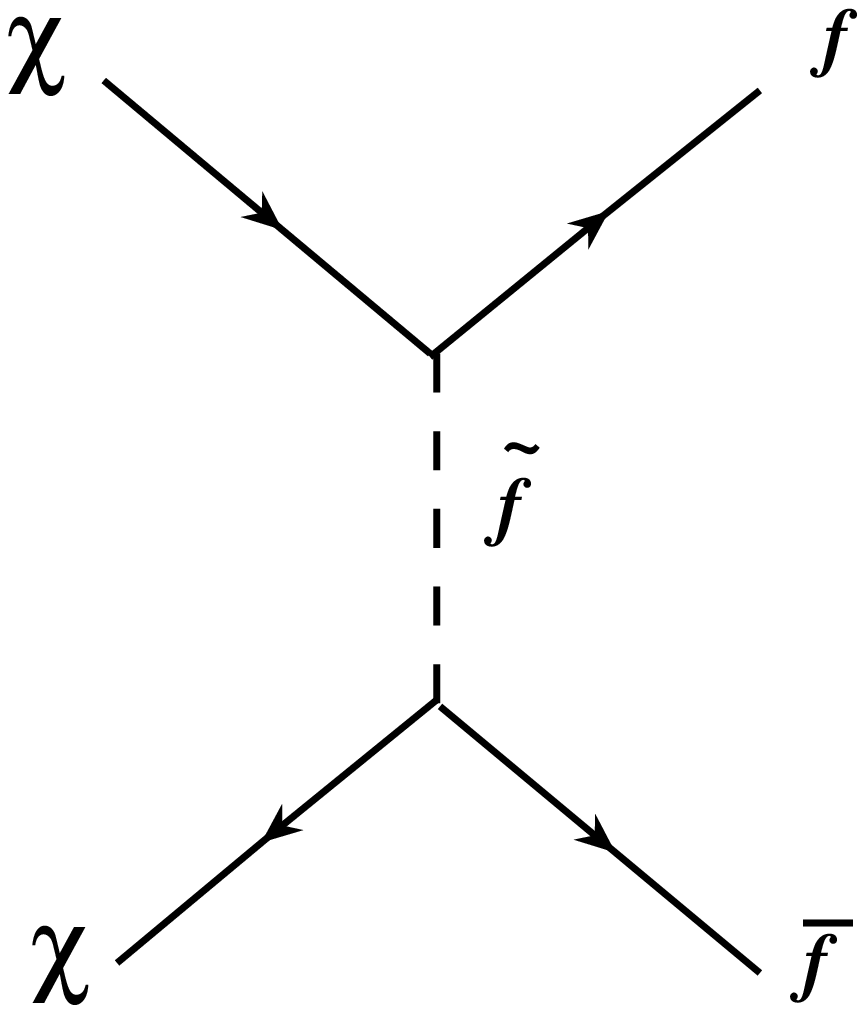}&
\includegraphics[width=0.225\textwidth]{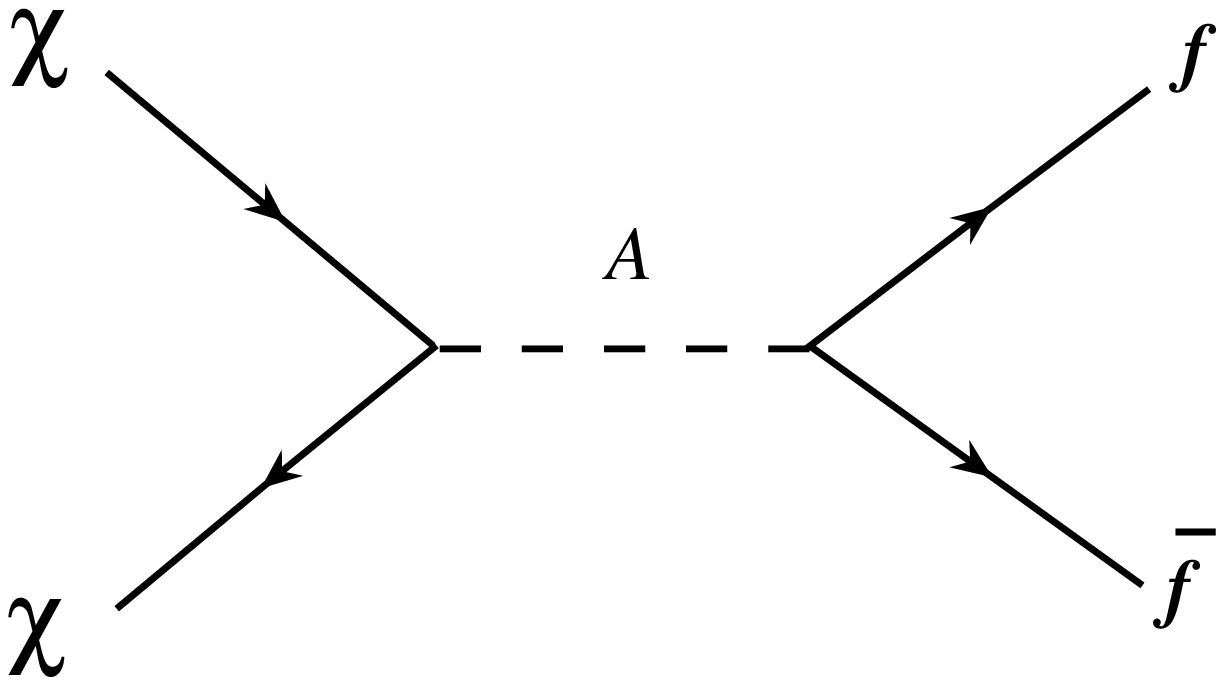}&
\includegraphics[width=0.225\textwidth]{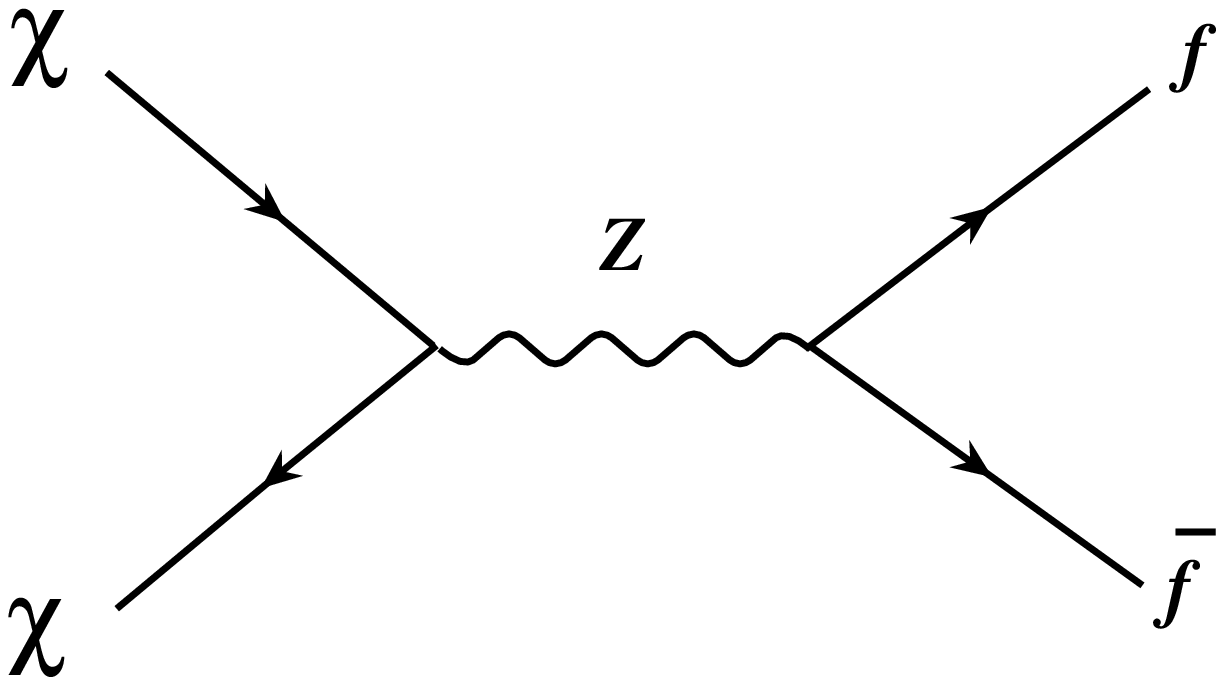}&
\includegraphics[width=0.145\textwidth]{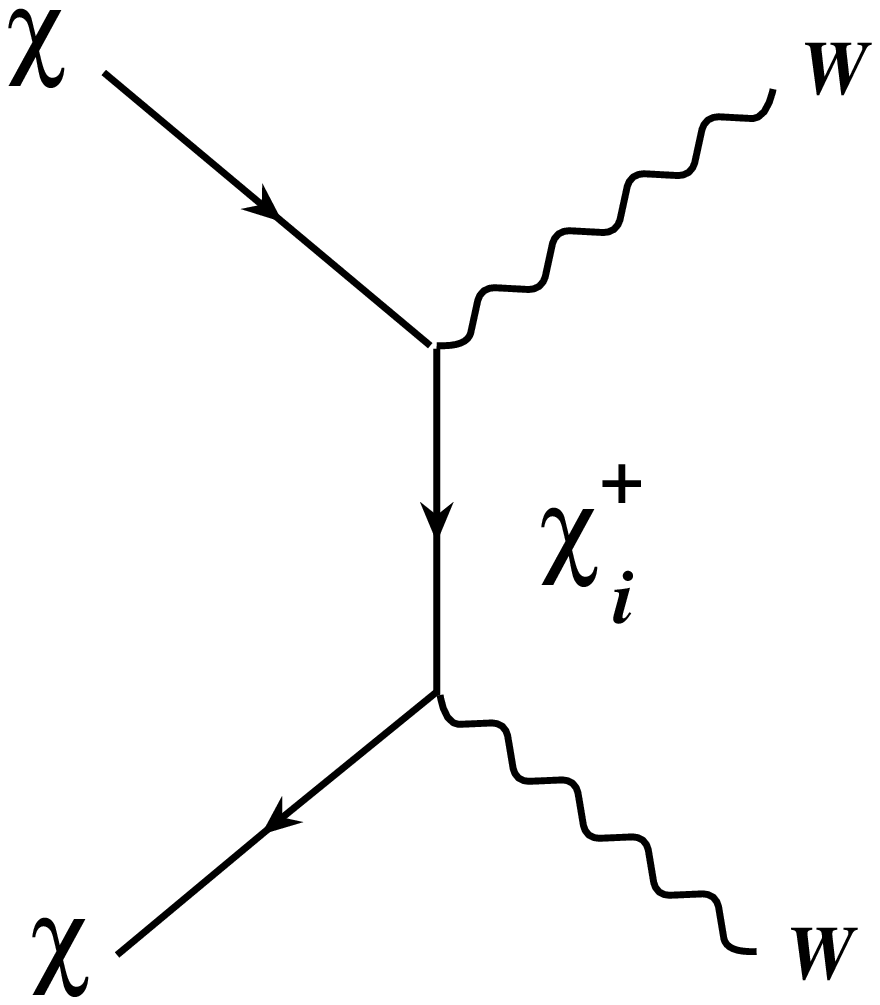}&
\includegraphics[width=0.145\textwidth]{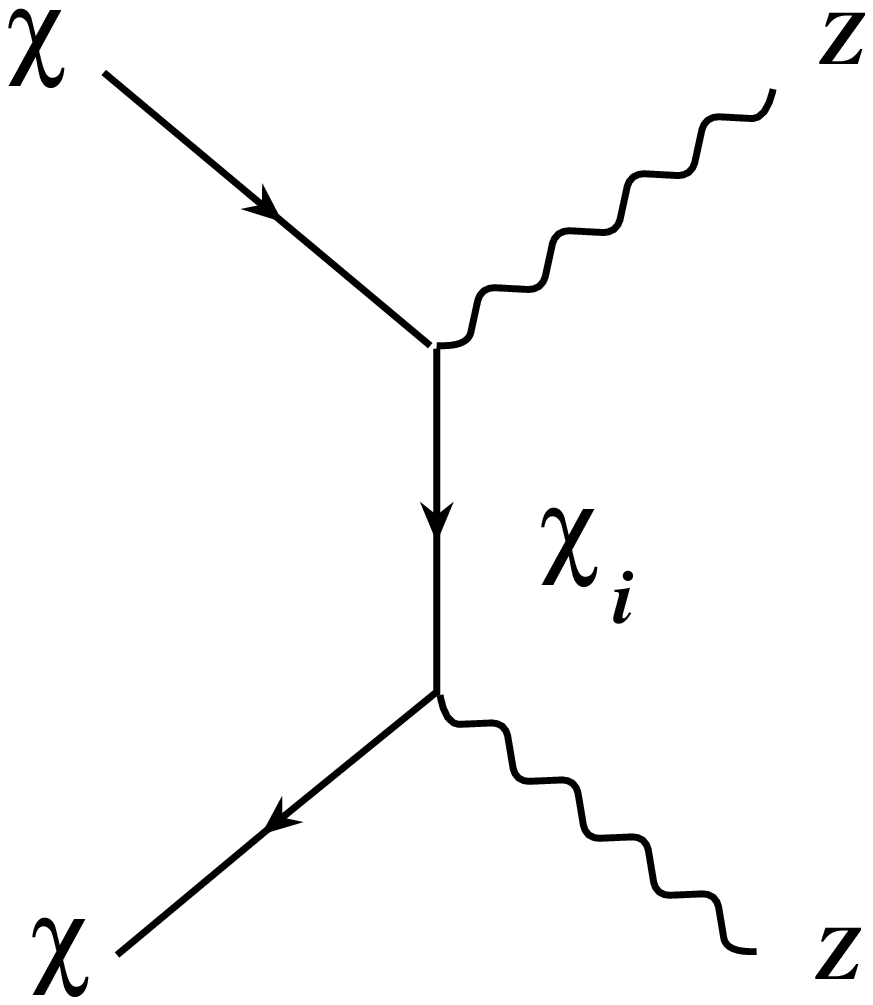}\\
 a) & b) & c) & d) & e)
\end{tabular}
 \caption{\small Feynman diagrams for the dominant channels contributing 
 to neutralino annihilation into SM particles.}  
\label{annihi-channels}
   \end{center}
\end{figure}

Dark matter can also be observed through its  annihilation products 
 in the galactic halo. In particular, the annihilation in the
Galactic Center (GC) where the dark matter density is important 
could lead to large fluxes and promising experimental signals, 
even if the exact behavior in the central region is poorly constrained. 
Unfortunately, the astrophysical uncertainties dominate largely the ones
coming from particle physics models, affecting considerably the prospects
of discovery in gamma indirect detection experiments.

The main annihilation processes entering in the calculation of gamma--ray
fluxes from the GC are depicted in Fig.~\ref{annihi-channels}. 
The large masses of the scalar fermions and their small Yukawa couplings 
usually suppress the contribution of the diagrams with $t$--channel sfermions 
exchange. The dominant cross sections are thus
$\sigma (  \chi \chi \xrightarrow{A}b \bar{b})  \propto [z_{11(2)}z_{13(4)}]^2
  $, $\sigma (  \chi \chi \xrightarrow{Z}t \bar{t} ) \propto [z^2_{13(4)}]^2$
  and  $\sigma ( \chi \chi \xrightarrow{\chi^+(\chi^0_j)}W^+W^-(ZZ) )  \propto
  [z_{13(4)}V_{12}]^2  \textrm{ and/or }  [z_{12}V_{11}]^2 $ 
  $([z_{13(4)}z_{j3(4)}]^2 )$, 
with $V_{ij}$ the chargino mixing matrice. 
Annihilation in these channels are favored for wino--like or Higgsino--like 
neutralino. The resulting observed differential gamma--ray flux at the
Earth coming from a direction forming an angle $\psi$ with respect
to the GC is
\begin{equation}
\frac{d\Phi_{\gamma}(E_{\gamma},\psi)}{d \Omega d E}= \sum_i \frac{1}{2}
\frac{dN^i_{\gamma}}{dE_{\gamma}} \langle \sigma_i v \rangle 
\frac{1}{4 \pi m_{\chi}^2}
\int_{\mbox{line of sight}}
\rho^2\left(r(l,\psi)\right) d l 
\label{eq:gamflux}
\end{equation}

\noindent
where the discrete sum is over all dark matter annihilation channels,
$dN^i_{\gamma}/dE_{\gamma}$ is the differential gamma--ray yield
and $\langle \sigma_i v \rangle$ is the annihilation cross section 
averaged over the velocity distribution. 
It is customary to isolate the dependence on the halo dark matter model
with respect to particle physics, defining the
dimensionless quantity (see Ref.~\cite{Turner:1986vr,Bergstrom:1997fj})
\begin{equation}
\bar{J}\left(\Delta \Omega \right) = \frac{1} {8.5\, \rm{kpc}} 
\left(\frac{1}{0.3\, \mbox{\small{GeV/cm}}^3}\right)^2
\int_{\Delta \Omega} \int_{\mbox{\small{line of sight}}}\rho^2\left(r(l,\psi)\right) d l d \Omega\,.
\label{gei}
\end{equation}

\noindent
in a solid angle $\Delta \Omega$ centered on $\psi=0$. 

As pointed out before, a crucial ingredient for the calculation of the 
annihilation fluxes is the density profile of dark matter around the core of the
GC. In the present work, we choose the intermediate NFW halo profile 
\cite{Navarro:1996he} ($\gamma=1$, $\bar{J}_{NFW}(\Delta \Omega = 10^{-3})\sim
10^3$). One can rescale fluxes to have results for other commonly used profiles
either with a stronger cusp like the one proposed by Moore et al.
\cite{Moore:1999gc} ($\gamma=1.5$, $\bar{J}_{Moore}(\Delta \Omega = 10^{-3})\sim
10^5$)  or shallower slope like the one proposed by Kravtsov et al.
\cite{Kravtsov:1997dp} ($\gamma=0.4$, $\bar{J}_{Kravtsov}(\Delta \Omega =
10^{-3})\sim 10$)\footnote{For $\gamma \geq 1.5,\  \bar{J}$ diverges and one has
to regularize the integral of eq. \ref{gei}.}.  The sensitivity of such
variations in the dark matter profile on the experimental prospects will be
illustrated later; see Figs.~\ref{fig:topmoorekra}c) and d).  In the literature,
some authors \cite{Bergstrom:1997fj} also consider as input parameter of the
theory a {\it boost factor} acting on $\bar{J}$, to take into account  possible
halo inhomogeneities (clumps for instance).

Recently several experiments have detected a significant ammount of gamma--rays
from the galactic center region. Observations by INTEGRAL \cite{Jean:2003ci} and
EGRET \cite{EGRET} have revealed $\gamma$--ray emission from this region
although  no corresponding sources have been identified so far. The VERITAS
\cite{Kosack:2004ri} and CANGAROO \cite{Cangaroo1}  collaborations using,
respectively, the Whipple 10 meters and CANGAROO--II  atmospheric Cerenkov
Telescopes (ACTs) have independently detected TeV $\gamma$--rays from the same
region. Finally, HESS \cite{Aharonian:2004wa} claims to have observed a  signal
corresponding to a WIMP in the multi--TeV energy range.  Here, we refrain from
interpreting all these signals as due to dark matter annihilation. Although an
explanation in terms of a heavy dark matter particle like the LSP neutralino
\cite{Mambrini:2005vk,deBoer:2004ab,Aharonian:2004wa} is
possible for each signal (except for INTEGRAL, see for instance 
Ref.~\cite{Boehm:2003bt}  for a light dark matter scenario proposal), these
measurements are not compatible with each other and cannot be explained by a
single scenario.  
Moreover, purely astrophysical interpretations of these
signals are  possible  \cite{Casse:2003fh,Bertone:2004ek}.  

In any case, considering the uncertainties in the computations and
 that alternative astrophysical interpretations are
 possible \cite{Casse:2003fh,Bertone:2004ek}, 
it is reasonable not to attribute these signals to a
neutralino and proceed with our prospective analysis in the 
 SUSY parameter space. Nevertheless, the EGRET signal  
($\sim 4\times 10^{-8}\ \gamma ~{\rm cm^{-2}s^{-1}}$) can be seen as an upper
bound even if one has to keep in mind that it may not arise exactly from the
galactic center \cite{Hooper:2002ru}. We will also consider the sensitivities of
the HESS \cite{HESS} and GLAST \cite{GLAST} experiments   (respectively
$10^{-12}\ \gamma~ {\rm cm^{-2}s^{-1}}$ with  a 100 GeV threshold and $10^{-10}\
\gamma~ {\rm cm^{-2}s^{-1}}$ with a 1 GeV threshold) as a probing test of our
models. The neutralino mass dependent integrated  sensitivities that we use in
our analysis can be found in Ref.~\cite{Morselli:2002nw}.

\subsection{Neutrino Indirect detection}

Dark matter particles of the halo can also be trapped in astrophysical bodies
(like the Sun) by successive elastic diffusion on its nuclei (hydrogen) during
the age of the target object ($\sim 10^{10}$ years). This leads to a
captured population which annihilates, 
producing neutrino fluxes that can be detected by a neutrino telescope,
signing the presence of dark matter in the storage object direction.
The annihilation rate at a given time $t$ can be written \cite{neutrinorate}:
\begin{equation}
\Gamma_A=\frac{1}{2}C_AN_{\chi}^2=\frac{C}{2}\tanh^2{\sqrt{CC_A}t},
\label{anihirate}
\end{equation}

\noindent
Where $C$ is the capture rate which depends on the local density $\rho_0$ and 
on the neutralino-proton elastic cross section.  
$\Gamma_A\approx \frac{C}{2}=cste$ when the neutralino population has
reached equilibrium, and $\Gamma_A\approx \frac{1}{2}C^2C_At^2$ in the
initial collection period. When accretion is
efficient, the annihilation rate follows the capture rate $C$ and thus the neutralino-quark elastic cross section, whereas only the differential spectrum
depends on the annihilation processes. The flux is then given by

\begin{equation}
  (\frac{d\Phi_{\nu}}{dE_\nu})=\frac{\Gamma_A}{4 \pi R^2}
  \sum_F B_F\frac{dN^F_\nu}{dE_\nu}(E_\nu)
\label{eq:nuflux}
\end{equation}

\noindent
where $F$ labels the annihilation final states and $R$ is the Sun-Earth distance.
As related to the local dark matter density, the astrophysical dependence
is weak, similarly to the direct detection case.
 One should notice that the collection of neutralinos is
time dependent such that the trapped population can have been enhanced if the
Sun has been flying in some clumps during its history.

The particle physics behavior is dominated by the capture rate
driven by $\sigma_{\chi -p}$. The dominant processes are shown on Fig.
\ref{diagsigma-chi-p} (spin dependent for the Sun because of the non zero 
hydrogen nucleus spin). The couplings have already been described in the 
section related to direct detection. 
The diagrams driving annihilation (see Fig. \ref{annihi-channels}) 
and their couplings have been discussed in the section devoted to 
gamma indirect detection.  For our prospect
we will consider the fluxes coming from the Sun which is favored for
neutralinos with a non negligible Higgsino component. 
Indeed the $Z$ exchange is
then allowed in the neutralino-quark diffusion and the resulting flux can be
high. The annihilation can also enhance the flux, especially by giving harder
neutrino spectra when the Higgsino and/or the wino fraction are not negligible
leading to $t\bar{t},W^+W^-$ final states instead of $b\bar{b}$ for a dominant
bino neutralino \cite{Mynonuniv,Myuniv}.

The Earth could be another possible source but the resulting fluxes are beyond reach of
detection \cite{Myuniv}. The neutralino annihilations in the galactic center can
also lead to neutrino fluxes ($i=\nu$ in equation \ref{eq:gamflux}) but the
gamma flux expectations are much more promising with regard to experiment
sensitivities \cite{Bertone:2004ag}.

Present experiments like MACRO \cite{Macro}, BAKSAN \cite{Suvorova}, SUPER
K \cite{SuperK} and AMANDA \cite{Amanda} (which size and place disfavors     
detection of horizontal flux coming from the Sun) give limits on possible fluxes around $10^4\ \mu\
{\rm km^{-2}yr^{-1}}$. Future neutrino telescopes like ANTARES \cite{Antares} or a ${\rm
  km}^3$ size like ICECUBE \cite{Ice3} will be able to probe respectively around
$10^3$ and $10^2\ \mu ~ \mathrm{km}^{-2} ~ \mathrm{yr}^{-1}$. 
We used neutralino mass dependent sensitivities of reference
\cite{bailey} for ANTARES and \cite{Ice3Edsjo} for ICECUBE.

\subsection{Positron Indirect detection}

Neutralino annihilations in the halo can also give rise to measurable positron
fluxes.  Positrons being charged particles interact during their
propagation such that the directional information is lost. Furthermore those
interactions imply that the observed positrons do not come from far 
away in the
galaxy. In addition to the variability of the density profile which is a
possible source of uncertainties at the production level, the understanding of
the propagation taking into account interactions with magnetic fields,
inverse Compton and synchrotron processes is the most relevant and difficult
question to control in order to be able to understand measurement or/and to
estimate positron spectra.
The positron flux results from the steady state solutions
of the diffusion-loss equation for the space density of cosmic rays
per unit energy, $dn/d\varepsilon$:
\begin{equation}
0(=  \frac{\partial}{\partial t}\frac{dn}{d\varepsilon})=\vec{\nabla}\cdot
  \left[K(\varepsilon,\vec{x})\vec{\nabla}\frac{dn}{d\varepsilon}\right]+
  \frac{\partial}{\partial \varepsilon}\left[b(\varepsilon,\vec{x})
  \frac{dn}{d\varepsilon}\right]+Q(\varepsilon,\vec{x}),
  \label{eq:diffloss}
\end{equation}
where $K$ is the diffusion constant (assumed to be constant in space
throughout a ``diffusion zone'', but it may vary with energy), 
$b$ is the energy loss rate and
$Q$ is the source term (see \cite{Baltz:1998xv} for details).
We take \cite{Moskalenko:2002yx}
\begin{equation}
  K(\varepsilon)=
  3.3\times 10^{28}\left[3^{0.47}+\varepsilon^{0.47}\right]\ {\rm cm}^2\;{\rm s}^{-1}.
  \label{eq:K}
\end{equation} 

and \cite{Longair}

\begin{equation}
  b(\varepsilon)_{e^+}= 10^{-16}\varepsilon^2\;{\rm s}^{-1},
  \label{eq:bofe}
\end{equation}
which results from inverse Compton scattering on both the cosmic
microwave background and diffuse starlight.
The diffusion zone is a slab of thickness $2L$ ($L$= 4 kpc to fit observations
of the cosmic ray flux, see \cite{Donato:2003xg} and references therein. Variations of the propagation parameters
may modified the computation results by around an order of magnitude (see
\cite{Hooper:2004bq}). The source term $Q=f(\rho(r),\langle \sigma v\rangle)$ can
be modified if one considers the presence of clumps in the (quite local) dark matter
distribution and a possible resulting multiplying boost factor $b \lesssim 10$ 
\cite{Hooper:2003ad}. 
The particle physics dependence also enters in the source term and comes from
the supersymmetric parameter influence
on annihilation cross section (see Fig. \ref{annihi-channels} and section 2.3). 

The HEAT experiment, 
in three flights which have taken place in 1994, 1995 and 2000,
observed a flux of cosmic positrons in excess of the predicted rate, 
peaking around 10 GeV \cite{heat}. This signal can be accommodated by neutralino 
annihilation but requires a boost factor \cite{Baltz:2001ir,Hooper:2004bq}. 
Furthermore the HEAT measurement uncertainties
in the 30 GeV bin are quite large. We thus consider in this work the estimated
 fluxes with regard to the future 
experiments AMS-02 and PAMELA. The exact positron spectrum depends on
annihilation final states, dark matter distribution and propagation parameters
(see \cite{Hooper:2004bq}) but as a reasonable approximation for our
prospect, one can consider the spectra being peaked around
$M_{\chi}/2$. At those energy, we checked that the background of reference \cite{Moskalenko:2001ya} can be
fitted by   $E^2 d\Phi_{e^+}/d\Omega dE \simeq 1.16 \times E^{-1.23}$. 
Following references \cite{Feng} and \cite{Baer:2004qq} we require as a
benchmark condition : $
\frac{\phi^{e^+}_{\chi}}{\phi^{e^+}_{Bckgd}}|_{m_{\chi}/2} \sim 0.01 $ (See
\cite{Baer:2005bu} for more precise criteria)

\subsection{Antiproton Indirect detection}

Another possible signal of dark matter may be the detection of antiproton
fluxes produced by neutralino annihilation. To calculate those fluxes we
need to solve a propagation equation for antiprotons 
\cite{Bergstrom:1999jc,Moskalenko:2002yx,Moskalenko:2001ya,Maurin:2001sj,Maurin:2002hw,Lionetto:2005jd}.  This
includes spatial diffusion in the disk and the halo, $K_x$ scaling 
with the rigidity (momentum per unit of charge, $R=p/Z$) as $K_0 R^{\delta}$. 
The galactic wind, with a speed $V_c$, imply convection effects deflecting 
antiproton away from the disk. Collisions with interstellar matter 
(hydrogen and helium) and Coulomb losses modify the energy distribution. Reacceleration by Fermi shocks on magnetic fields could be taken into account
 by a diffusion coefficient $K_p$ related to the spatial diffusion 
$K_x$ and the Alven velocity of disturbances in the plasma, $V_A$. 
We used the diffusion model \cite{Bergstrom:1999jc} of the DarkSusy package 
with the diffusion--convection \cite{Moskalenko:2001ya} option,

\begin{equation}
\vec{\nabla}\cdot
  \left[K_x\vec{\nabla}\frac{dn}{d\varepsilon}\right]-\vec{\nabla}\cdot
  \left[\vec{V_c}
  \frac{dn}{d\varepsilon}\right]-n^H v_{\bar{p}} (\varepsilon)
\sigma^{in}_{H}(\varepsilon) \frac{dn}{d\varepsilon}
+Q(\varepsilon,\vec{x})=0,
  \label{eq:antiproton}
\end{equation}

\noindent
with
$\delta=0.6,\ V_c=10\ {\rm km.s^{-1}}, \ K_0=25\times 10^{27}\ {\rm cm^2 ~
s^{-1}}$. $n^H$ is the interstellar Hydrogen density number, $v_{\bar{p}}$ is 
the antiproton velocity and $\sigma^{in}_{H}$ is the inelastic
antiproton--hydrogen cross section.
Antiprotons propagate on longer distance than positrons. 
The resulting flux is thus slightly more sensitive to the dark 
matter distribution in the galaxy, and especially in its central region. 
The antiproton flux can be expressed as

\begin{equation}
\phi_{\bar{p}}(R_0,T)=b\ \langle \sigma v \rangle \sum_i
\frac{dN^i}{dT}B^i\left(\frac{\rho_0}{m_{\chi}}\right)^2\ C_{{\rm
prop}}(T)
\end{equation}

\noindent
where $T$ is the $\bar{p}$ kinetic energy, $C_{{\rm prop}}(T)$ contains
propagation effect, and $b$ is a possible boost factor resulting from halo
clumpyness.
Experiments like BESS and CAPRICE measured the antiproton flux. 
The signal can be fitted by the astrophysical background antiproton flux and 
seems to be peaked at 1.76 GeV around $2\times 10^{-6}\
\bar{p}\ {\rm cm}^{-2}s^{-1}sr^{-1}$. The measurement at 37.5 GeV seems to
suggest an excess compared to the models. We estimated the antiproton
fluxes from
neutralino annihilation at those two energy for which the diffusion
dependence
is weaker than for the low energy part of the spectrum. Following 
\cite{Baer:2005bu}, we will 
show as a benchmark region where $\phi_{\bar{p}}(R_0,1.76)> 2\times
10^{-7} \
\bar{p}\ {\rm cm}^{-2}s^{-1}sr^{-1}$ checking also the value at 37.5
GeV.

\section{Collider searches}

\subsection{Constraints}

\subsubsection{The mass spectrum constraints.}

We have implemented in our analysis the lower bounds on the masses of SUSY particles and of the lightest Higgs boson.
In the squark and slepton sector parameters leading to tachyons are excluded. 
We applied the LEP2 lower bound limit on the mass of the lightest 
chargino  $m_{\chi^+_1} > 103.5$ GeV \cite{charginolimit}. 
Typically, the most constraining bound comes from the lightest 
Higgs boson mass limit. 
In the decoupling regime ($m_A \gg M_Z$, valid in all our parameter space),
$m_h > 114.4$ GeV \cite{Higgslimit}. 
It is well known than the theoretical prediction of the
Higgs mass is very sensitive to the value of the top mass.
The radiative corrections used for the calculation of the higgs mass are
very well described in \cite{Allanach:2004rh}.
To take into account this sensitivity in the analysis, we have used  $m_t = 175$
 GeV but we illustrate the dependance of our result on the top mass
(178 to 182 GeV) in Fig. \ref{fig:topmoorekra}b
\cite{Abazov:2004cs}.
 
\subsubsection{The $b \rightarrow s \gamma$ branching ratio.}

One observable where SUSY particle contributions might be large is the radiative
flavor changing decay $b \rightarrow s \gamma$ \cite{Degrassi:2000qf}. 
In the Standard Model this
decay is mediated by loops containing the charge 2/3 quarks and $W-$bosons.
In SUSY theories additional contributions come from loops involving charginos and
stops, or top quarks and charged Higgs bosons. 
The measurements of
the inclusive decay $B \rightarrow X_s \gamma$ at CLEO \cite{cleo} and BELLE 
\cite{belle}, lead to restrictive bounds on the branching ratio $b\to s\gamma$.
The experimental value for the branching ratio of the process
$b\to s\gamma$ is $(3.52 \pm 0.30) \times 10^{-4}$ \cite{Heavy}.
Including theoretical errors \cite{Gambino:2004mv} 
$(0.30 \times 10^{-4})$ coming from
its prediction by adding the two uncertainties in quadrature, we impose
$2.33\times 10^{-4}\leq BR(b\to s\gamma)\leq 4.15\times10^{-4}$,
 at the 3$\sigma$ level. 
Typically, the $b \rightarrow s \gamma$ is more important for $\mu < 0$,     
but it is also relevant for $\mu > 0$, particularly when tan$\beta$ is large.

\subsubsection{The anomalous moment of the muon.}

We have also taken into account the SUSY contributions to
the anomalous magnetic moment of the muon, 
$\delta a_{\mu}= a_{susy} - a_{SM}$ \cite{Degrassi:1998es}.
We used in our analysis the recent experimental results 
 \cite{g-2}, as well as the most recent
theoretical evaluations of the Standard Model contributions
\cite{newg2}.
An excess of about 2.7 sigmas between experiment and theory is found
when $e^+e^-$ data are used to estimate $a_{SM}$, leaving room for
a SUSY contribution of $a_{susy}=( 25.2 \pm 9.2)\times 10^{-10}$,
or, at the two sigma level, $6.8 < a_{susy} \time 10^{10} < 43.6$. 
Such a contribution
favors $\mu > 0$ and rather light sleptons and gauginos. However,
this slight discrepancy is smaller if tau data are used instead
to evaluate $a_{SM}$. We therefore do not restrict the parameter space
with the $\delta a_{\mu}$ constraint, but show the relevant contour
$a_{SUSY}=6.8 \times 10^{-10}$ instead.

\subsubsection{The $B_s \to \mu^+ \mu^-$ branching ratio.}

Finally, we have considered the limit \cite{bmumuexp} on the
$B_s \to \mu^+ \mu^-$ branching ratio \cite{Babu:1999hn}.
The upper bound on this process  B($B_s \to \mu^+ \mu^-$) $< 2.9\times 10^{-7}$ does not constrain the parameter space of mSUGRA.
 However  it has been stressed recently that for non-universal soft terms 
the constraint can be very important \cite{ko,ko2}, especially for large $\tan \beta$ and low values of the Higgs masses. 
There is  also a strong correlation between the $B_s \to \mu^+ \mu^-$ branching ratio and  cross sections for direct \cite{ko2} and 
indirect \cite{Mambrini:2005vk} detection of dark matter.

\subsection{LHC}

The LHC is a $pp$ collider with center of mass energy of
$\sqrt{s}=$ 14 TeV which is expected to start in 2007. 
Hadronic colliders produce mainly colored particles like
squark pairs $\tilde q \tilde q$, squark anti-squark
$\tilde q \tilde{q}^*$,  gluino pairs $\tilde g \tilde g$ or
associated squark--gluino production $\tilde q \tilde g$ :

\begin{center}
\begin{tabular}{l}
$q\overline{q}, ~ g g \xrightarrow{} \tilde{q} \tilde{q}^*$ 
\\
$qq\xrightarrow{} \tilde{q} \tilde{q}$\\
$q\overline{q}, ~g g \xrightarrow{} \tilde{g} \tilde{g}$  
\\
$qg \xrightarrow{} \tilde{q} \tilde{g}$\\
\end{tabular}
\end{center}

The $\tilde q \tilde q$ final state requires initial state
of the form $q \overline{q}$ or $g g$ whereas squark pair
are only produced from $qq$ state. Gluino pairs come from
$q \overline{q}$ and $gg$ states and the squark--gluino are only
produced via quark--gluon collisions. 
Cross sections for squark and gluino productions are very high at LHC, 
e.g. for $m_{\tilde q} = m_{\tilde g} = 500$ GeV,
 $\sigma(\tilde q- \tilde g)\sim$ 62 pb. For an integrated
luminosity of 100 $\mathrm{fb}^{-1}$, 
corresponding to one year of LHC running at high 
luminosity, 6.2 millions squark-gluino pairs are thus expected to be produced,
leading to a "fast" (assuming detectors are well understood) discovery
and to hints on the underlying SUSY model. Off course, for heavier 
spectrum, cross sections will be lower, but in any case, the production of 
squarks and gluino at the LHC, if kinematically allowed, should be important.

The decays of squarks and gluinos lead to multi-jets + isolated leptons +
missing $E_T$ signals. We consider the exclusion limits of
reference\cite{Charles:2001ka} which establish that squarks and gluinos could be
detected up to $m_{\tilde{q} - \tilde{q}} \sim 2-2.5$ TeV for the first
two generations of squarks, which nearly corresponds 
to the parton-parton kinematics limit is roughly $14/3$ TeV. 
The detection of the third
generation of squarks (sbottom $\tilde b_1$ and stop $\tilde t_1$) appears
to be more difficult in hadronic collider due to their special decay 
modes \cite{Boehm:1999tr}.

\subsection{Sparticle production in $e^+ e^-$ colliders.}

We also analyzed the prospects for producing SUSY particles and heavy
Higgs bosons at high--energy and high luminosity $e^+ e^-$ colliders
 \cite{ILC}.
In this exploratory study we will assess the accessibility of certain
production modes simply through the corresponding total cross section,
without performing any background studies. However, in most cases the
clean experimental environment offered by $e^+ e^-$ colliders
should allow discovery of a certain mode, given a sample of a few
dozen events. Difficulties might arise in some narrow regions of parameter space, 
which we will point out in the following discussion. 
We have taken the example of a future International Linear Collider (ILC)
with center of mass energy of 1 TeV and an integrated luminosity of
500 $\mathrm{fb^{-1}}$. We will consider a given channel to be visible
if its total cross section exceeds $\sigma_{\mathrm{min}}=0.1$ fb,
which correspond to a sample of 50 signal events per year.

\begin{figure}[t]
\begin{center}
 \begin{tabular}{ccccc}
\psfrag{ep}[c][c]{\scriptsize $e^+$}
\psfrag{em}[c][c]{\scriptsize $e^-$}
\psfrag{Z}[c][c]{\scriptsize $Z,\gamma$}
\psfrag{H}[c][t]{\scriptsize $\tilde{e},\tilde{\tau}$}
\psfrag{A}[t][rb]{\scriptsize $\tilde{e},\tilde{\tau}$}
\includegraphics[width=0.23\textwidth]{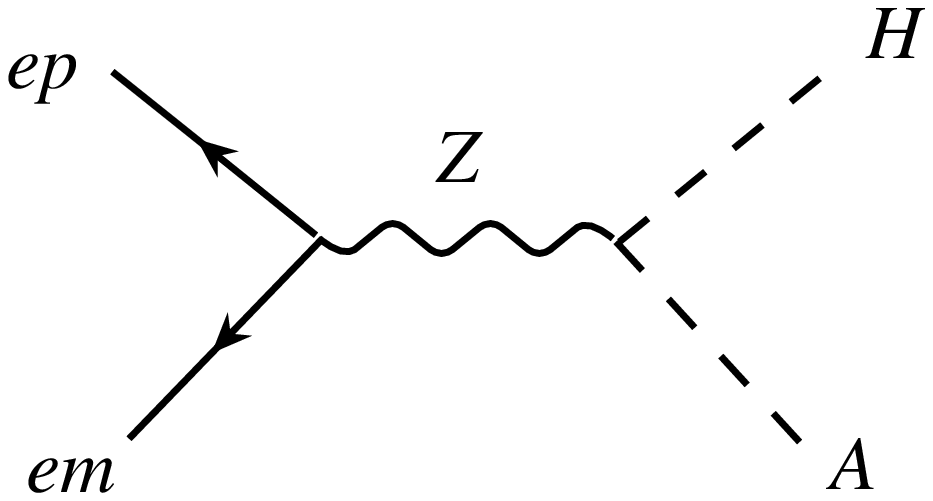}&
\psfrag{ep}[c][c]{\scriptsize $e^+$}
\psfrag{em}[c][c]{\scriptsize $e^-$}
\psfrag{ci}[c][c]{\scriptsize $\chi^0_i$}
\psfrag{sel}[c][c]{\scriptsize $\tilde{e}$}
\psfrag{sel}[c][c]{\scriptsize $\tilde{e}$}
\includegraphics[width=0.13\textwidth]{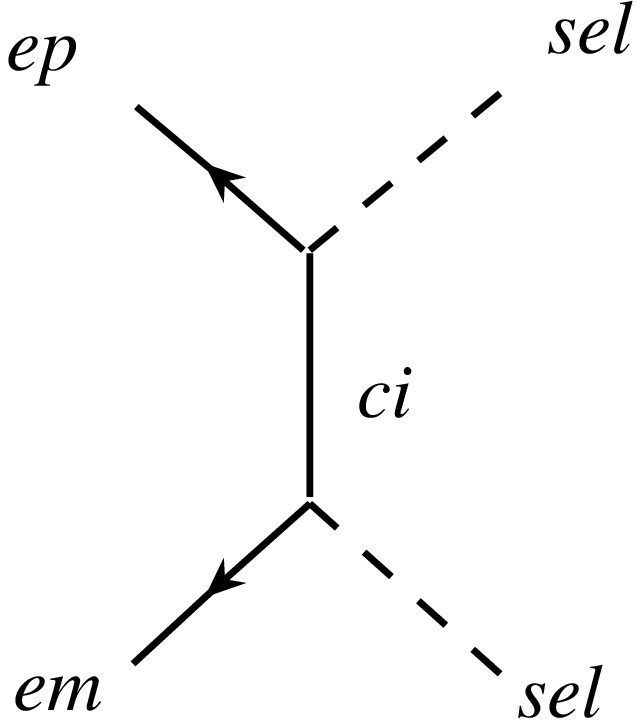}&
\psfrag{ep}[c][c]{\scriptsize $e^+$}
\psfrag{em}[c][c]{\scriptsize $e^-$}
\psfrag{Z}[lc][c]{\scriptsize $Z( + \gamma)$}
\psfrag{ci}[c][t]{\scriptsize $\chi^0_i (\chi^+_i)$}
\psfrag{cj}[t][rb]{\scriptsize $\chi^0_j (\chi^-_j)$}
\includegraphics[width=0.23\textwidth]{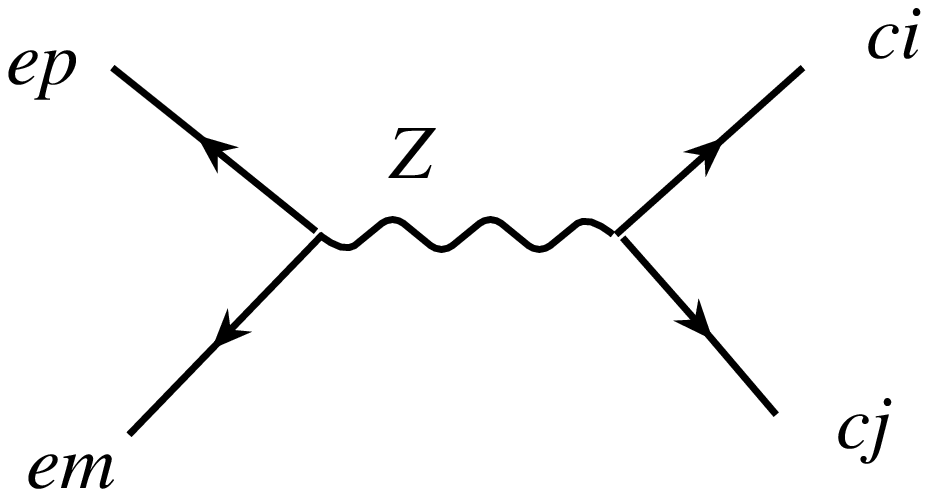}&
\psfrag{ep}[c][c]{\scriptsize $e^+$}
\psfrag{em}[c][c]{\scriptsize $e^-$}
\psfrag{sel}[c][r]{\scriptsize $\tilde{e}(\tilde{\nu})$}
\psfrag{ci}[c][t]{\scriptsize $\chi^0_i (\chi^+_i)$}
\psfrag{cj}[c][rb]{\scriptsize $\chi^0_j (\chi^-_j)$}
\includegraphics[width=0.13\textwidth]{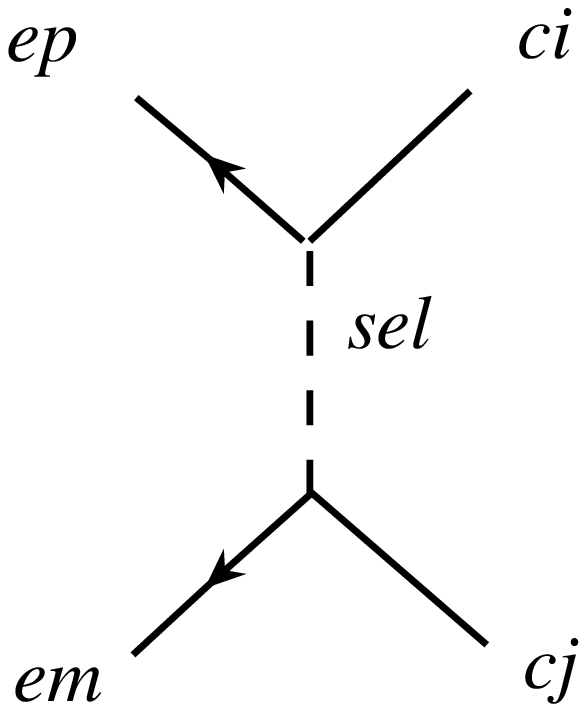}&
\psfrag{ep}[c][c]{\scriptsize $e^+$}
\psfrag{em}[c][c]{\scriptsize $e^-$}
\psfrag{Z}[c][c]{\scriptsize $Z$}
\psfrag{H}[c][c]{\scriptsize $H$}
\psfrag{A}[c][c]{\scriptsize $A$}
\includegraphics[width=0.23\textwidth]{plots/epemZHA_1.ps}\\
 a) & b) & c) & d) & e)
\end{tabular}
 \caption{\small Production processes for a Linear Collider.}  
\label{LCchannels}
   \end{center}
\end{figure}

\noindent 
In our study, we will consider the following production processes,
shown on Fig. \ref{LCchannels}.

\begin{center}
\begin{tabular}{l}
$e^+e^-\xrightarrow{}\tilde{l}\tilde{l}^*$ (mainly $\tilde{\tau}\tilde{\tau}$
and $\tilde \nu \tilde \nu$)\\
$e^+e^-\xrightarrow{}{\chi^+} {\chi^-}$\\
$e^+e^-\xrightarrow{}{\chi} {\chi^0_2}$\\
$e^+e^-\xrightarrow{}HA$,\\
\end{tabular}
\end{center}

Concerning the sleptons, pairs of 
$\tilde e_{R,L}^{\pm}$ are
produced via $s$--channel photon and $Z$ boson exchange and the $t$--
channel exchange of the four neutralinos $\chi^0_i$. Since the
electron--Yukawa coupling is suppressed, only the gaugino fraction 
of the neutralinos exchanged contributes to the process. 
Thus the influence of the soft breaking gaugino masses $M_1$ $M_2$ and
$\mu$ through $M_{H_u}$ will be important in
the production cross section. The main final state will be the lightest
state, $\tilde e_R$, as in supergravity models, the 
$\tilde e_R - \tilde e_L$ mass difference can be important.
For the third generation of slepton, the production proceeds only via
$\gamma$ and $Z$ boson exchange. In this case, we will only concentrate
on the production of the lightest state, 
$e^+ e^- \rightarrow \tilde \tau_1 \tilde \tau_1$ which offers the largest
 possibilities of discovery. A look at the formulas in the appendix
of \cite{abdelsug} shows a strong dependence of the cross section on the
selectron velocity $\beta$ : only sleptons with masses of several GeV
below the kinematical limit can be observed\footnote{We can also see it
at the natural cross section suppression of spin 1 $\rightarrow$ spin 0 spin 0
processes}. Note that it is also
possible to produce and observe sleptons through their decay even if 
$m_{\tilde l} > \sqrt{s}/2$ \cite{Datta:2002mz}.

Due to the couplings and the kinematics, the sleptons will mainly decay 
into their leptonic partners and the gaugino--like neutralinos or
charginos (if allowed). In other words, the regions of low values of $\mu$
 with Higgsino--like $\chi^0_{1,2}$ and $\chi^+_1$ 
will be a blind region for the detection of sleptons. Whereas the
lighter $\tilde e_R$ will predominantly decay following
$\tilde e^{\pm}_R \rightarrow e^{\pm} \chi^0_1$, the heavier left handed
$\tilde e_L$ will decay into wino--like chargino $\chi^{\pm}_1$ or neutralino
$\chi^0_2$ because these processes occur via the SU(2) coupling, much
stronger than the $\mathrm{U(1)_{Y}}$ involved in  
$\tilde e^{\pm}_R \rightarrow e^{\pm} \chi^0_1$.

The charginos are produced through $s$--channel photon and $Z$ boson
exchange as well as $t$--channel sneutrino exchange 
(see Fig. \ref{LCchannels}). Note that the sneutrino channel contributes
with an opposite sign (see \cite{abdelsug}) to the $s$--channel diagrams.
The production will thus be maximized in regions of heavy sneutrinos
and for Higgsino--like charginos ($|\mu| \ll M_2$). For light sneutrino 
 the destructive interference can affect considerably the
cross section whereas productions of Higgsino--like charginos are mainly
insensitive to $m_{\tilde \nu}$ (the $\tilde \nu e \chi{\pm}_1$ coupling
vanish in this case). In any case, the cross section is usually rather large,
making productions possible for masses up to the kinematical threshold region.

For $M_2,~ |\mu| < $ scalar/sleptons masses, the chargino is the lighter charged sparticle and
mainly decay into $\chi^0_1~W$, with the $W$ decaying into a $ff'$ pair with
a known branching ratio. For small slepton masses, virtual slepton exchange
can enhance other processes leading to only 
$\tau^{\pm} \nu_{\tau} \chi^0_1$ final states \cite{Djouadi:2001fa}.
For large values of tan$\beta$, charged Higgs boson exchange contribution
can also enhance the branching fraction for the $\tau$ final state.

The production of the lightest neutralinos $\chi^0_{1,2}$ 
occurs via $s$--channel $Z$ boson exchange and $t$-- or $u$--channel
$\tilde e_L, ~ \tilde e_R$ exchanges. 
A gaugino--like neutralino does
not couple to the $Z$ boson. However, a high Higgsino fraction leads to an enhancement of the 
$Z \chi^0_1 \chi^0_2$ coupling and a suppression of the 
$e \tilde e_{L,R} \chi^0_{1,2}$ one proportional to the gaugino fraction of 
the neutralinos. Except in the extreme Higgsino limit, the cross 
section is much smaller than the chargino one (whose nature ensures 
a reasonable production rate for gaugino--like of Higgsino--like charginos
through $Z$ exchange). 

The decay modes of the $\chi^0_2$ depends strongly on the SUSY parameter
space, and can be completely leptonic (if the two--body decay
$\chi^0_2 \rightarrow l^{\pm} \tilde l^{\mp}$ is the main process) 
or hadronic (if $\chi^0_2 \rightarrow h \chi^0_1$ is dominant).
However, at an $e^+e^-$ collider, hadronic $\chi^0_2$ decays 
are as easy to sign as the leptonic ones. The only difficulty will be
in regions of the parameter space where the mass difference 
$m_{\chi^0_2} - m_{\chi^0_1}$ is small, where $\chi^0_2$ decays almost 
exclusively into quasi--invisible modes.

If the pseudoscalar mass is sufficiently heavy (around $\sim 200$ GeV
depending on tan$\beta$) the model is in the so--called decoupling regime
\cite{Djouadi:2005gi}, where the masses of the scalar $H$ and pseudoscalar $A$ 
(and even $H^{\pm}$ for larger $m_A$) are almost degenerate. In
this limit, both the (tree--level) coupling of the $A$ and the $H$ 
to massive vector bosons are suppressed, as the $ZAh$ one. 
The only important Higgs production process becomes thus the associated
$HA$ production through $Z$ boson exchange in the $s$--channel
(see Fig. \ref{LCchannels}) \cite{Djouadi:2005gi}. 
In any case, the cross section is suppressed
by the kinematical $\beta^3$ factor near the threshold.

If $m_A < 2 m_t$ or $\tan^2\beta> m_t / m_b$, the heavy scalar Higgs boson
$H$ and pseudoscalar $A$ will mainly decay into 
$b \overline{b}$ and $\tau \overline{\tau}$ pairs \cite{Djouadi:2005gi}.   
If the $t \overline{t}$ channel is kinematically open and for lower values of
tan$\beta$, the process $A/H \rightarrow t \overline{t}$ dominates the
 decay modes. In some region of the parameter space, decays into
SUSY particles are possible. These modes will be more difficult to
analyze in the framework of an $e^+ e^-$ collider but the signals
should be clear enough to be detectable \cite{Djouadi:1996jc}.

We draw attention to the fact that we did not include ISR in our calculation
but check its weak relative importance for the focus of the present work with
regard {\it e.g} to astrophysical uncertainties.

\section{Prospects for Discovery}

\begin{figure}[t]
\begin{center}
\begin{tabular}{cc}
\includegraphics[width=0.45\textwidth]{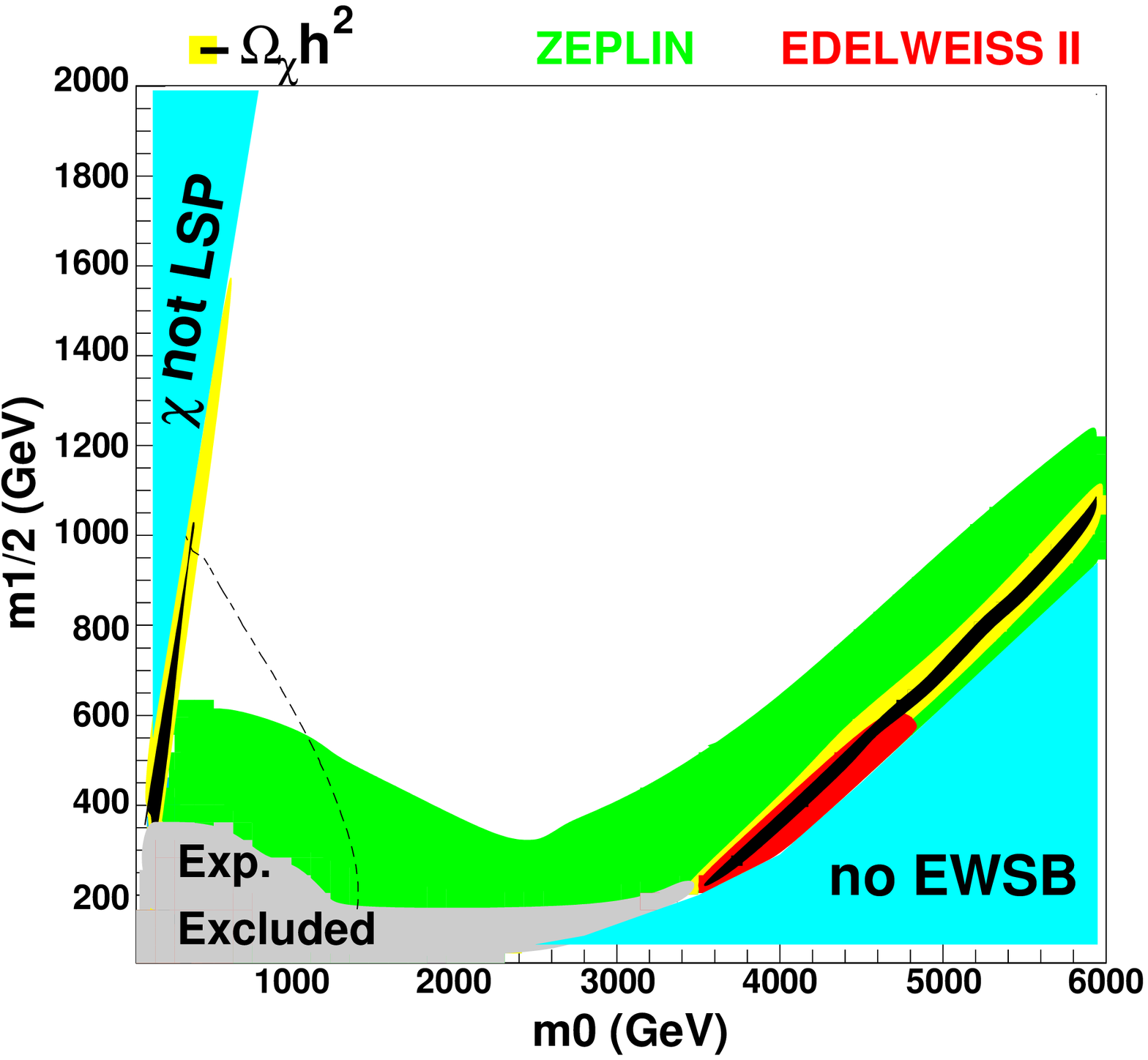}&
\includegraphics[width=0.45\textwidth]{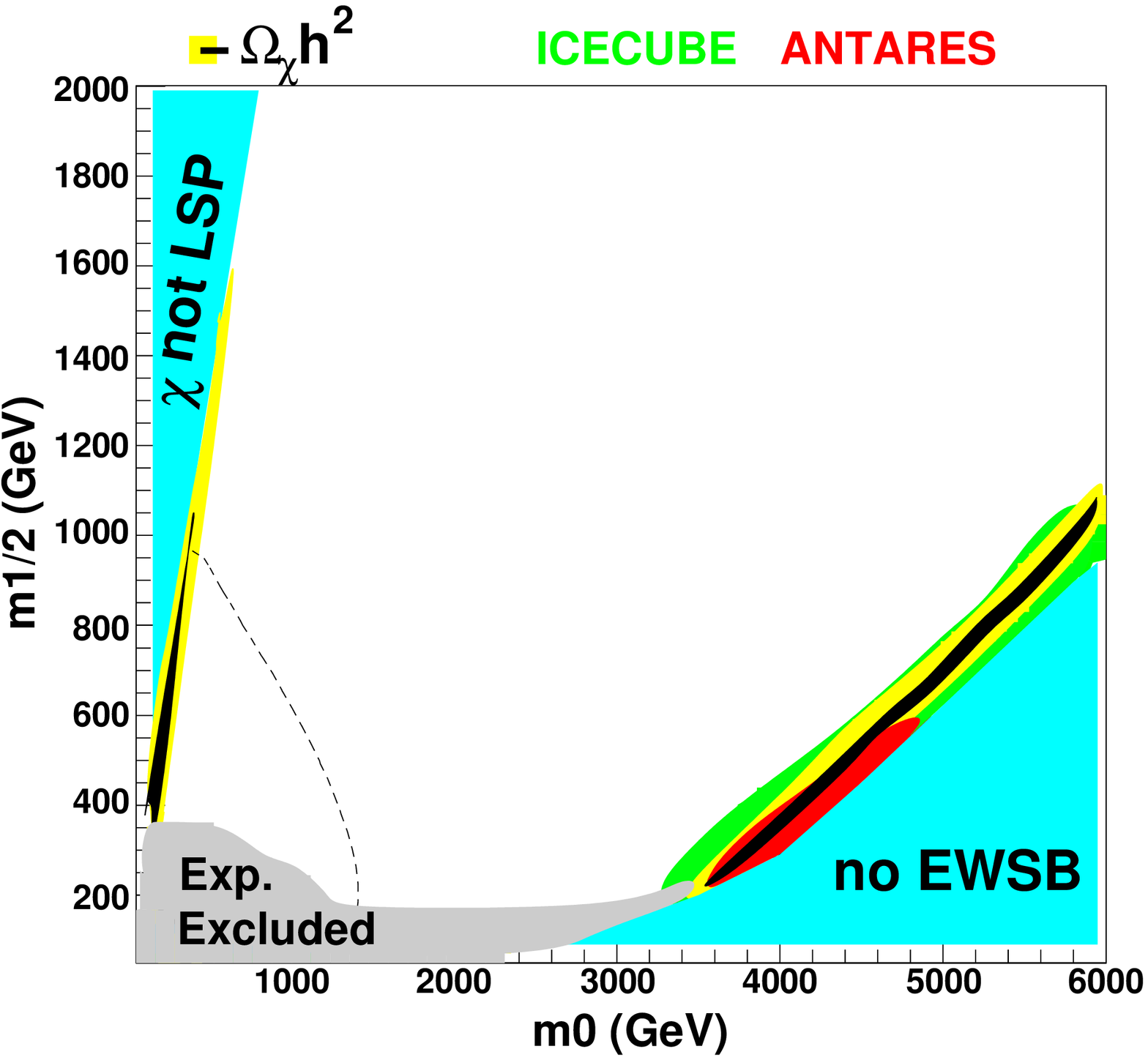}\\
a) Direct Detection & b) $\nu$ Indirect Detection (Sun) \\
\includegraphics[width=0.45\textwidth]{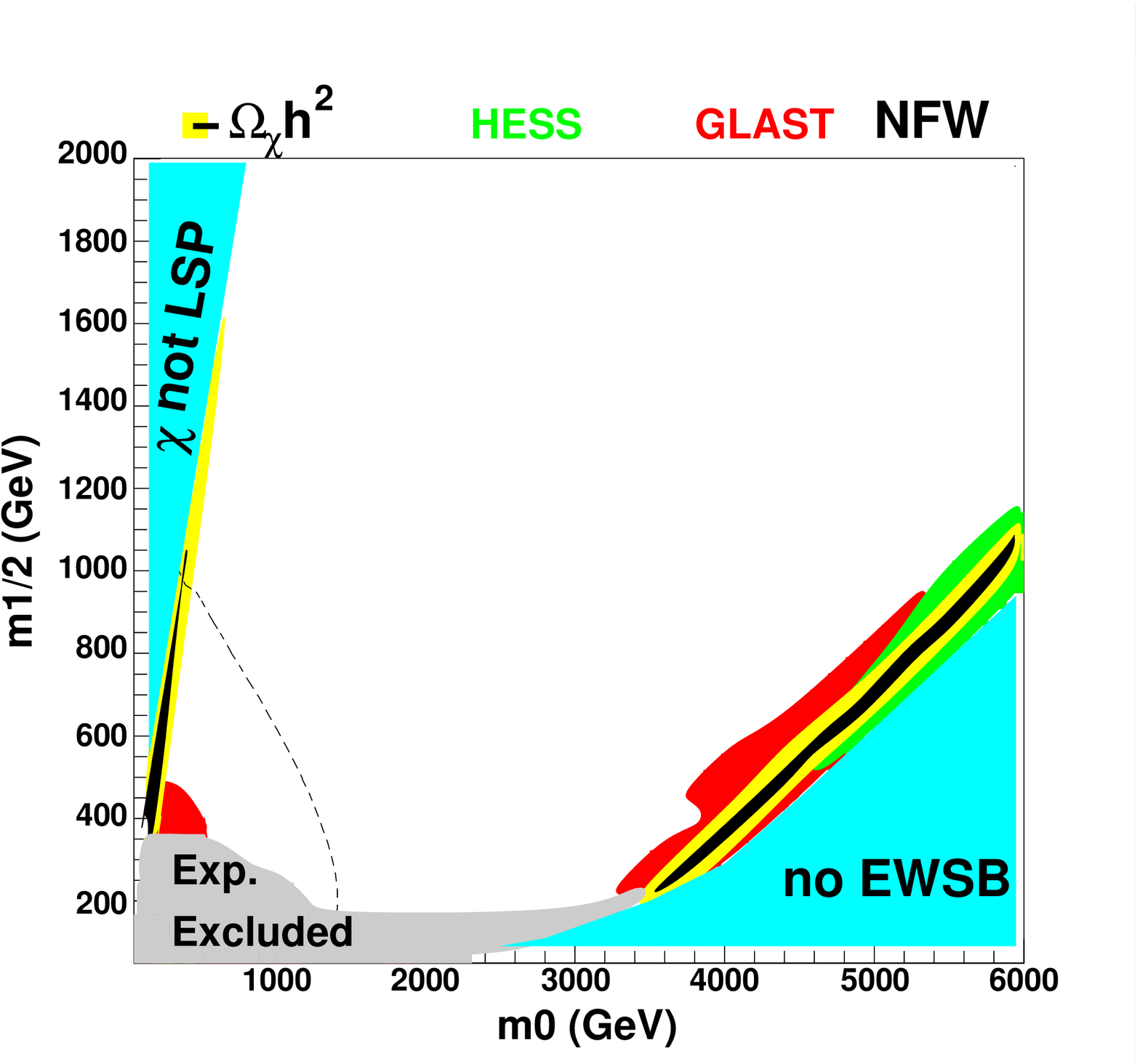}&
\includegraphics[width=0.45\textwidth]{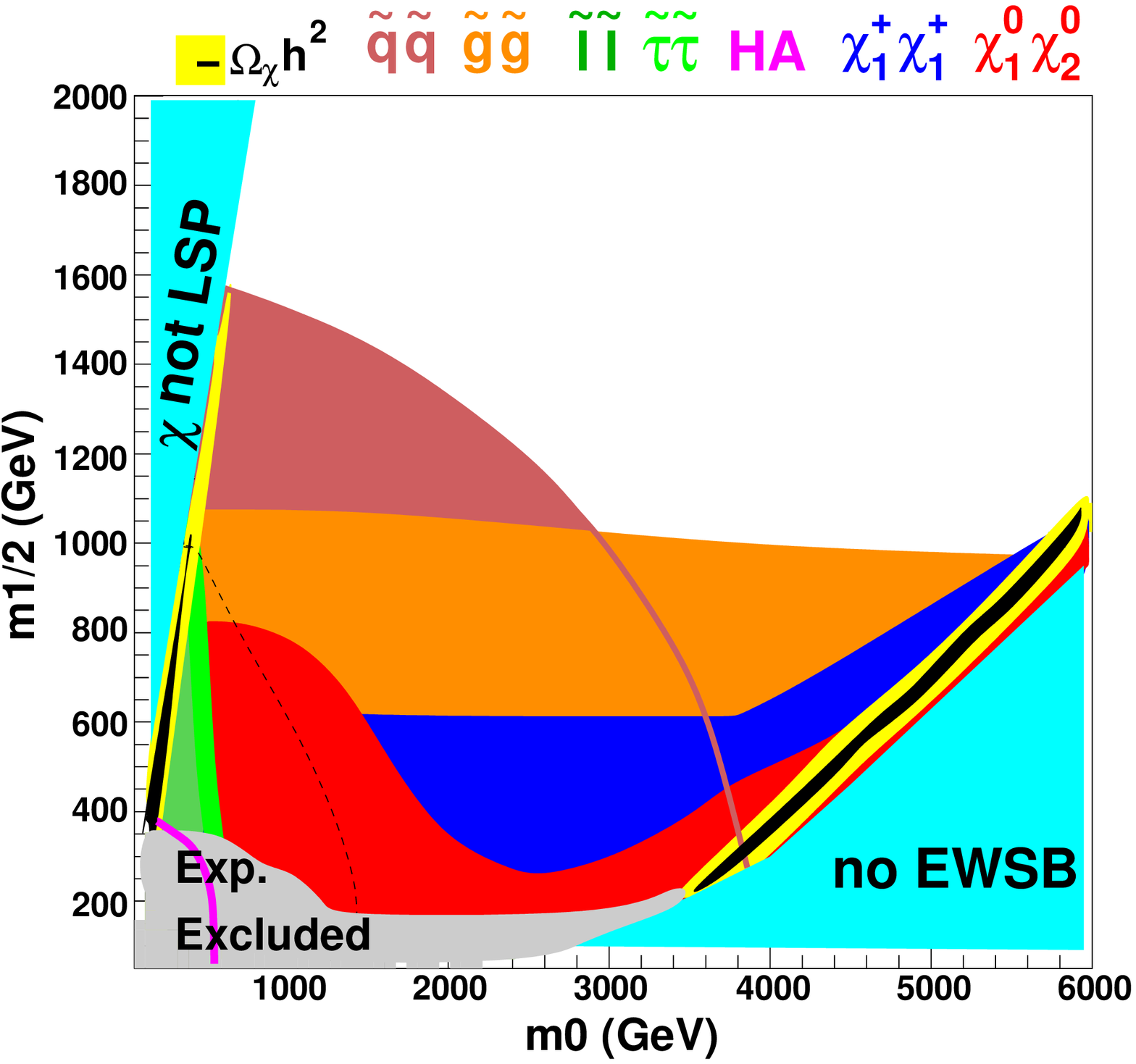}\\
c) $\gamma$ Indirect Detection (GC) & d) Collider production (LHC,ILC)\\
\includegraphics[width=0.45\textwidth]{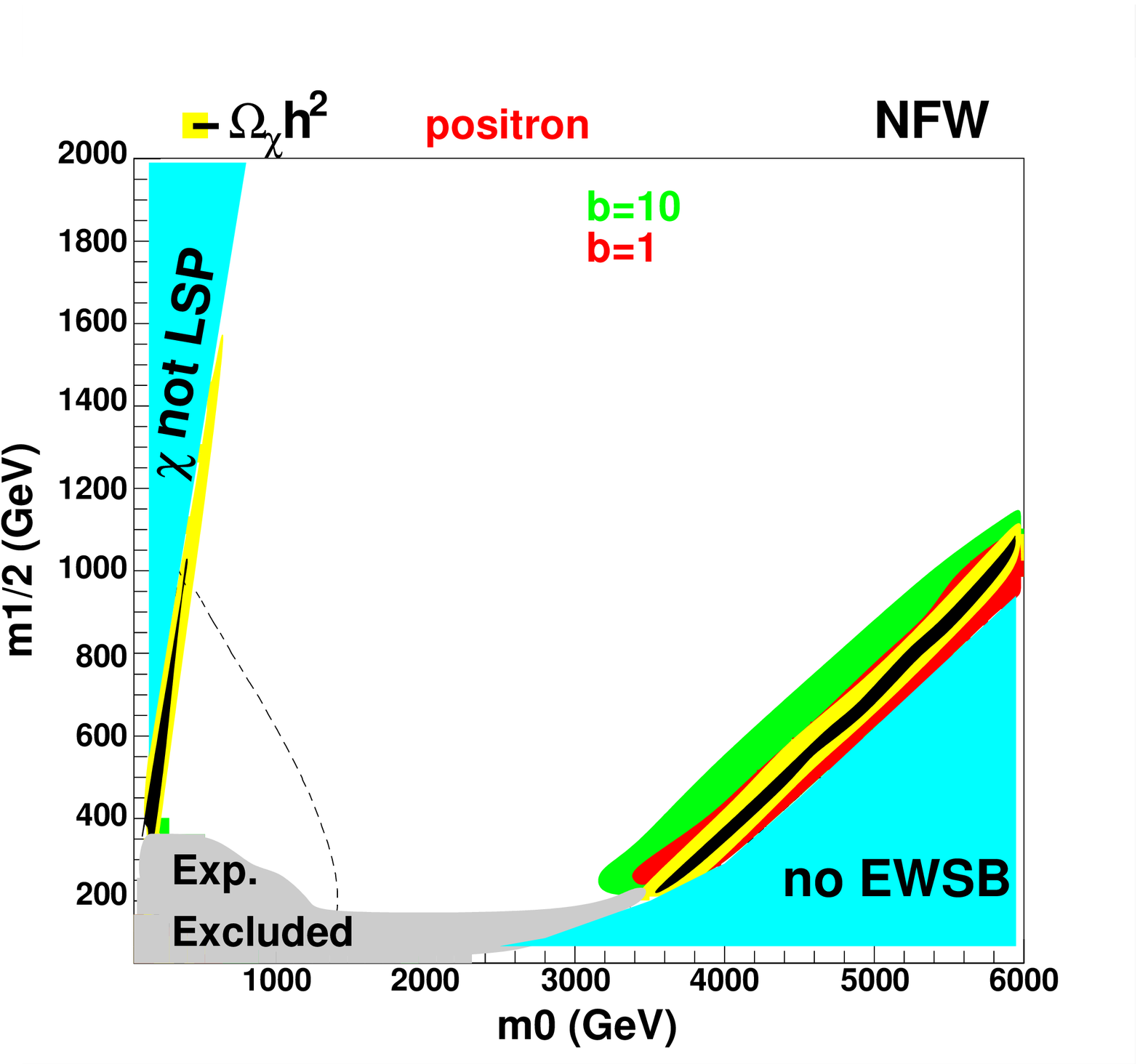}&
\includegraphics[width=0.45\textwidth]{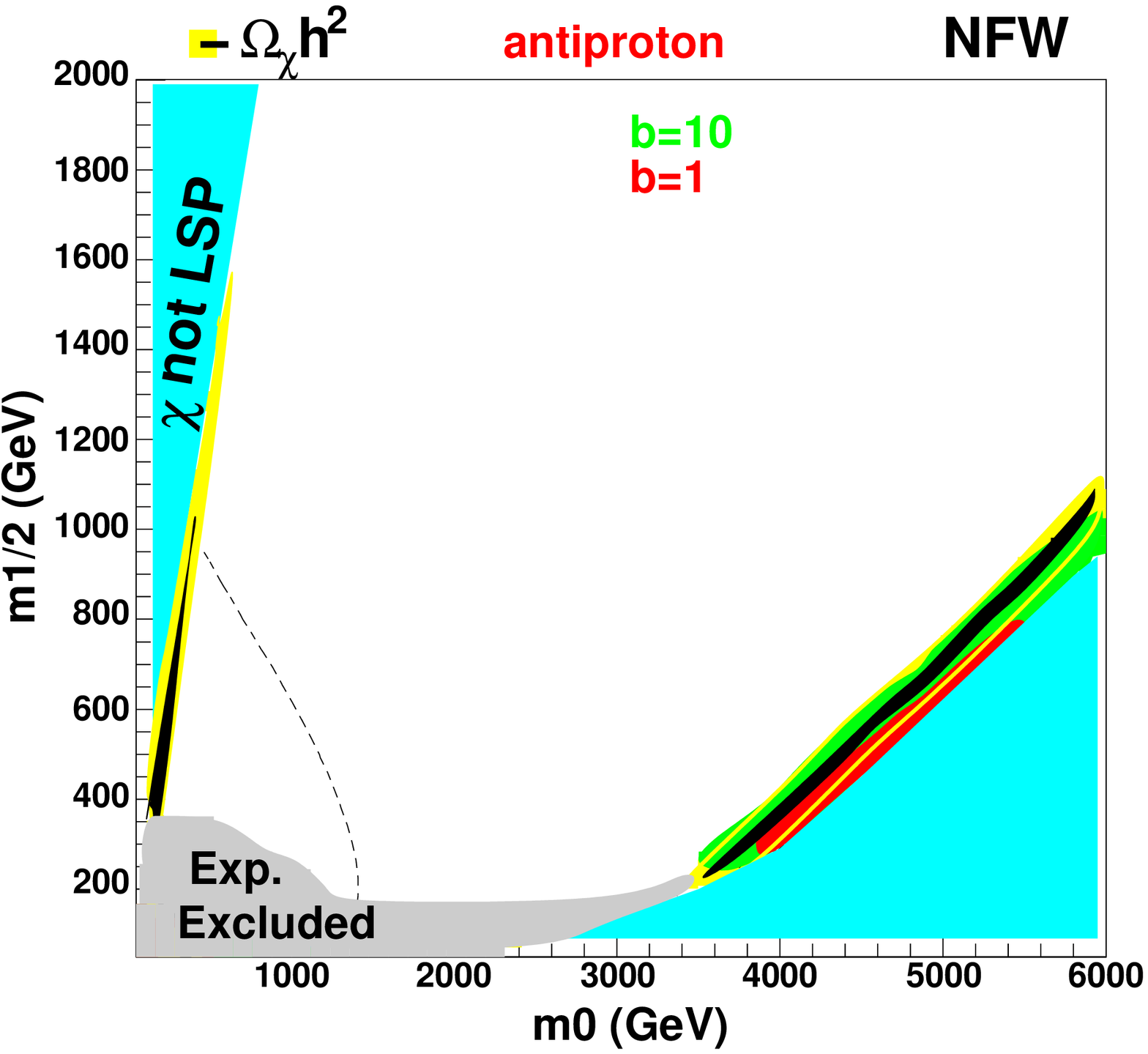}\\
e) $e^+$ Indirect Detection (halo) &f) $\bar{p}$ Indirect Detection (halo)\\
&\\
\end{tabular}
\caption{ MSUGRA Universal $A_0=0$, $\tan{\beta}=35$, $\mu>0$}
\label{fig:Utb35}
\end{center}
\end{figure}

\begin{figure}[t]
\begin{center}
\begin{tabular}{cc}
\includegraphics[width=0.45\textwidth]{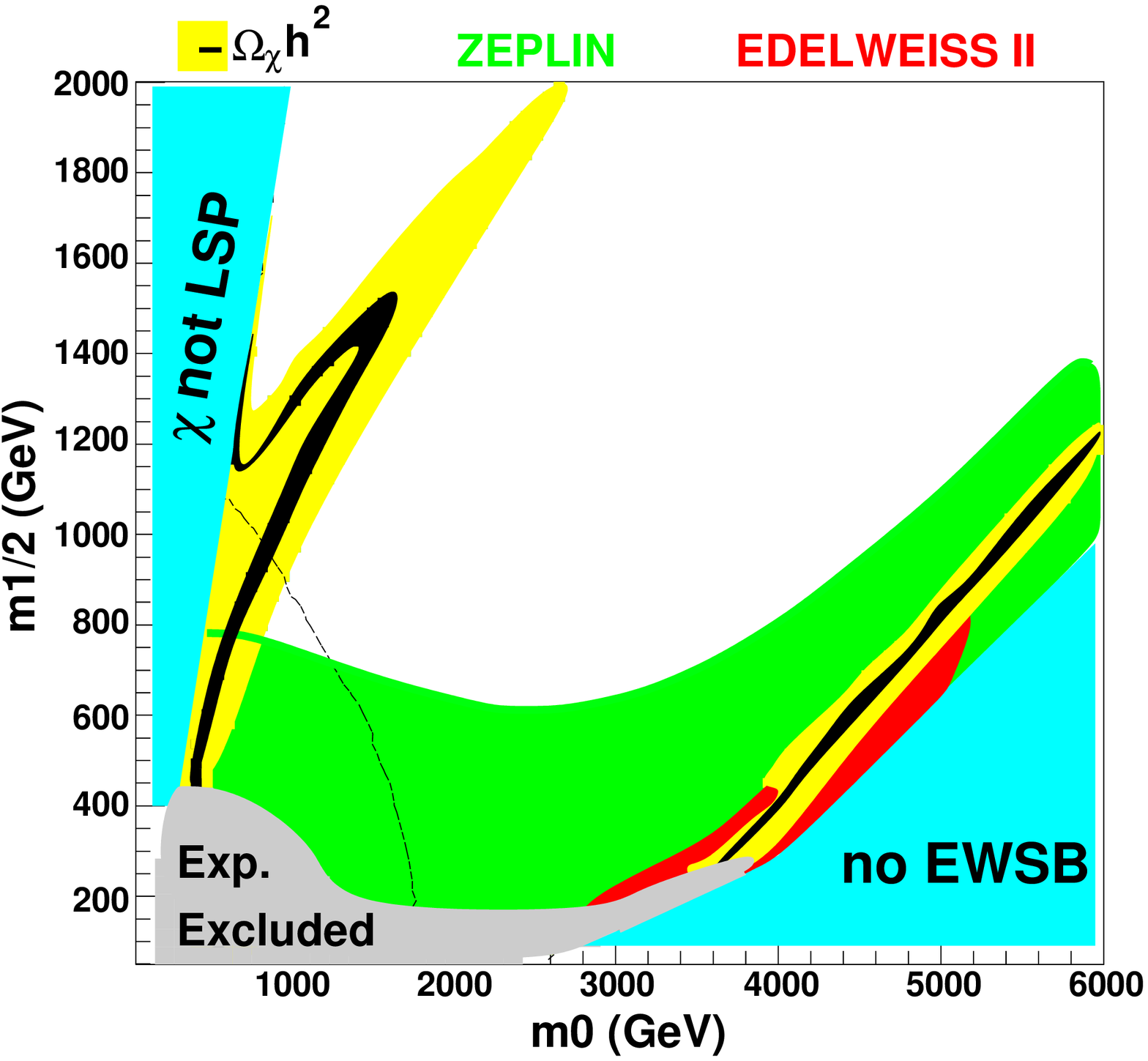}&
\includegraphics[width=0.45\textwidth]{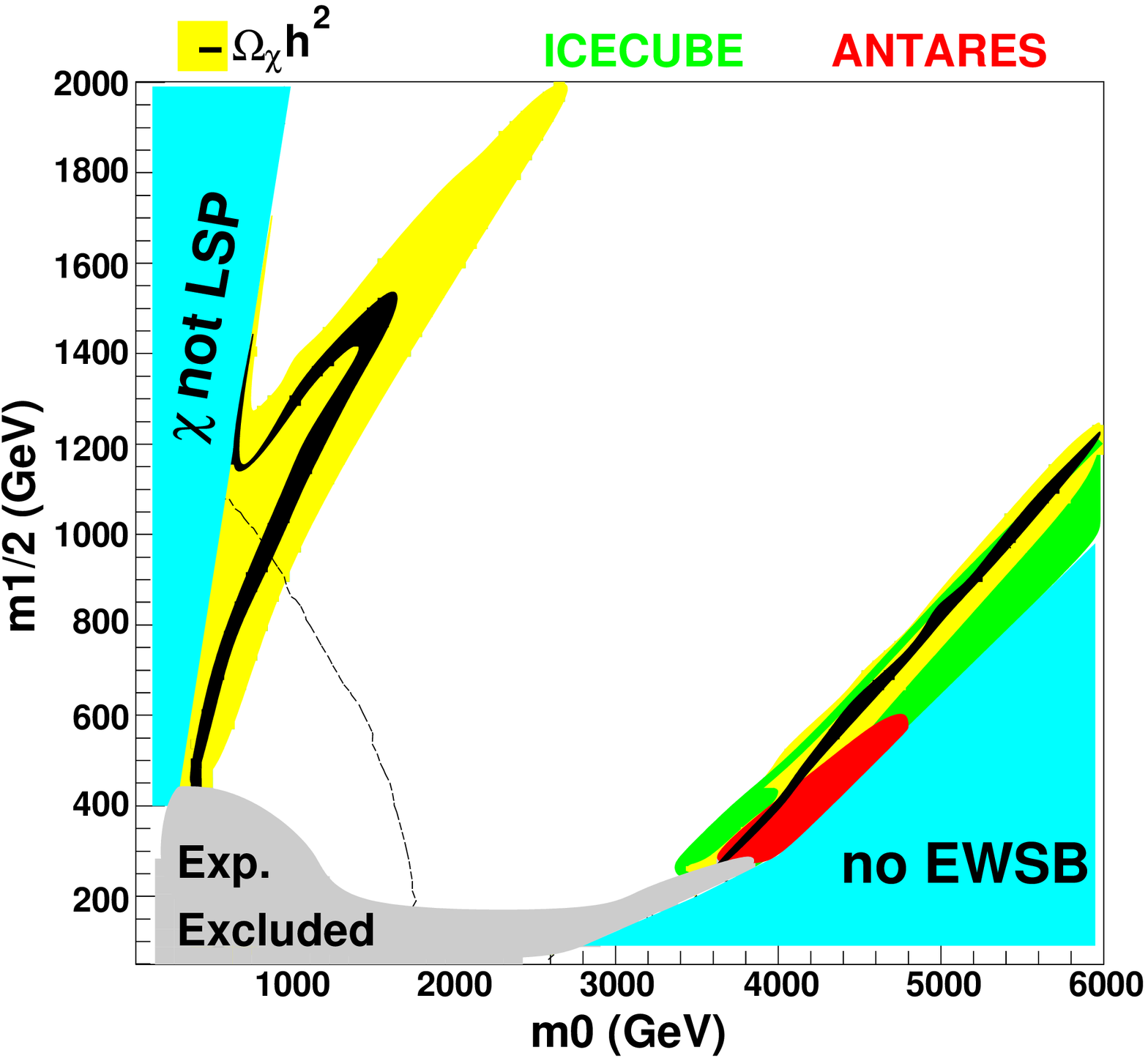}\\
a) Direct Detection & b) $\nu$ Indirect Detection (Sun) \\
\includegraphics[width=0.45\textwidth]{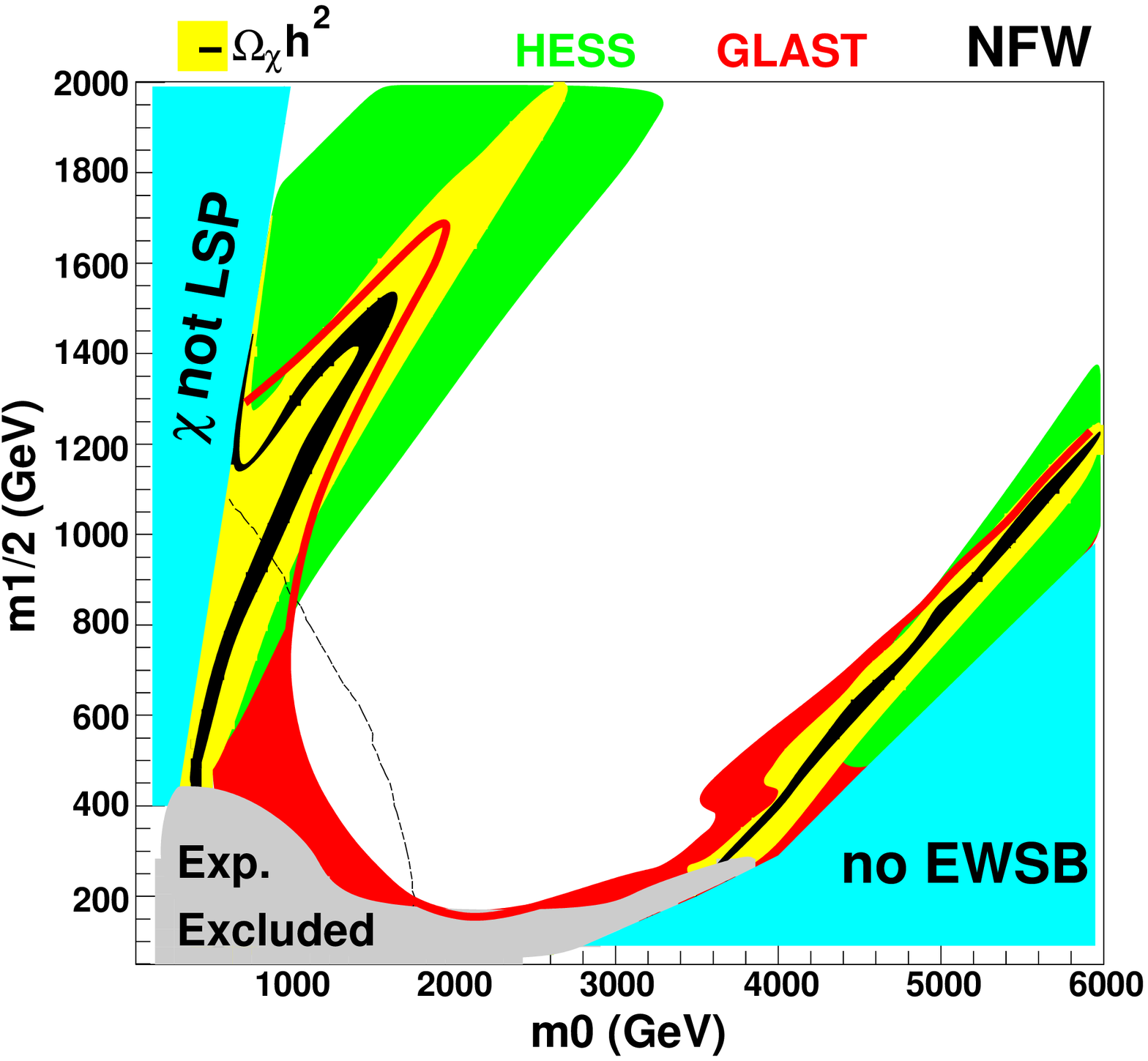}&
\includegraphics[width=0.45\textwidth]{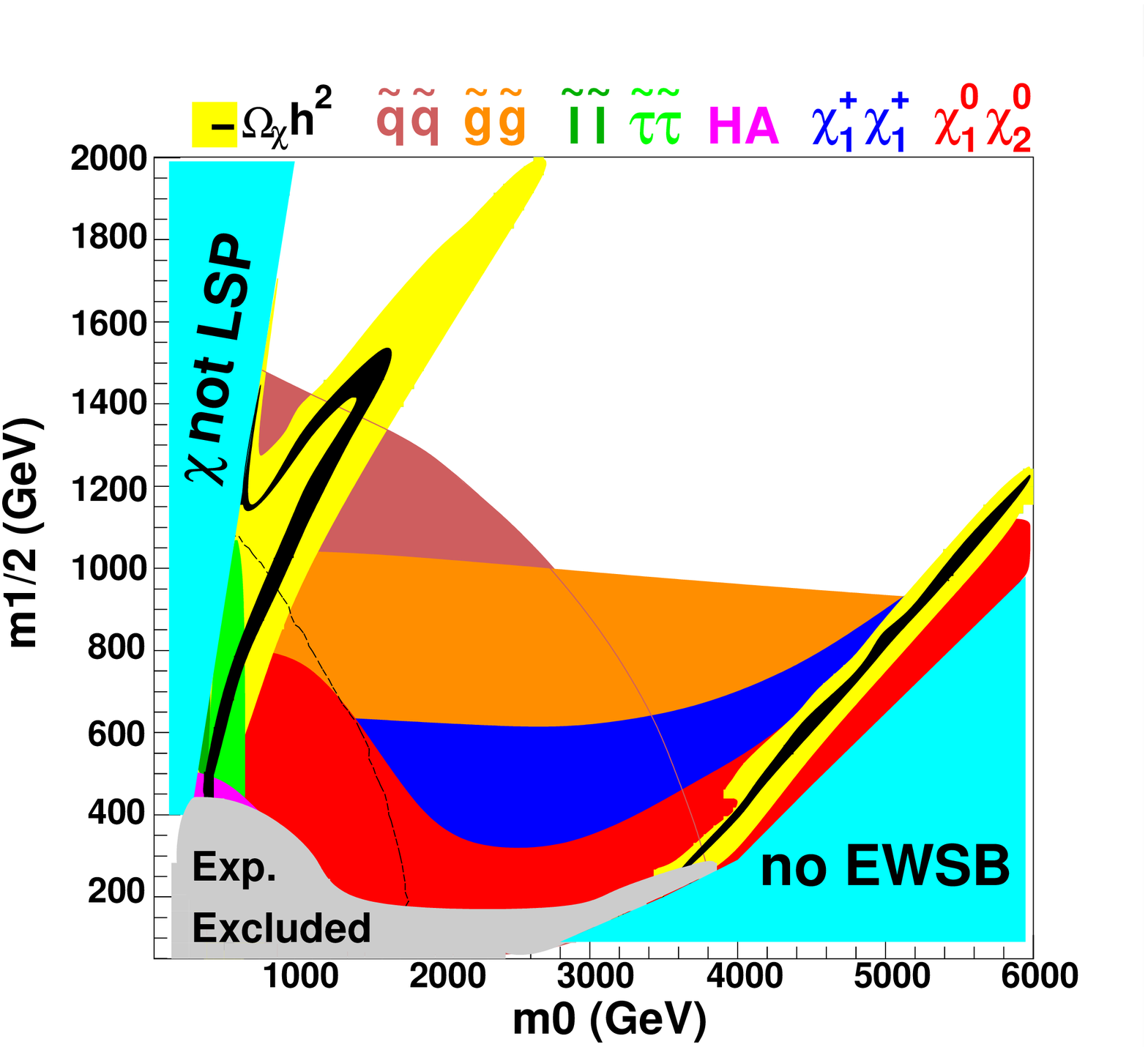}\\
c) $\gamma$ Indirect Detection (GC) & d) Collider production (LHC,ILC)\\
\includegraphics[width=0.45\textwidth]{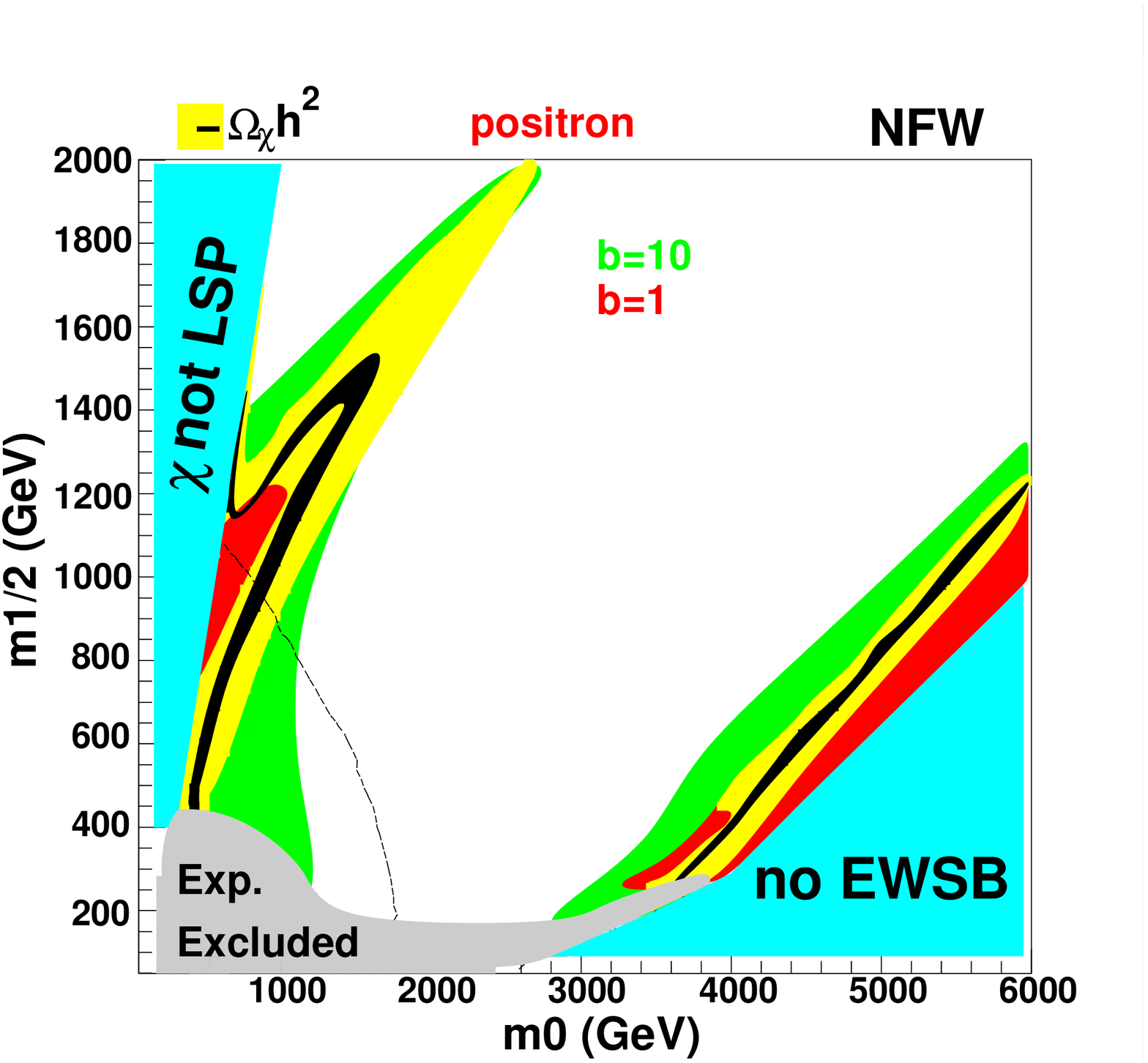}&
\includegraphics[width=0.45\textwidth]{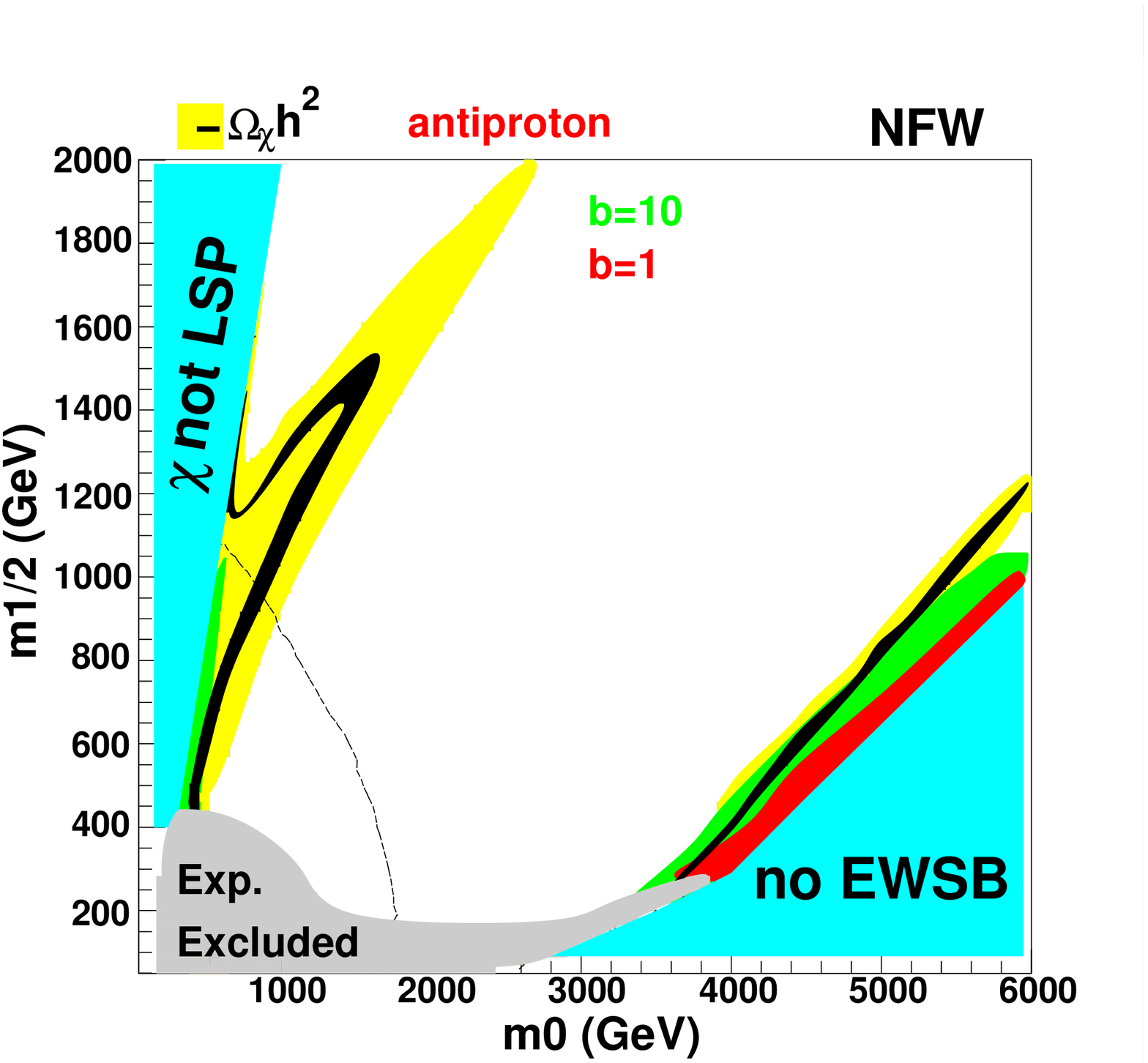}\\
e) $e^+$ Indirect Detection (halo) &f) $\bar{p}$ Indirect Detection (halo)\\
&\\
\end{tabular}
\caption{ MSUGRA Universal $A_0=0$, $\tan{\beta}=50$, $\mu>0$}
\label{fig:Utb50}
\end{center}
\end{figure}

Using the theoretical, experimental and cosmological constraints discussed
in the previous sections, we perform a full scan of the ($m_0$, $m_{1/2}$)
plane for a given value of tan$\beta$ and $A_0$, fixing the Higgsino 
parameter $\mu$ to be positive. The results are illustrated in Figs. 
\ref{fig:Utb35} to \ref{fig:MHdtb35} which show the regions allowed
by the different constraints we imposed in universal (Fig. \ref{fig:Utb35}, 
\ref{fig:Utb50}),
gaugino non--universal (Fig. \ref{fig:M2tb35}, \ref{fig:M3tb35}),
and scalar non--universal (Fig. \ref{fig:MHutb35}, \ref{fig:MHdtb35})
scenarios. We present also the regions of the parameter space 
which will be accessible in a near future for typical experiments  
of the different kinds of detection discussed above. 
The influence of other external free parameters ($m_t$ and galactic profiles)
is illustrated in Fig. \ref{fig:topmoorekra}.

The areas excluded or disfavored by the experimental constraints 
are shown in grey. For the anomalous moment of the
muon, the black dashed lines corresponds to  $\delta a_{\mu}=6.8\times 10^{10}$  which decreases in the direction of increasing $m_0$. 
The cosmologically favored relic density range 
$0.03<\Omega_{\chi}{\rm h}^2<0.3$ is shown in yellow 
(very light grey) and the WMAP~\cite{WMAP} constraint,
Eq.(1.1) is the
internal black region inside the yellow (very light grey) area.
Our starting parameter space is the Universal mSUGRA/CMSSM plane, where one
assumes  a unified gaugino and scalar mass at the GUT scale ($m_{1/2}$ and $m_0$
respectively). We first choose $A_0=0,\tan{\beta}=35,\mu>0$
and perform a full scan of the ($m_0$, $m_{1/2}$) plane :
$0< m_0 < 6000$ GeV, $0 < m_{1/2} < 2000$ GeV
. We will then point out the effects of non universal mass terms in 
gaugino and Higgs sectors (wino
mass $M_2|_{GUT}$, gluino $M_3|_{GUT}$, up-type Higgs mass $M_{H_u}|_{GUT}$ and
down-type Higgs mass $M_{H_d}|_{GUT}$) as well as $\tan{\beta}$ and $m_t$.

\subsection{Universal case}

For intermediate values of $\tan{\beta}$, there are mainly two regions 
leading to a favored neutralino relic abundance. The first one is at
low $m_0$ where the lighter stau $\tilde \tau_1$ is almost degenerate
 with the neutralino, and the $\tilde \tau_1 \chi$ as well as 
$\tilde \tau_1 \tilde \tau_1$ co--annihilations are efficient enough 
to reduce the relic density.  The second one is along
the boundary where the electroweak symmetry breaking cannot be achieved
radiatively ( Hyperbolic Branch/Focus Point (HB/FP): high $m_0$ corresponding
to a low $\mu$). 
In this region the neutralino is mixed bino-Higgsino,
enhancing $\chi \chi$ annihilation through $Z$ exchange and 
$\chi \chi_1^+,\chi \chi_2^0$ coannihilations. 
Those two regions are generically thin and fine tuned.
Direct detection experiments are then favored for light Higgs scalar $H$   
(mainly low $m_0,m_{1/2}$) or around the HB/FP region 
 where the Higgsino
 fraction is sufficient to increase significantly the scattering cross section
on the nucleus and allow an observation in an experiment like ZEPLIN 
(see Fig. \ref{fig:Utb35}a). 
Concerning indirect detection with neutrino telescopes,
 a significant signal from the Sun requires a large Higgsino fraction to
enhance the spin dependent interaction $\chi q\xrightarrow{Z} \chi q$ .
 This can only take place in the HB/FP branch where a 
${\rm km^3}$ size detector like
ICECUBE will be able to probe models satisfying the WMAP constraint 
(see Fig. \ref{fig:Utb35}b).

Gamma indirect detection of neutralino in the galactic center
requires efficient annihilation cross section. The possible processes are either 
$\chi\chi \xrightarrow{A}b\bar{b}$ which is favored in region
where the pseudo--scalar $A$ is light (low $m_0,m_{1/2}$),
or/and when the $\chi\chi A$ coupling 
($\propto z_{11(2)}z_{13(4)}$) is enhanced through the Higgsino fraction 
in the HB/FP branch. In this region, the annihilation process 
$\chi\chi \xrightarrow{Z}t\bar{t}$ is also favored since the  $\chi\chi Z$ 
coupling is proportional to $z_{13(4)}^2$.
 This zone is within reach of the HESS telescope and will be covered by future satellite like GLAST as we can clearly
 see in Fig \ref{fig:Utb35}c. This figure should be compared to
 Fig. \ref{fig:topmoorekra}c and \ref{fig:topmoorekra}d to keep in mind 
the importance of the halo profile assumption we made.
The positron and antiproton fluxes have essentially the same particle physics
 dependence than the gamma--ray fluxes through the annihilation cross section
 factor $\langle \sigma v \rangle$. 
The favored region for positron and antiproton are thus also located where the 
neutralino annihilation is strong and an experiment of the type of
PAMELA should be able to detect any signal from this region for a
sufficiently large boost factor 
(see Figs. \ref{fig:Utb35}e and \ref{fig:Utb35}f). 

Prospects for producing SUSY particles and heavier Higgs bosons at
future colliders is shown on Fig. \ref{fig:Utb35} d). 
LHC will be efficient in the parameter space where the 
particles charged under SU(3) are light : light squarks $\tilde{q}$ 
( low $m_0$ values $\lesssim 2-2.5 $ TeV) and/or light gluinos $\tilde{g}$  
(small $M_3$ {\it i.e}  $m_{1/2}\lesssim 1000$). 
A future 1 TeV Linear Collider
 can probe the slepton sector for light $\tilde{l}$ 
($m_0\lesssim 700$ GeV, $m_{1/2}\lesssim 1000$ GeV).
 The $\chi \chi_2^0$ (mainly bino and wino respectively) production is 
also favored for low $m_0$ through selectron exchange but
decreases when $m_{\tilde{e}}$ (mainly $m_0$) increases up to $m_0\sim 2000$
GeV where the Higgsino fraction of the neutralinos allows the $Z$ exchange 
along the EWSB boundary. The chargino production follows first the kinematics
limit of wino chargino production ($m_{1/2}\sim 600$ GeV,
$2 m_{\chi^+_1}\simeq 2 \times 0.8 \times m_{1/2}\simeq 1$ TeV) and then reaches
 higher $m_{1/2}$ values thanks to the Higgsino component of $\chi^+_1$ along 
the EWSB boundary at high $m_0$. The region which can give a sufficiently high
rate of $HA$ production is restricted to the lower left corner of the plane
and is already experimentally excluded.

Non zero value for the trilinear coupling $A_0$ term mainly
affects third generation sfermion masses through its splitting. Thus, it has not
direct consequences for dark matter searches like direct detection and neutrino
indirect detection for which essentially first generation of quark--squarks 
coupling from proton scattering are
involved. Annihilation can be enhanced with a positive non--zero
value of the trilinear coupling through
$\tilde{\tau},\tilde{b},\tilde{t}$ exchange which can be of interest for
$\gamma,e^+,\bar{p}$ indirect detection.\\
It can also favored $\chi\tilde{\tau}$ ($\chi\tilde{b}$,$\chi\tilde{t}$)
coannihilations processes. 
Those region are difficult for dark matter searches but can be
reach for favorable astrophysics scenario.
If the extreme case, $\chi\tilde{q}$ coannihilation region leads to 
signal difficult to be detected at LHC (essentially missing $E_T$ and few jets)
but on the other hand the possibility of lighter squarks 
($\tilde{b} \ {\rm or} \ \tilde{t}$) can favor the LHC perspectives, especially
in the low $m_0$ region. 
Such trilinear mixing also favors a discovery at a 1 TeV Linear Collider   
through the production of lighter stau at low $m_0$.
We should notice here also that $A_0=m_0$ pushes away focus point region.

The value of $m_t$ is also essential for the position or the existence of
this region which strongly depends on the top Yukawa coupling. We show the relic density and collider situation for
$m_t=178$ and 182 GeV on Fig.\ref{fig:topmoorekra}a and \ref{fig:topmoorekra}b
respectively. For $m_t=178$ GeV, one needs to extend the $m_0$ range up to
9 TeV to get the no EWSB boundary, but it is not enough for
$m_t=182$ GeV where we can find this region but for even larger values
of $m_0$ (around 20 TeV) re--opening the question of fine tuning.
As a consequence, the range shown on
Fig.\ref{fig:topmoorekra}b does not contain any region with interesting
relic density. Considering the collider capabilities, the gaugino
chargino/neutralino concerned regions are extended as the Higgsino region is
pushed away or absent.

High value of $\tan{\beta}$ ($\sim 50$) leads to light Higgses $A,H$. 
This can open a Higgs funnel which decreases the relic density. 
It enhances annihilation and favors $\gamma,e^+,\bar{p}$ indirect detection. 
A lighter scalar Higgs $H$ also increases direct detection rate. 
High  $\tan{\beta}$ also enhances the splitting
in Isospin$=1/2$ sfermion mass matrix  favoring LHC discovery {\it e.g}
 in case of lighter $\tilde{b}$ squark, whereas 
lighter stau $\tilde{\tau}$, $H,A$ favor their production in a
Linear Collider as it is clearly shown in Fig. \ref{fig:Utb50}.

\subsection{Gaugino sector}

\subsubsection{The wino mass : $M_2|_{GUT}$}

\begin{figure}[t]
\begin{center}
\begin{tabular}{cc}
\includegraphics[width=0.45\textwidth]{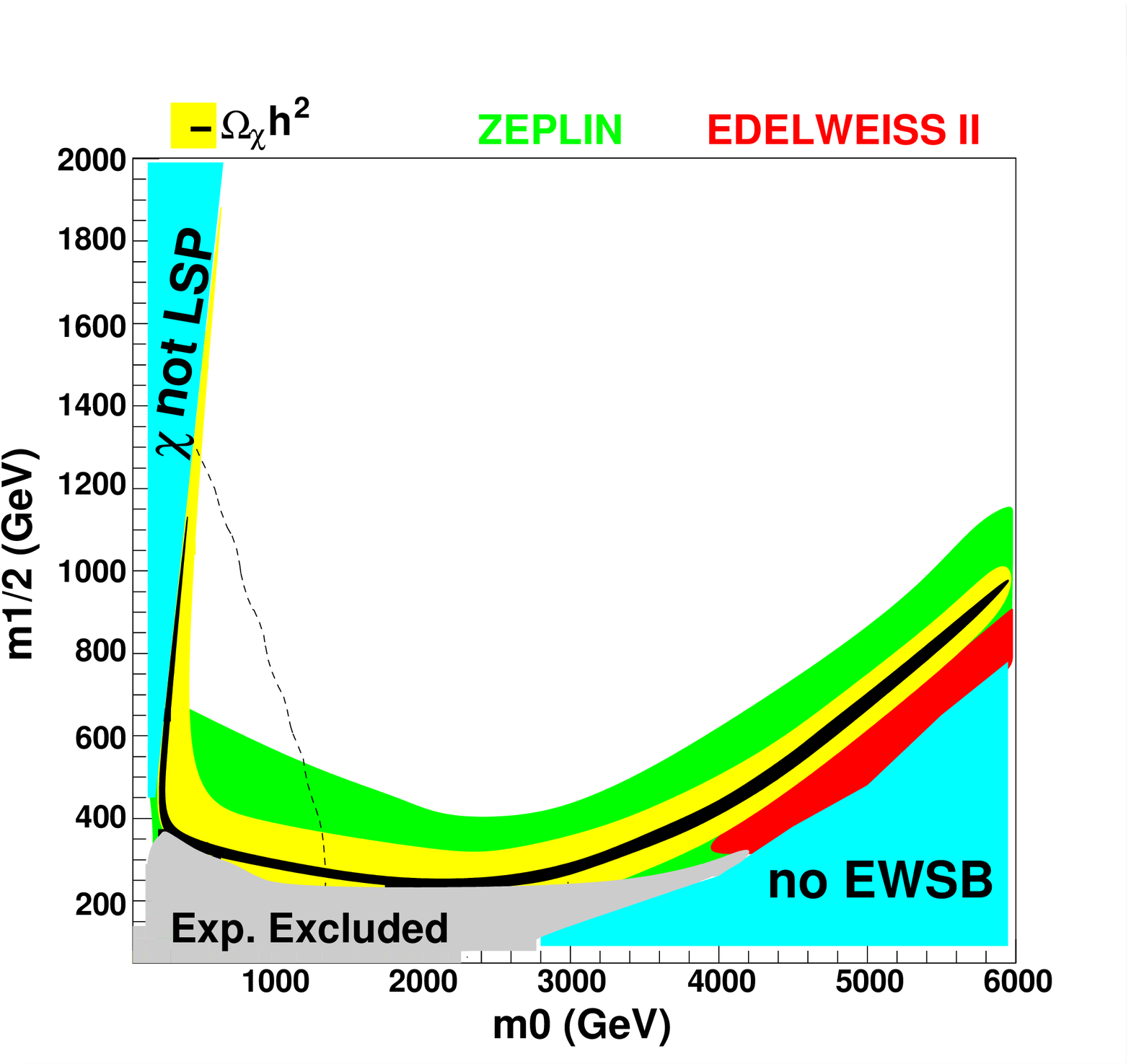}&
\includegraphics[width=0.45\textwidth]{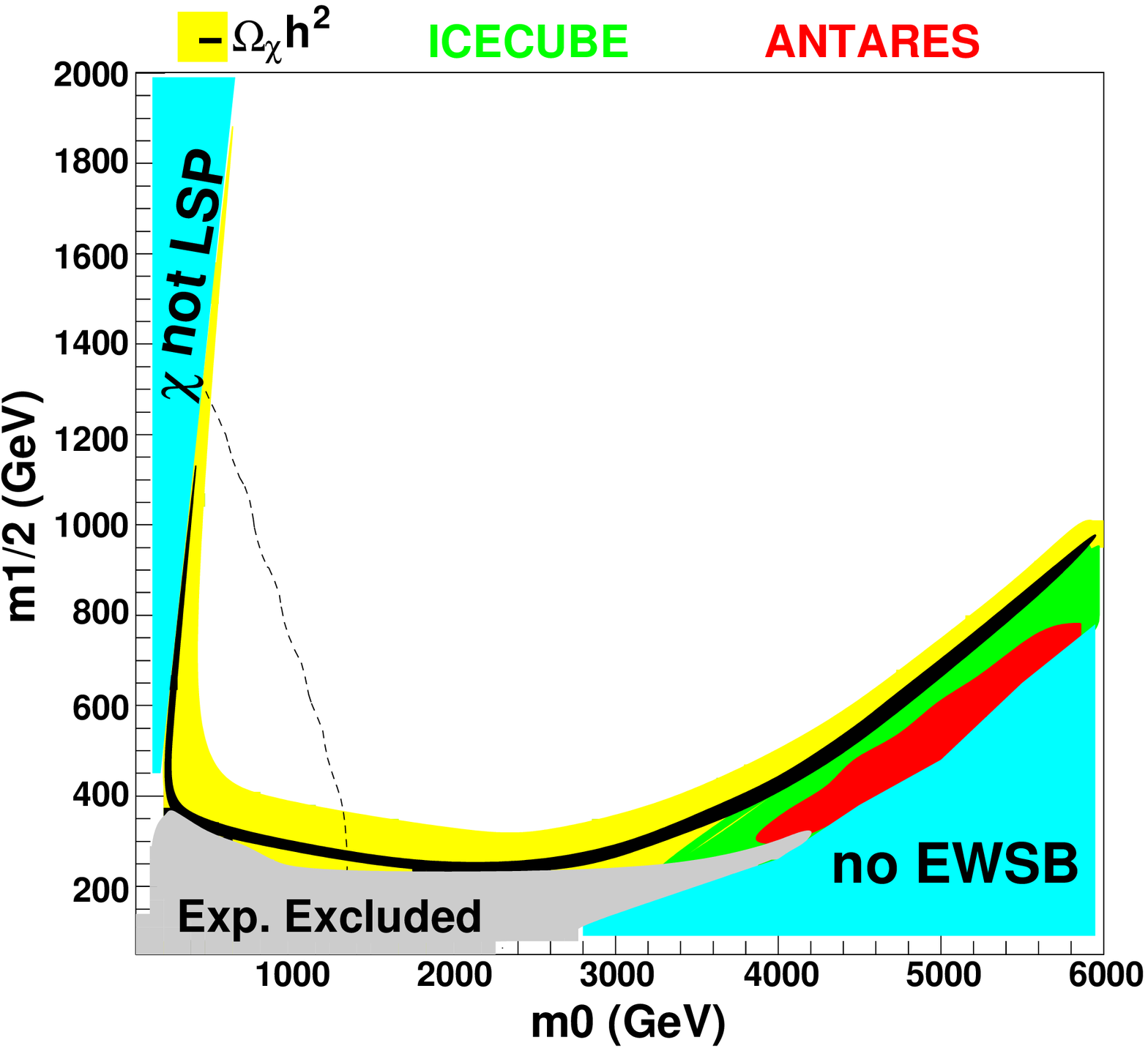}\\
a) Direct Detection & b) $\nu$ Indirect Detection (Sun) \\
\includegraphics[width=0.45\textwidth]{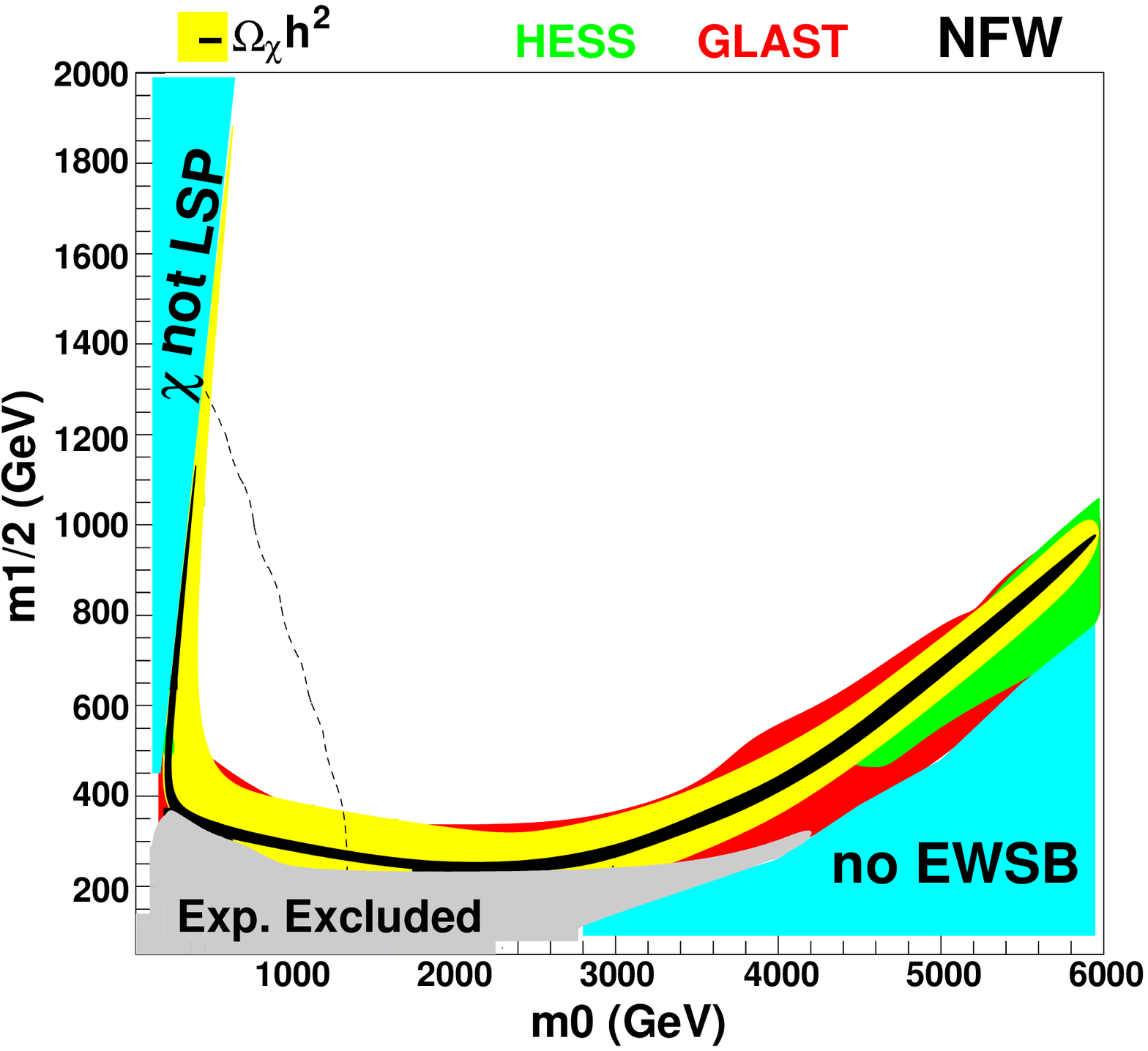}&
\includegraphics[width=0.45\textwidth]{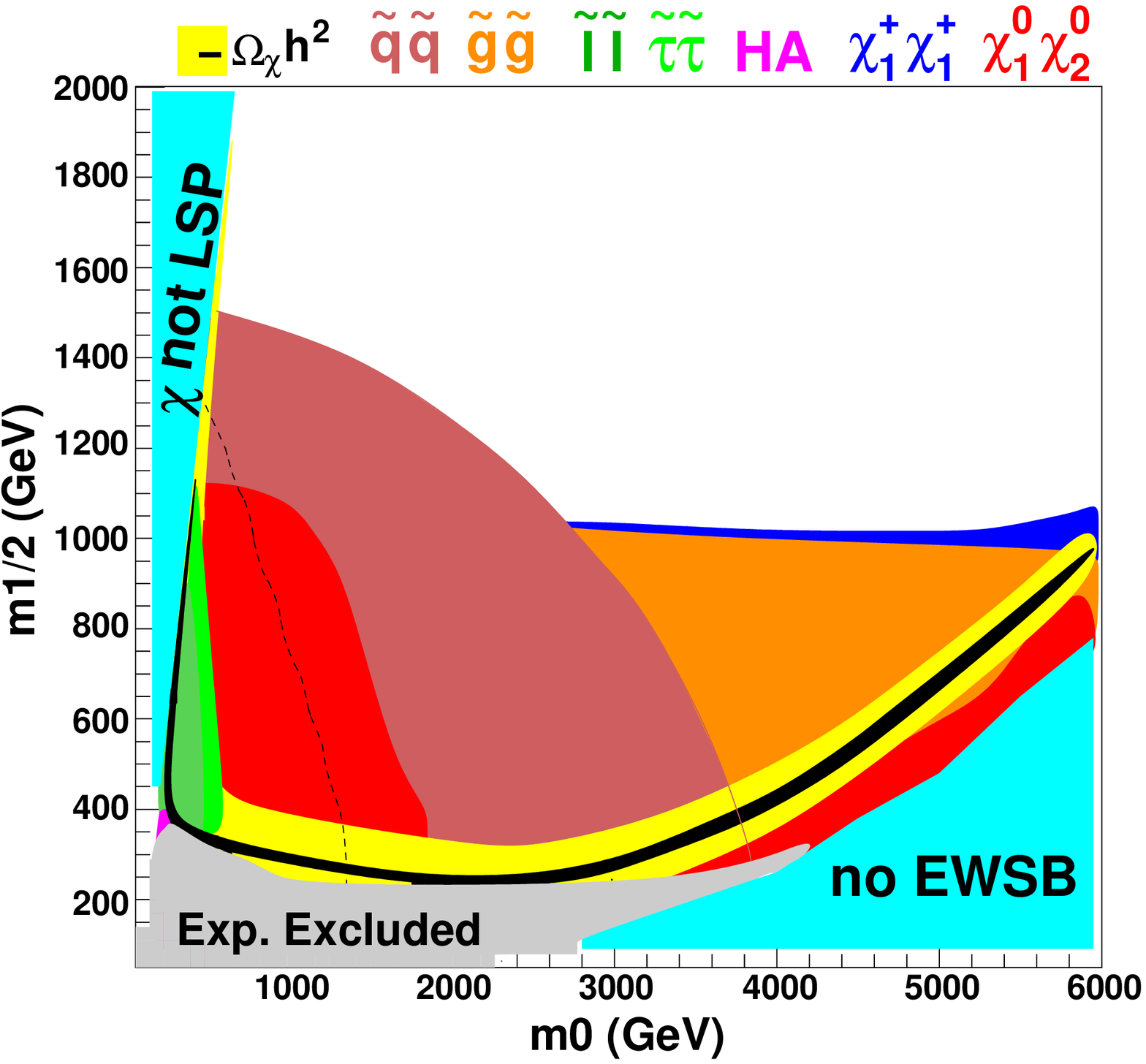}\\
c) $\gamma$ Indirect Detection (GC) & d) Collider production (LHC,ILC)\\
\includegraphics[width=0.45\textwidth]{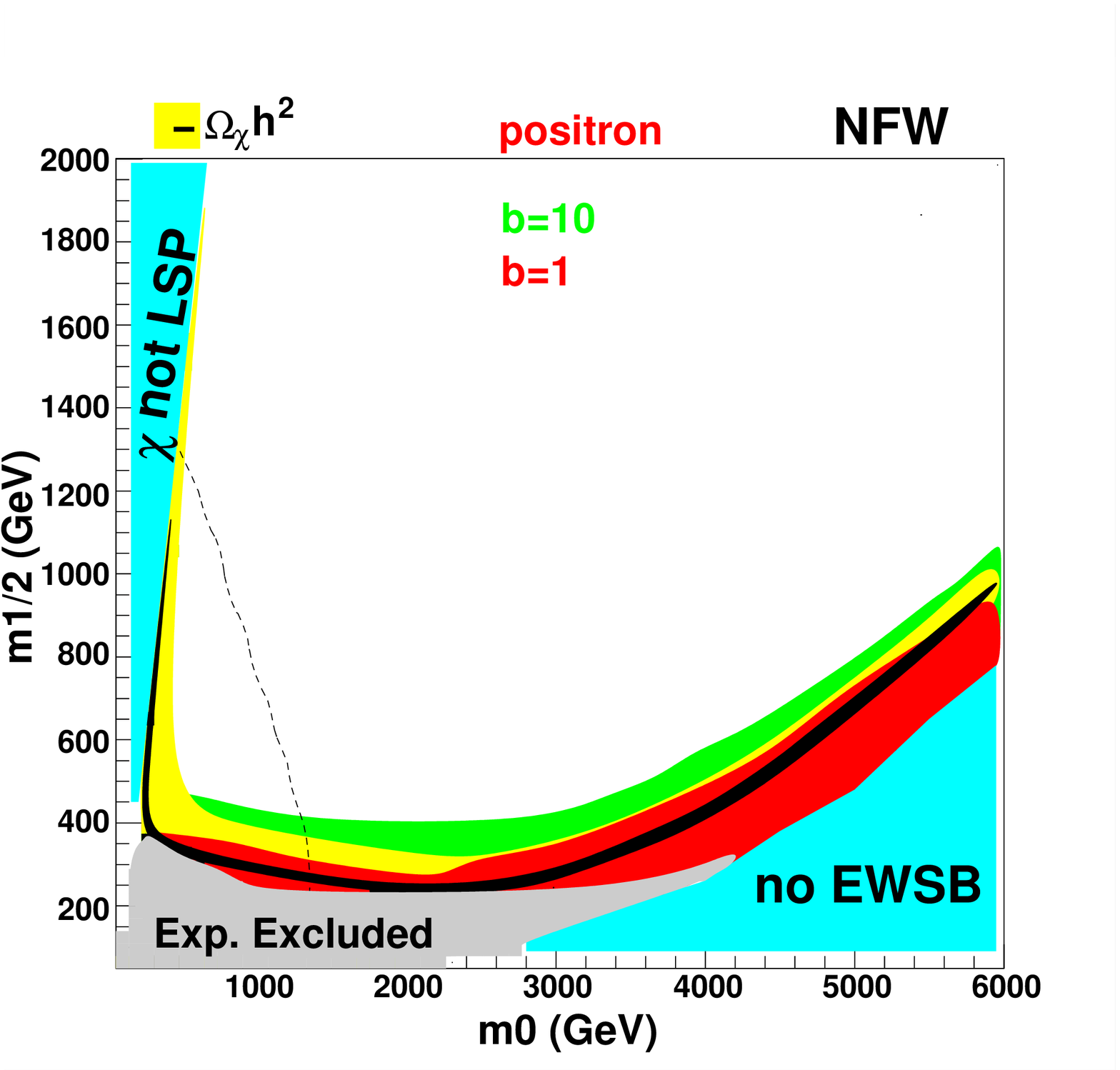}&
\includegraphics[width=0.45\textwidth]{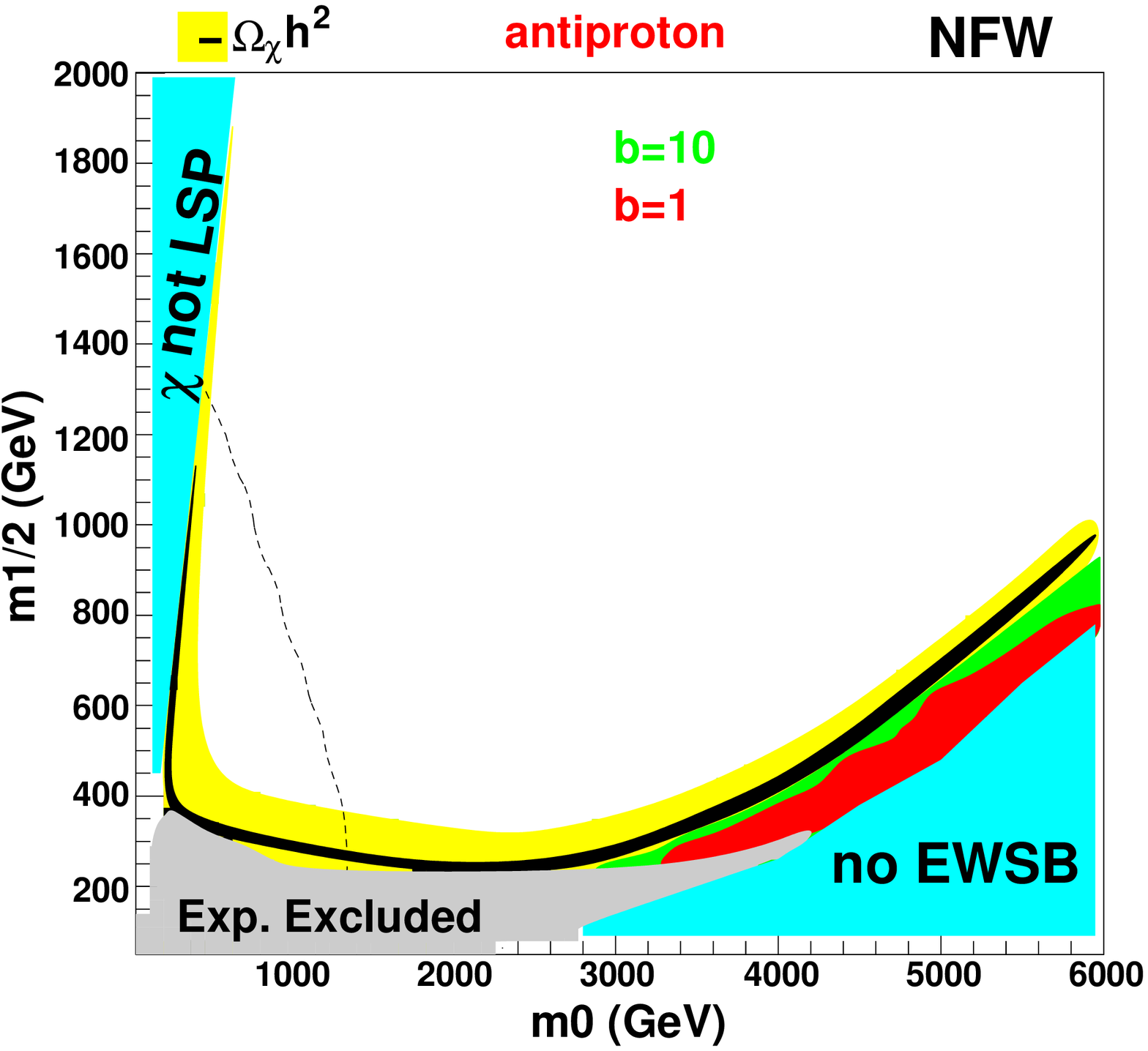}\\
e) $e^+$ Indirect Detection (halo) &f) $\bar{p}$ Indirect Detection (halo)\\
&\\
\end{tabular}
\caption{ MSUGRA non Universal $M_2|_{GUT}=0.6m_{1/2}$, $A_0=0$, $\tan{\beta}=35$, $\mu>0$}
\label{fig:M2tb35}
\end{center}
\end{figure}

We show in Fig. \ref{fig:M2tb35} the effects of non--universality of
the gaugino breaking mass term $M_2$. Other authors in the literature
has already underlined the phenomenological and cosmological effects of 
such pattern of the breaking mass terms \cite{Ellis:2003eg} 
Decreasing $M_2|_{GUT}$ increases the wino content of the neutralino which
decreases strongly the relic density \cite{BirkedalnonU,Mynonuniv,Birkedal-Hansen:2001is}. A near WMAP value
is obtained for an almost equal amount of bino and wino $i.e$ 
$M_1\simeq M_2$ at the SUSY breaking scale, requiring $M_2 \simeq 0.6 m_{1/2}$ 
at the GUT scale \cite{Mynonuniv}.
Direct detection is favored through better couplings in the diffusion
cross section (no $\tan{\theta_W}$ suppression with regard 
to bino coupling).
Gamma, positron and antiproton indirect detections are also made
easier because of the large fluxes coming from strong annihilation 
$\chi \chi \xrightarrow{}W^+W^-$ when $m_{\chi} > m_W$ 
and the enhancement of the $\chi \chi A$ coupling for the $s$--channel
$A$ exchange.
Concerning neutrino indirect detection, the wino component has no effect on
capture in the Sun but the annihilation can give harder neutrino spectrum
from $W^+W^-$ decays.
The situation at LHC is the
same as for the universal case. The Linear Collider perspective is very good
because of lighter neutralino and chargino through their wino component.
 The $\chi\chi^0_2,\chi^+\chi^-$ can thus be produced for higher values of 
$m_{1/2}$. 
One has to be aware that a smaller $M_2/m_{1/2}$ ratio at GUT scale can lead 
to $\chi \chi^+_1$ and $\chi \chi^0_2$ degeneracies which can affect the
detection procedure. 
It is important to keep in mind that the numerical computation  of
$\Omega h^2$ is very sensitive to the
 wino fraction in $\chi$. Experiment
 perspectives are thus weak in regions of parameter space 
satisfying WMAP constraints, with low $M_2$ at GUT scale. 
We show on Fig, \ref{fig:M2tb35} the
 results for $M_2/m_{1/2}=0.6$ at GUT scale but we mention that for
 $M_2/m_{1/2}=0.55$ the {\it whole} ($m_0,m_{1/2}$) plane has
 $\Omega_{\chi} h^2 <0.1$ but the regions accessible by dark matter experiments 
corresponds to regions where the relic density is too low 
($\Omega_{\chi} h^2 <0.03$). One has to increase $m_{1/2}(m_{\chi})$ to obtain
a relic density satisfying WMAP constraints with smaller $M_2/m_{1/2}$ ratio.

\subsubsection{The gluino mass : $M_3|_{GUT}$}

\begin{figure}[t]
\begin{center}
\begin{tabular}{cc}
\includegraphics[width=0.45\textwidth]{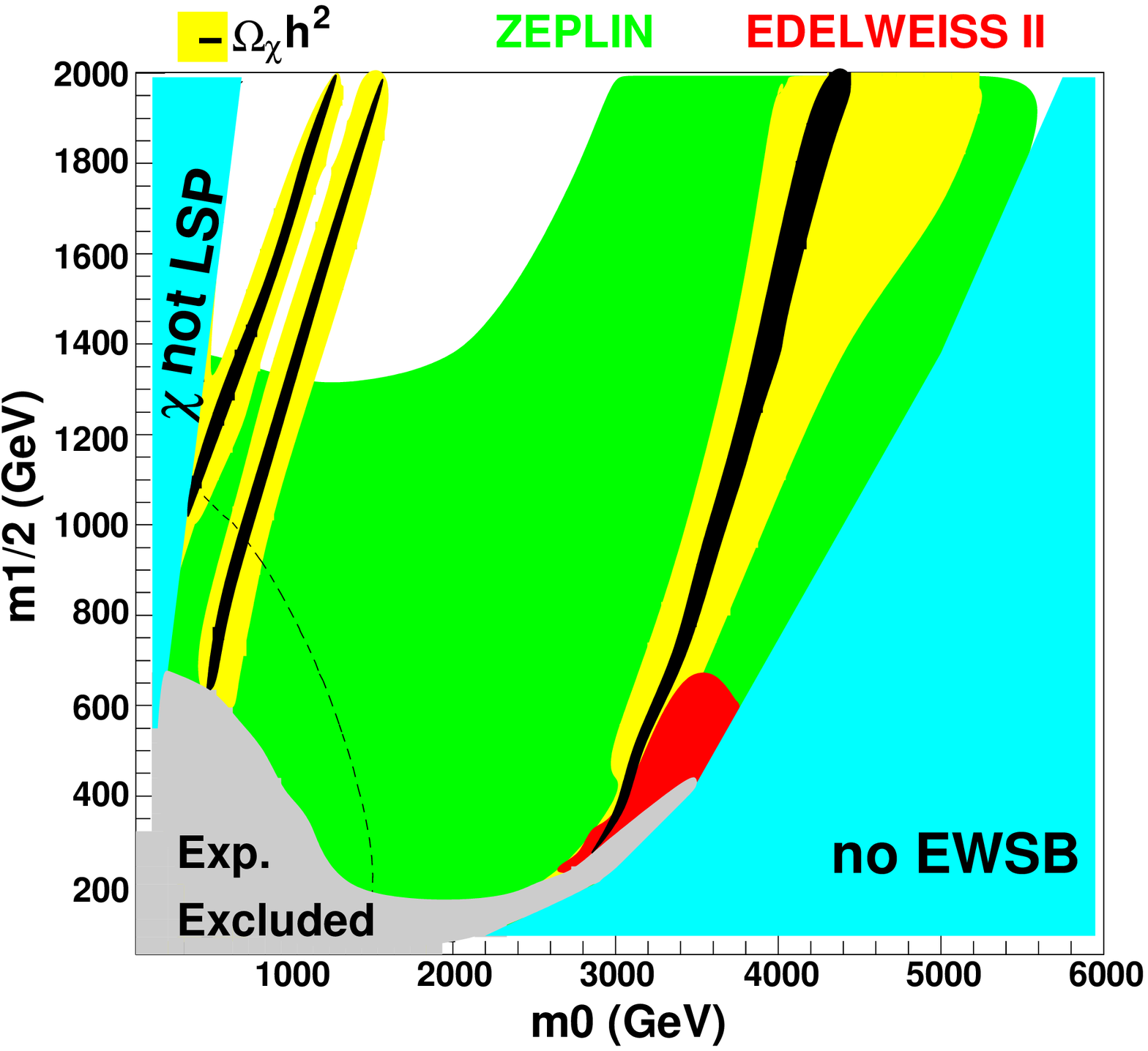}&
\includegraphics[width=0.45\textwidth]{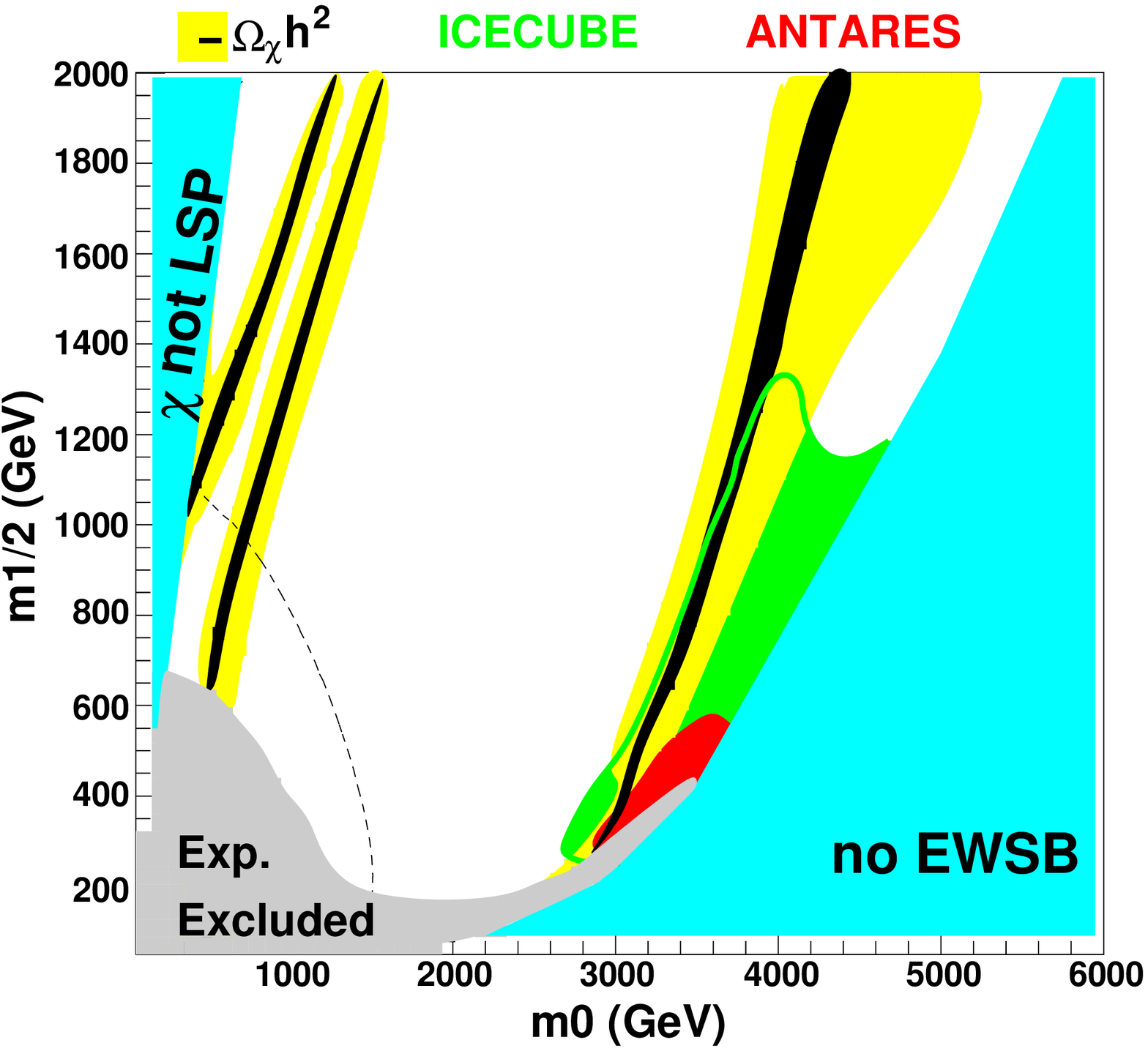}\\
a) Direct Detection & b) $\nu$ Indirect Detection (Sun) \\
\includegraphics[width=0.45\textwidth]{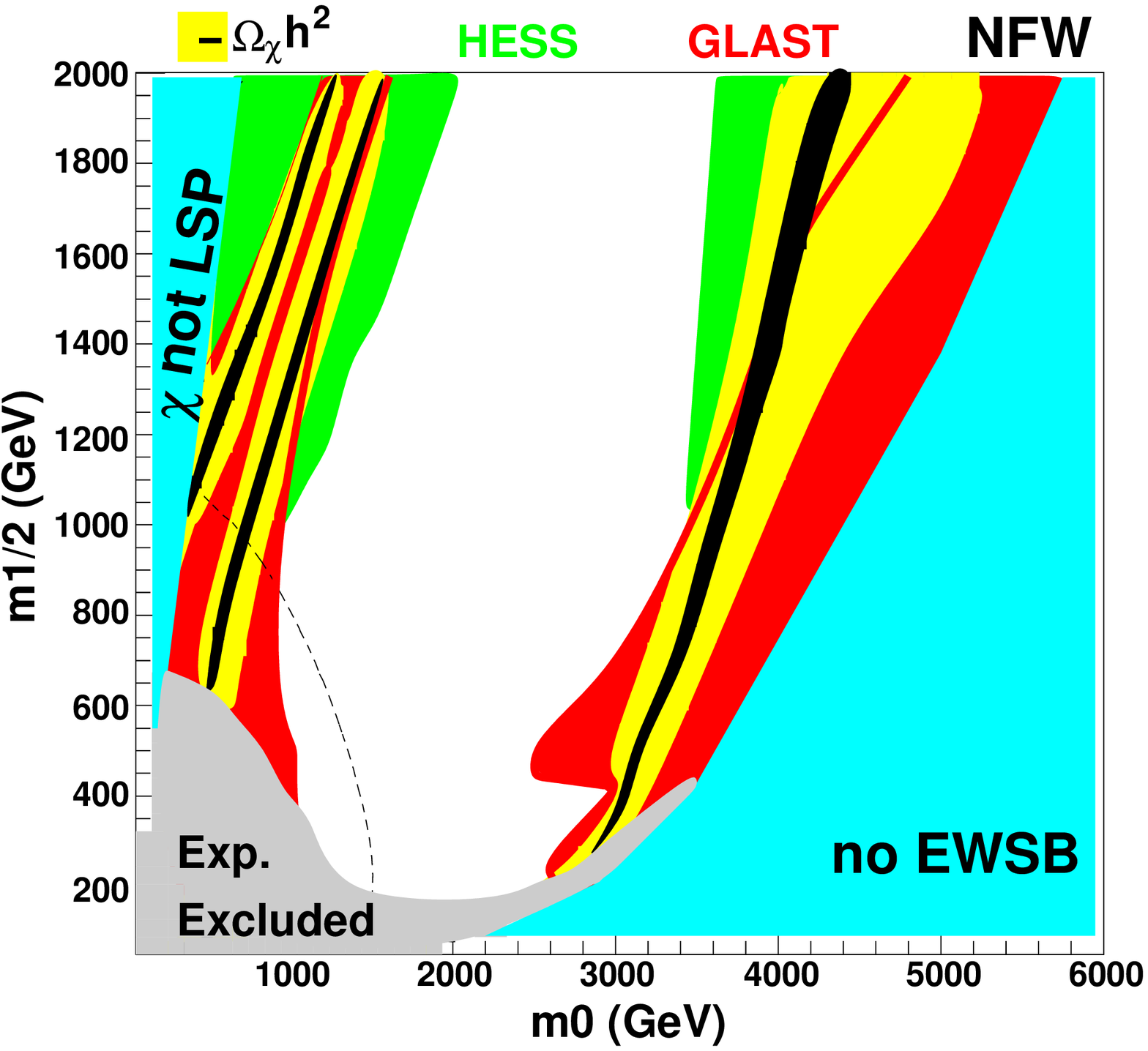}&
\includegraphics[width=0.45\textwidth]{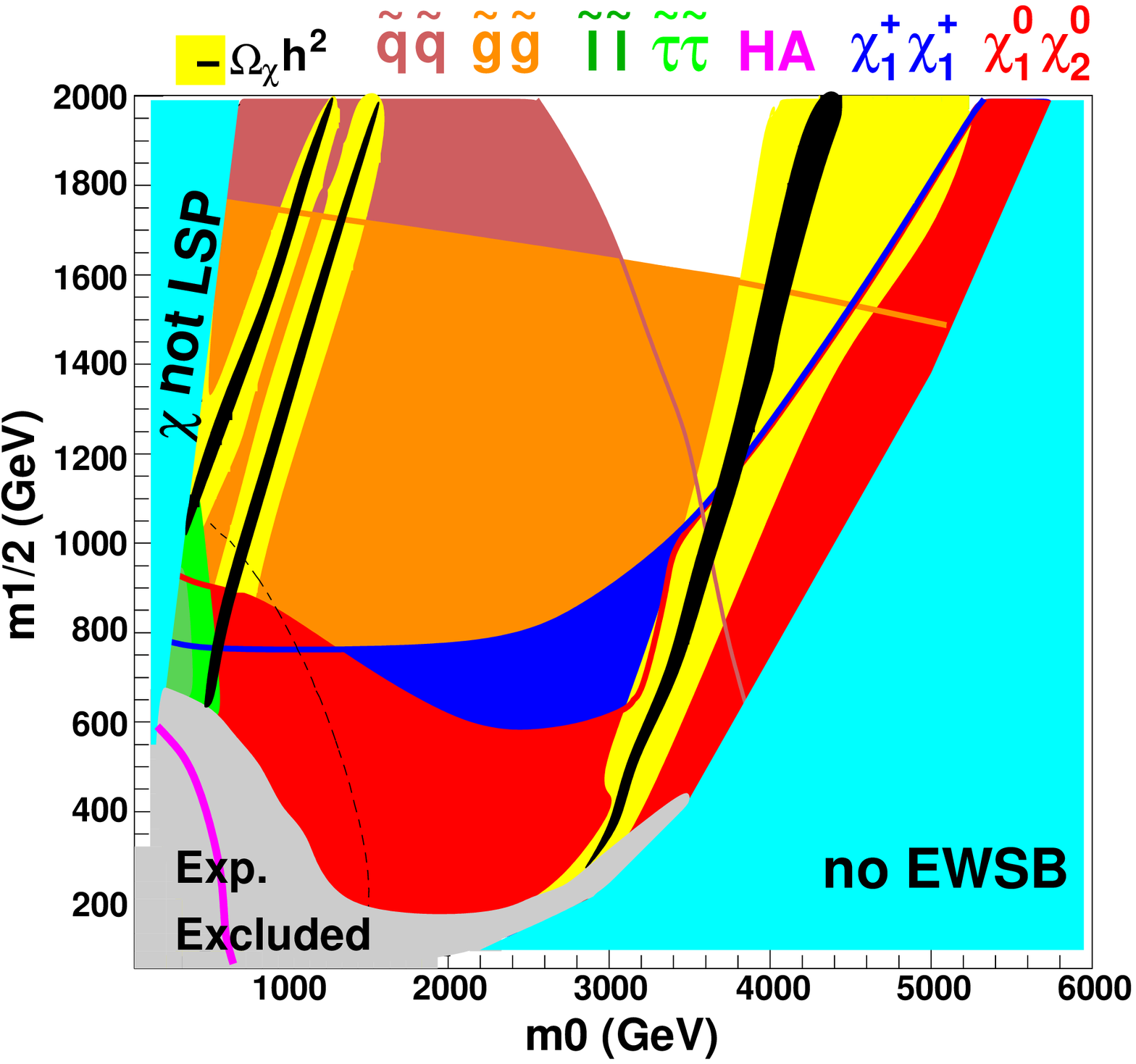}\\
c) $\gamma$ Indirect Detection (GC) & d) Collider production (LHC,ILC)\\
\includegraphics[width=0.45\textwidth]{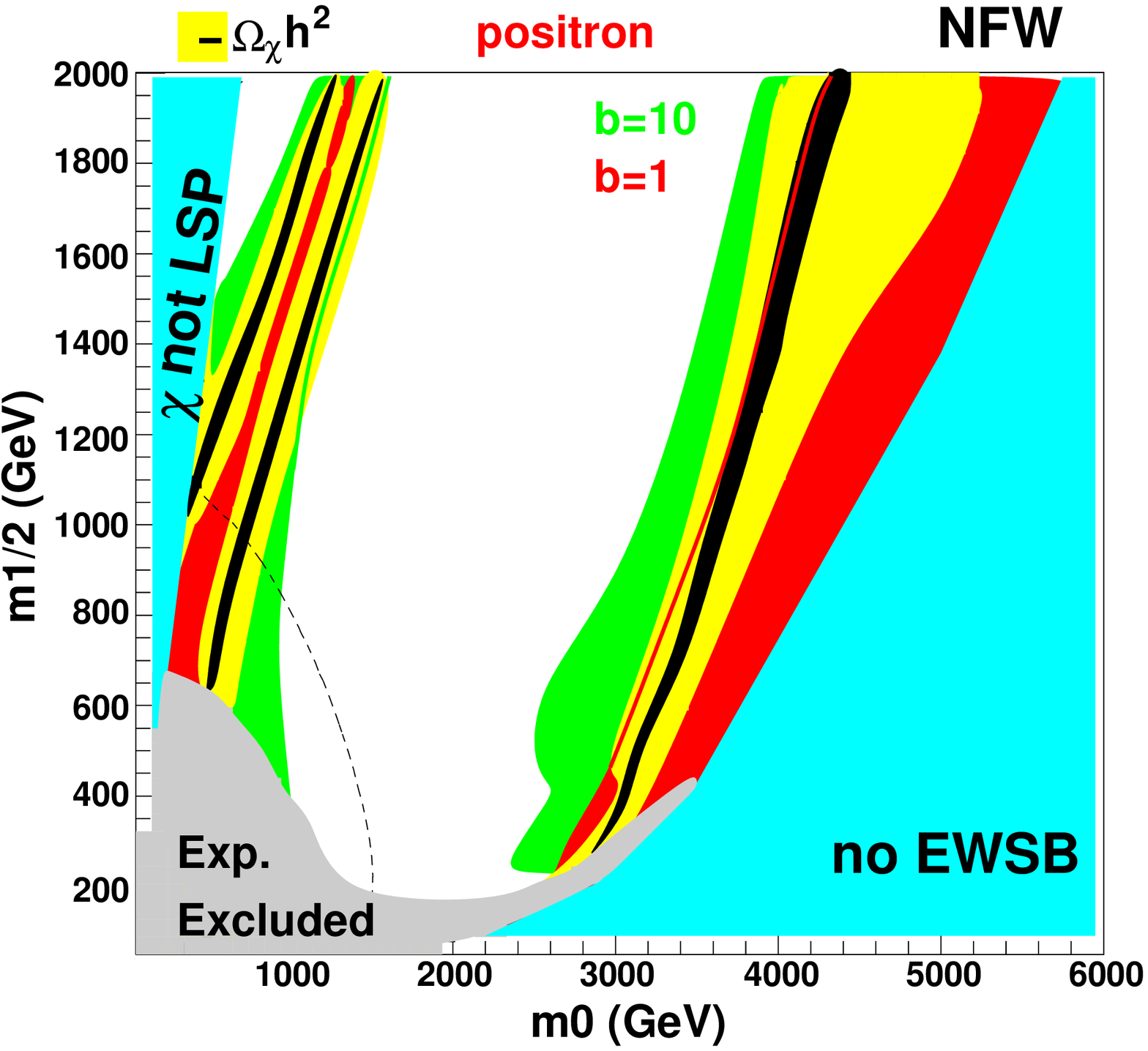}&
\includegraphics[width=0.45\textwidth]{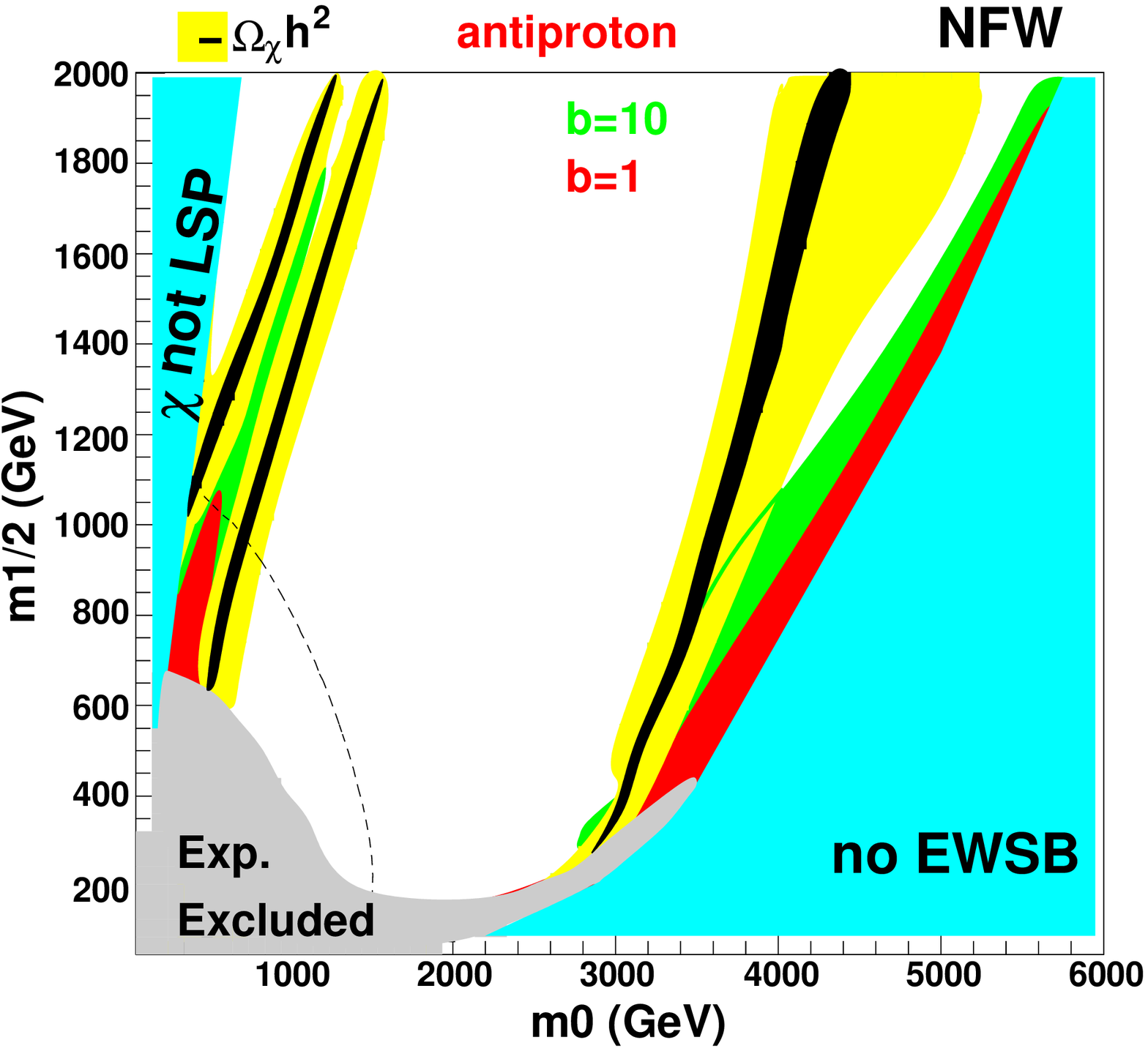}\\
e) $e^+$ Indirect Detection (halo) &f) $\bar{p}$ Indirect Detection (halo)\\
&\\
\end{tabular}
\caption{ MSUGRA non Universal $M_3|_{GUT}=0.6m_{1/2}$, $A_0=0$, $\tan{\beta}=35$, $\mu>0$}
\label{fig:M3tb35}
\end{center}
\end{figure}

\begin{figure}[t]
\begin{center}
\begin{tabular}{cc}
\includegraphics[width=0.45\textwidth]{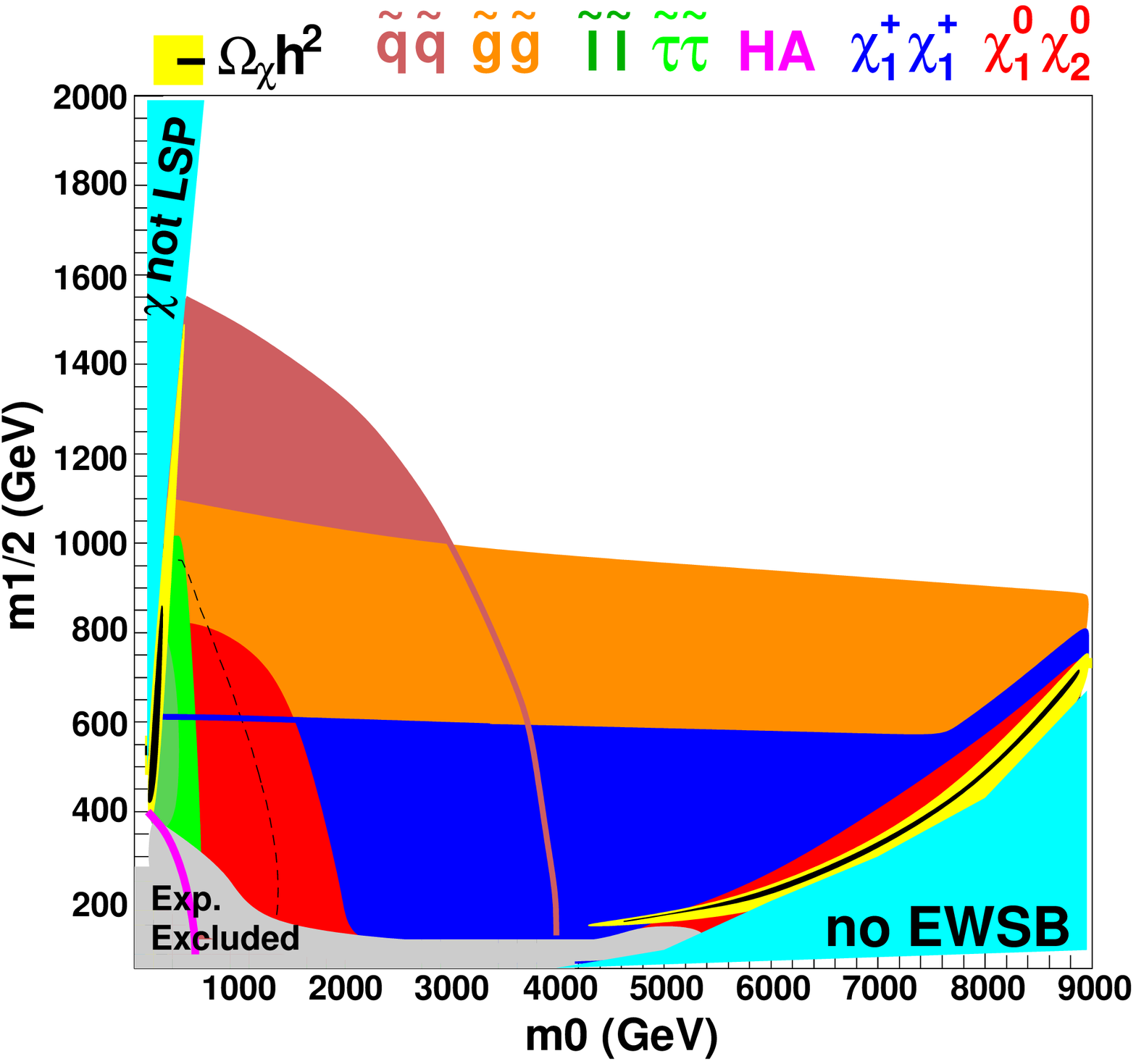}&
\includegraphics[width=0.45\textwidth]{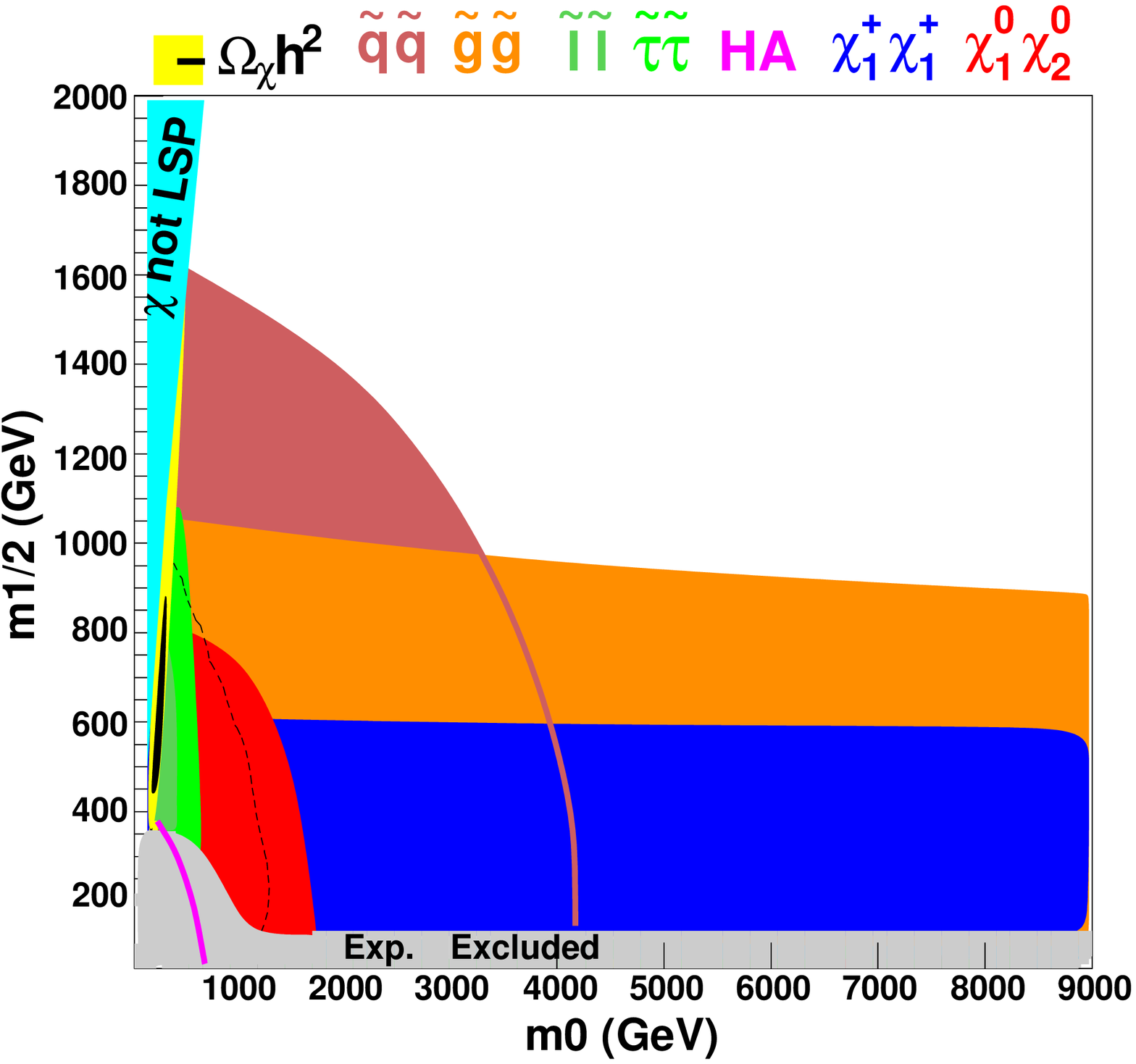}\\
a) $m_t=178$  GeV & b) $m_t=182$  GeV\\
\includegraphics[width=0.45\textwidth]{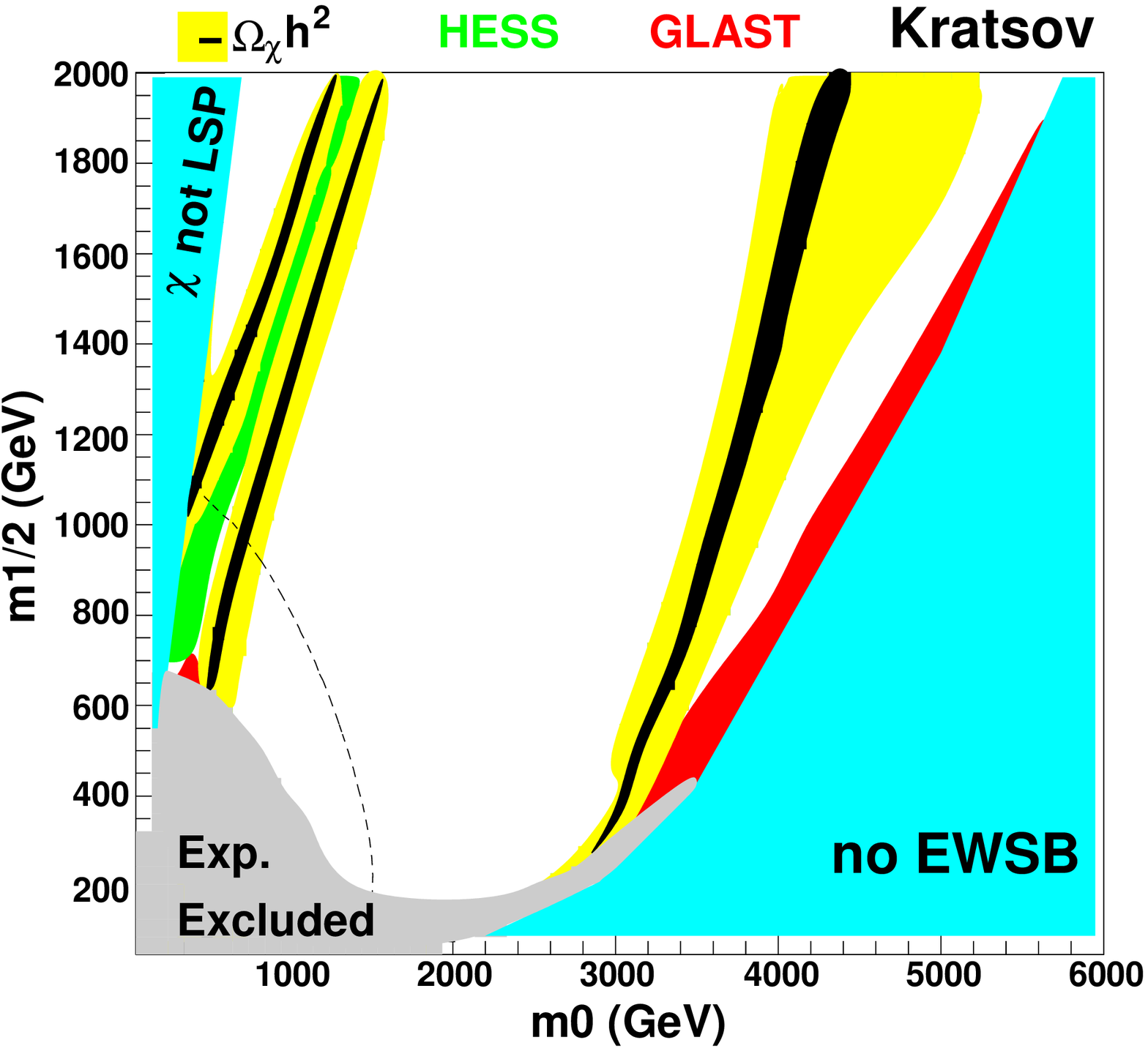}&
\includegraphics[width=0.45\textwidth]{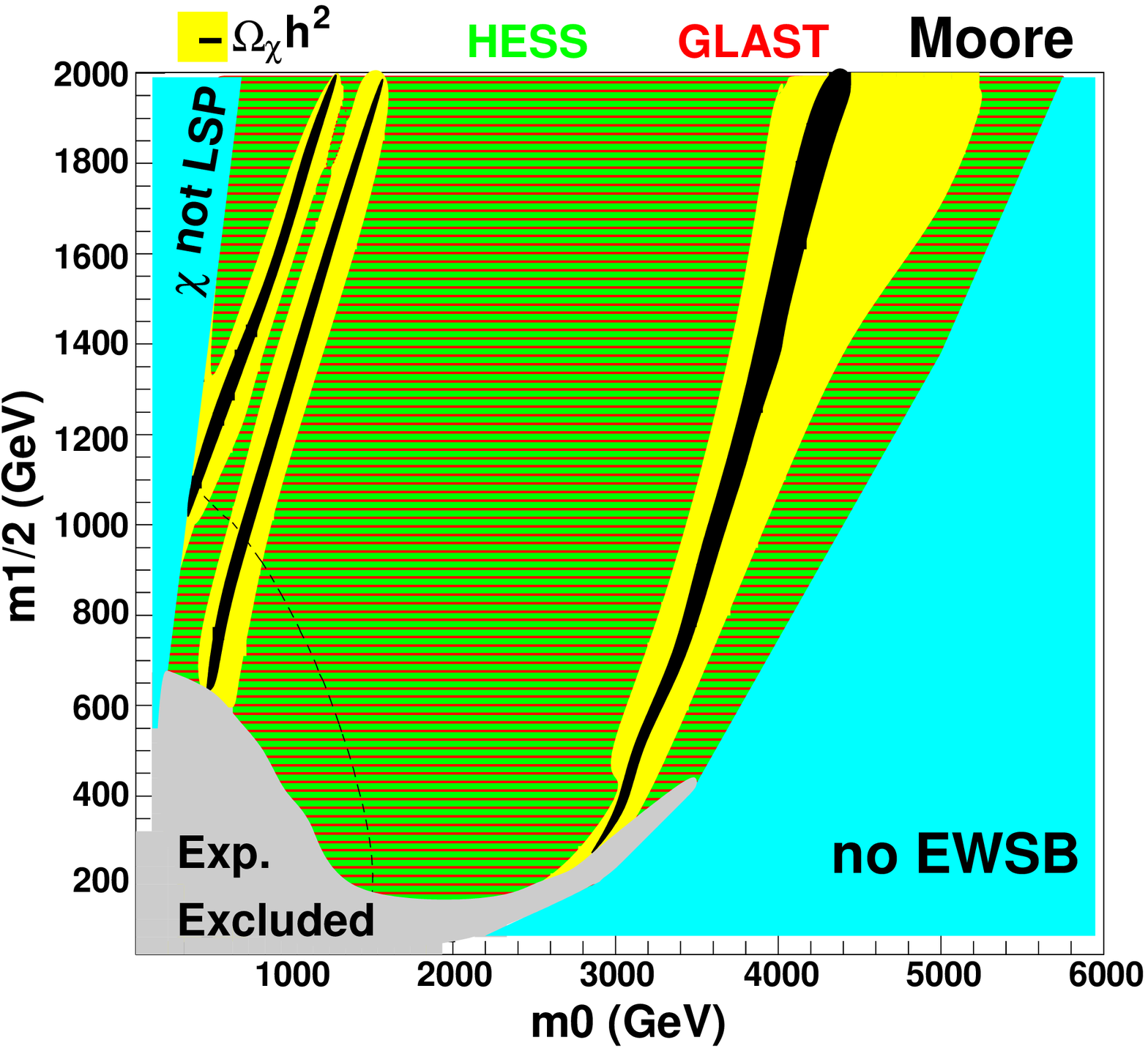}\\
c) $\gamma$ Indirect Detection : Kravtsov profile & d)  $\gamma$ Indirect
Detection : Moore profile\\
&\\
\end{tabular}
\caption{  Effect of $m_t$=178 GeV in a) and $m_t$=182 GeV in b) on
  LHC and ILC performances in the universal case (to be compared with Fig. \ref{fig:Utb35}d)). Halo profile influence on $\gamma$ indirect detection. $A_0=0$,
  $\tan{\beta}=35$, $\mu>0$  for non universal gluino mass 
  $M_3|_{GUT}=0.6m_{1/2}$ with a Kravtsov profile in c) and Moore profile in d)
  (to be compared with the NFW profile case Fig. \ref{fig:M3tb35}c))}
\label{fig:topmoorekra}
\end{center}
\end{figure}

Fig. \ref{fig:M3tb35} shows the effects of non--universality of
the gaugino breaking mass term $M_3$. 
The gluino mass parameter influences considerably the MSSM spectrum through
 the Renormalization Group Equations (see for instance \cite{Martin:1997ns} for
a review on the subject). Decreasing $M_3$ decreases squark masses, 
increases the up-type Higgs mass $M^2_{Hu}$  at low energy where it
becomes less negative, and  decreases the down-type Higgs mass $M^2_{Hd}$
  which implies lighter $m_{A,H}$ and an increase
of the Higgsino content of neutralinos and charginos. That can be easily understood 
looking at the approximate tree level relations :

\begin{equation}
\mu^2\simeq-M^2_{H_u}-1/2M^2_Z \ \ {\rm and} \ \ m^2_A\simeq
M^2_{H_d}-M^2_{H_u}-M^2_Z
\label{eq:treelevelmumA}
\end{equation}

\noindent
As a result, relic density constraints are more easily satisfied than in the
universal case : both $\chi \chi \xrightarrow{A}b\bar{b}$ annihilation 
(higher coupling {\it and} lighter $A$ which can open the $A$ funnel) 
and focus point region with the $\chi \chi \xrightarrow{Z}t\bar{t}$ annihilation
process are enhanced.
Direct detection gets advantage of higher couplings $z_{11}z_{13}$ and lighter
scalar Higgs $H$.
The higher Higgsino fraction favors neutrino indirect detection through the
coupling in $\chi q \xrightarrow{Z} \chi q$ of the capture rate.
Gamma, positron and antiproton indirect detection are favored by the
annihilation  enhancement. 
LHC gets strong potentiality enhancement because the squarks 
(especially the $\tilde{t_1}$) and gluinos are lighter than in the Universal 
case.
Finally, the $HA$ production at a 1 TeV Linear Collider is kinematically  
enhanced, $H$ and $A$ being lighter than in mSUGRA.
$\chi^+\chi^-$ $\chi\chi^0_2$ production are also favored because a lower value 
of $\mu$. As in the non--universal wino mass case, smaller  $M_3|_{GUT}/m_{1/2}$ values can lead to $\chi \chi^+_1$ and $\chi \chi^0_2$ degeneracies, now through
 the Higgsino component, which constrains the detection but those regions
 have too small relic density driven by coannihilation to be really favored.
These points are well illustrated in Fig. \ref{fig:M3tb35} for the 
ratio $M_3/m_{1/2}=0.6$ at GUT scale \cite{Mynonuniv}.
Those kinds of models with light gluino mass are very interesting for SUSY
detection and all neutralino dark matter detections and can be found in some
effective string inspired scenarios 
\cite{Binetruy:2003yf,Bertone:2004ps,Falkowski:2005ck}.

\subsection{Higgs sector}

\subsubsection{Up-type Higgs mass : $M_{H_u}|_{GUT}$}

\begin{figure}[t]
\begin{center}
\begin{tabular}{cc}
\includegraphics[width=0.45\textwidth]{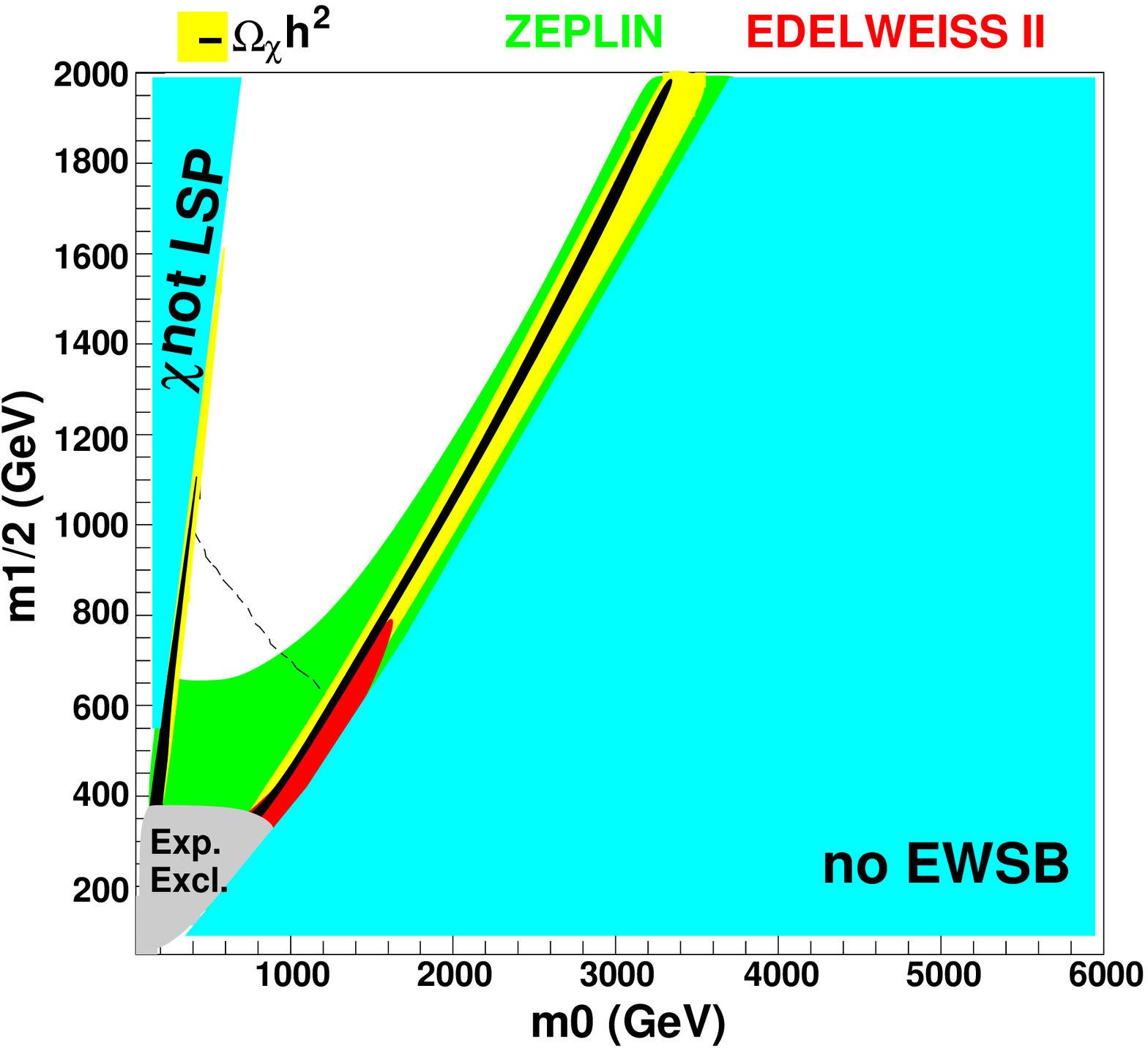}&
\includegraphics[width=0.45\textwidth]{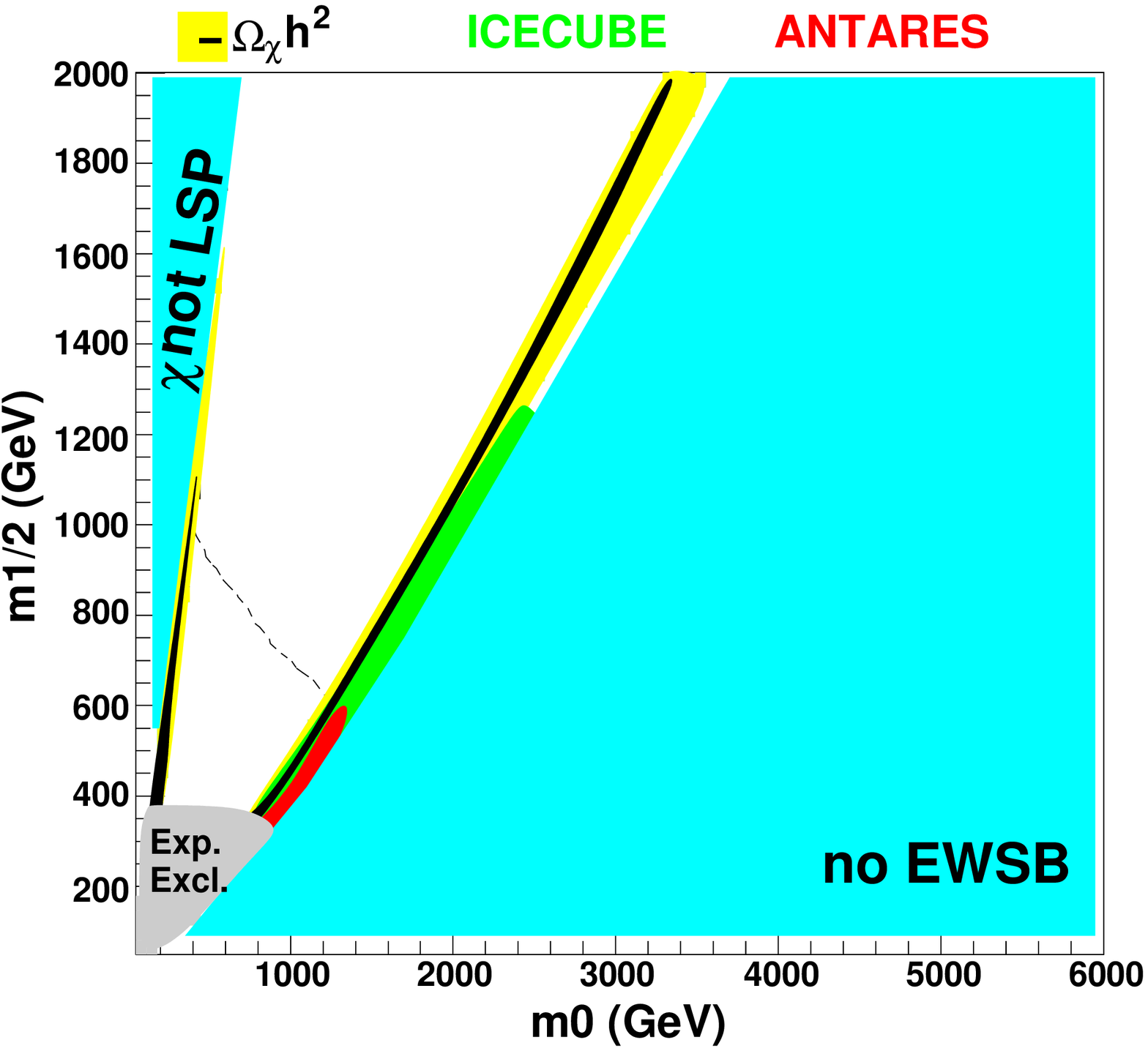}\\
a) Direct Detection & b) $\nu$ Indirect Detection (Sun) \\
\includegraphics[width=0.45\textwidth]{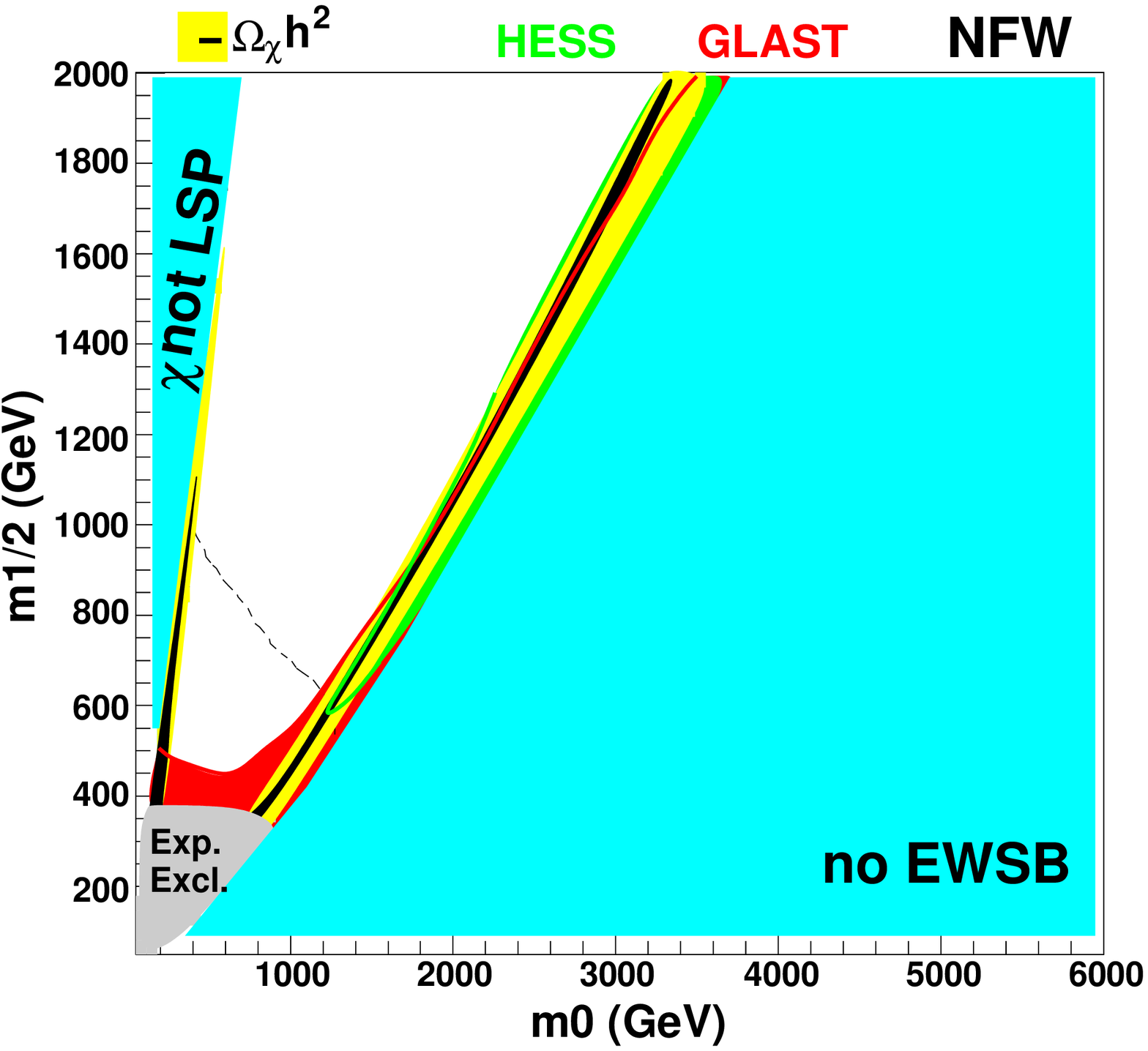}&
\includegraphics[width=0.45\textwidth]{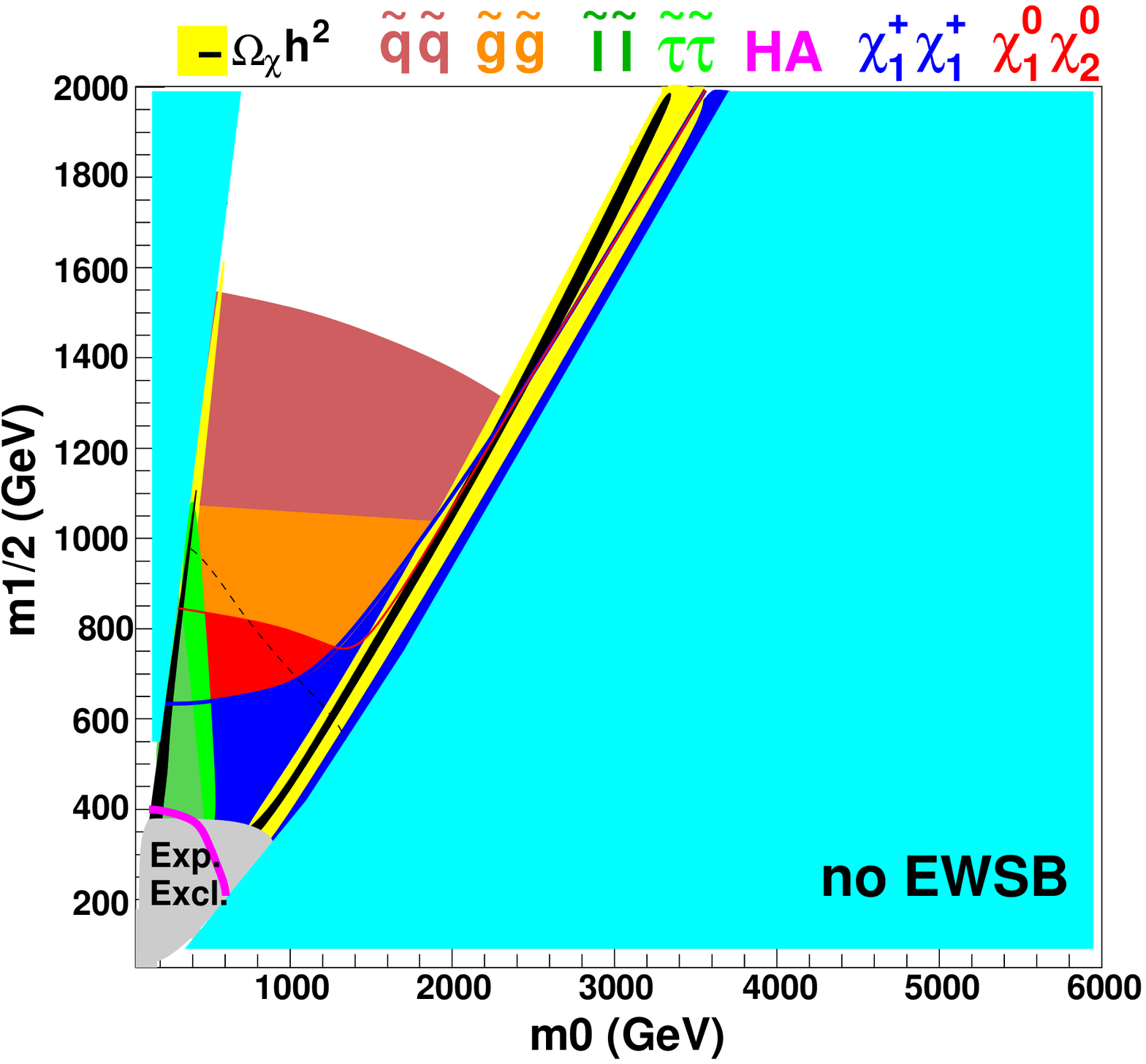}\\
c) $\gamma$ Indirect Detection (GC) & d) Collider production (LHC,ILC)\\
\includegraphics[width=0.45\textwidth]{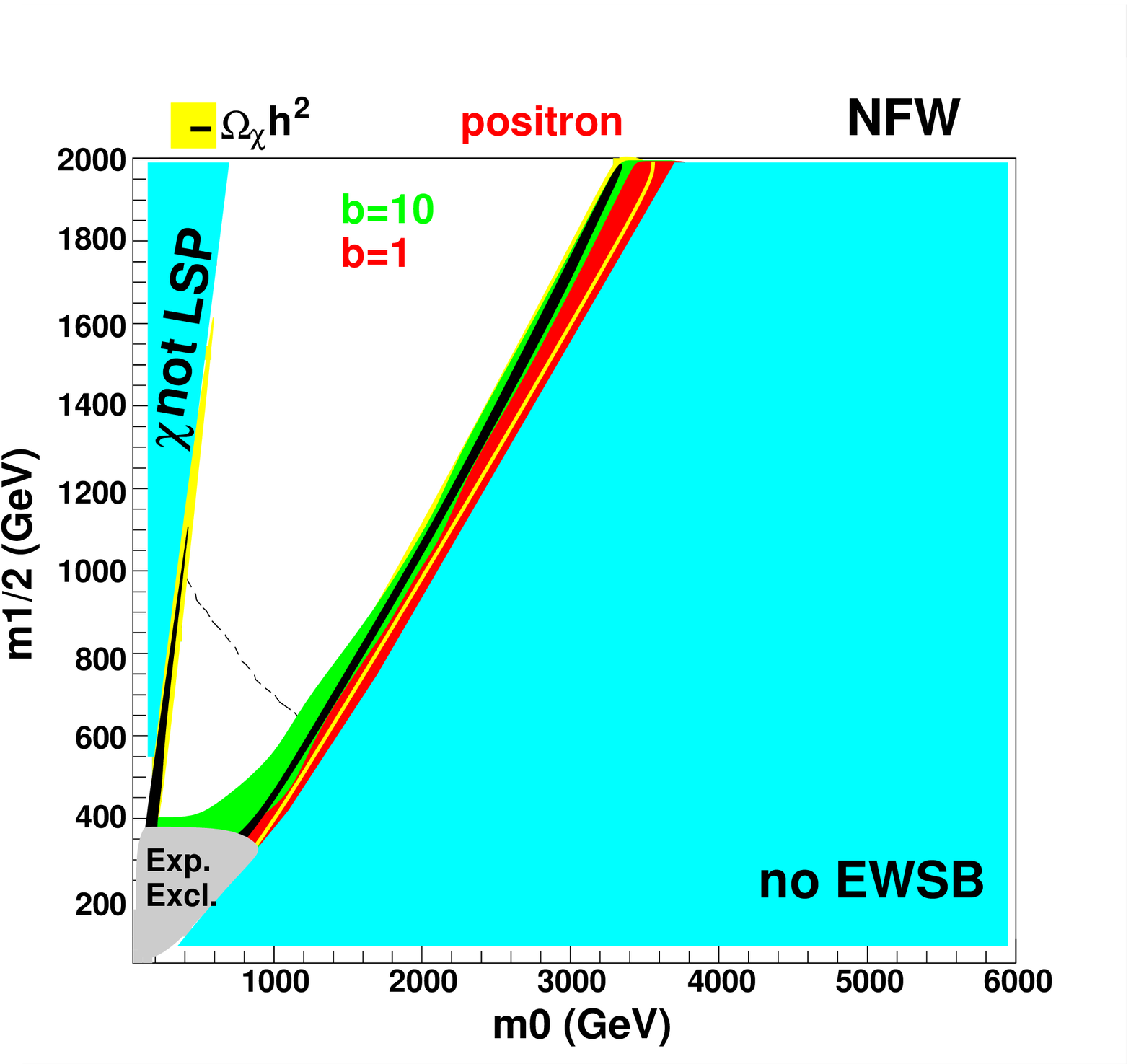}&
\includegraphics[width=0.45\textwidth]{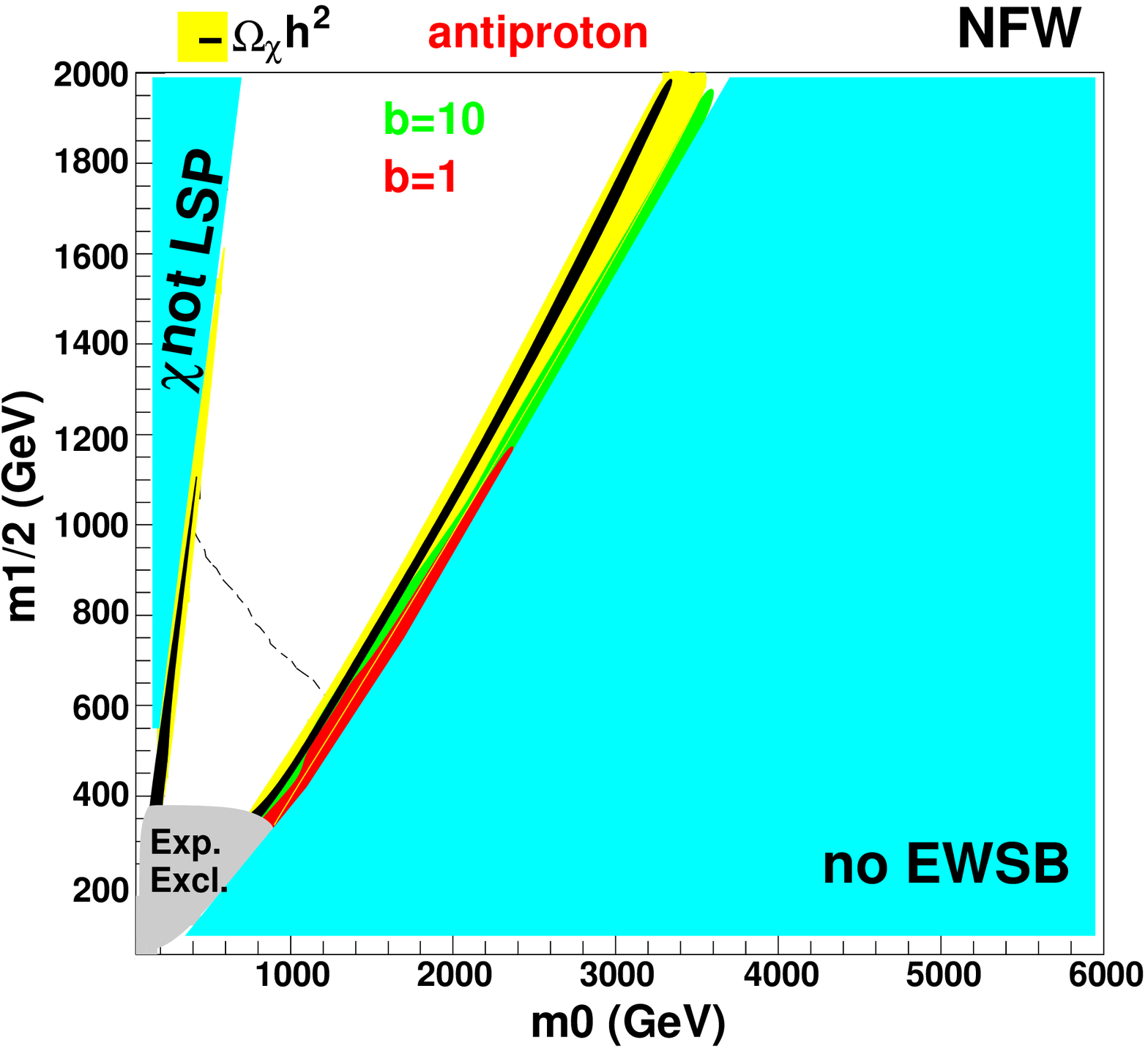}\\
e) $e^+$ Indirect Detection (halo) &f) $\bar{p}$ Indirect Detection (halo)\\
&\\
\end{tabular}
\caption{MSUGRA non Universal $M_{Hu}|_{GUT}=1.2m_{0}$}
\label{fig:MHutb35}
\end{center}
\end{figure}

Fig. \ref{fig:MHutb35} shows the prospects of detection 
for $M_{H_u}/m_{0}=1.2$ at the GUT scale.
Increasing the up-type Higgs mass $M_{H_u}$ at the GUT scale has some common
 effects with the case described above when decreasing gluino mass as can 
be explicitly seen from Eq. \ref{eq:treelevelmumA}. 
One has to notice that the sensitivity on this parameter is high, leading   
to a thiner region with interesting results and wider ``no EWSB'' area 
compared to the previous non universal gluino mass case. As was done in the 
gaugino sector \cite{Mynonuniv} we varied continuously the non--universality
 in the Higgs sector at GUT scale ($M_{H_u}/m_{0}$ and $M_{H_d}/m_{0}$) 
and found for the up-type mass that the relevant value of the ratio 
 leading to WMAP relic density is around 1.2 for $\tan{\beta}=35$.
With respect to the universal case, the mixed bino-Higgsino region is more
important but the pseudo--scalar Higgs $A$ is still too much heavy to
open the on--shell $A-$pole channel.
 All kinds of dark matter detections are thus possible in the resulting 
mixed bino-Higgsino region as well as chargino production in a
future Linear Collider where $HA$ pairs can also be produced. 
 LHC gets no enhancement in squark and gluino production but covers a 
wide part of the not excluded remaining plane.

\subsubsection{Down-type Higgs mass : $M_{H_d}|_{GUT}$}

\begin{figure}[t]
\begin{center}
\begin{tabular}{cc}
\includegraphics[width=0.45\textwidth]{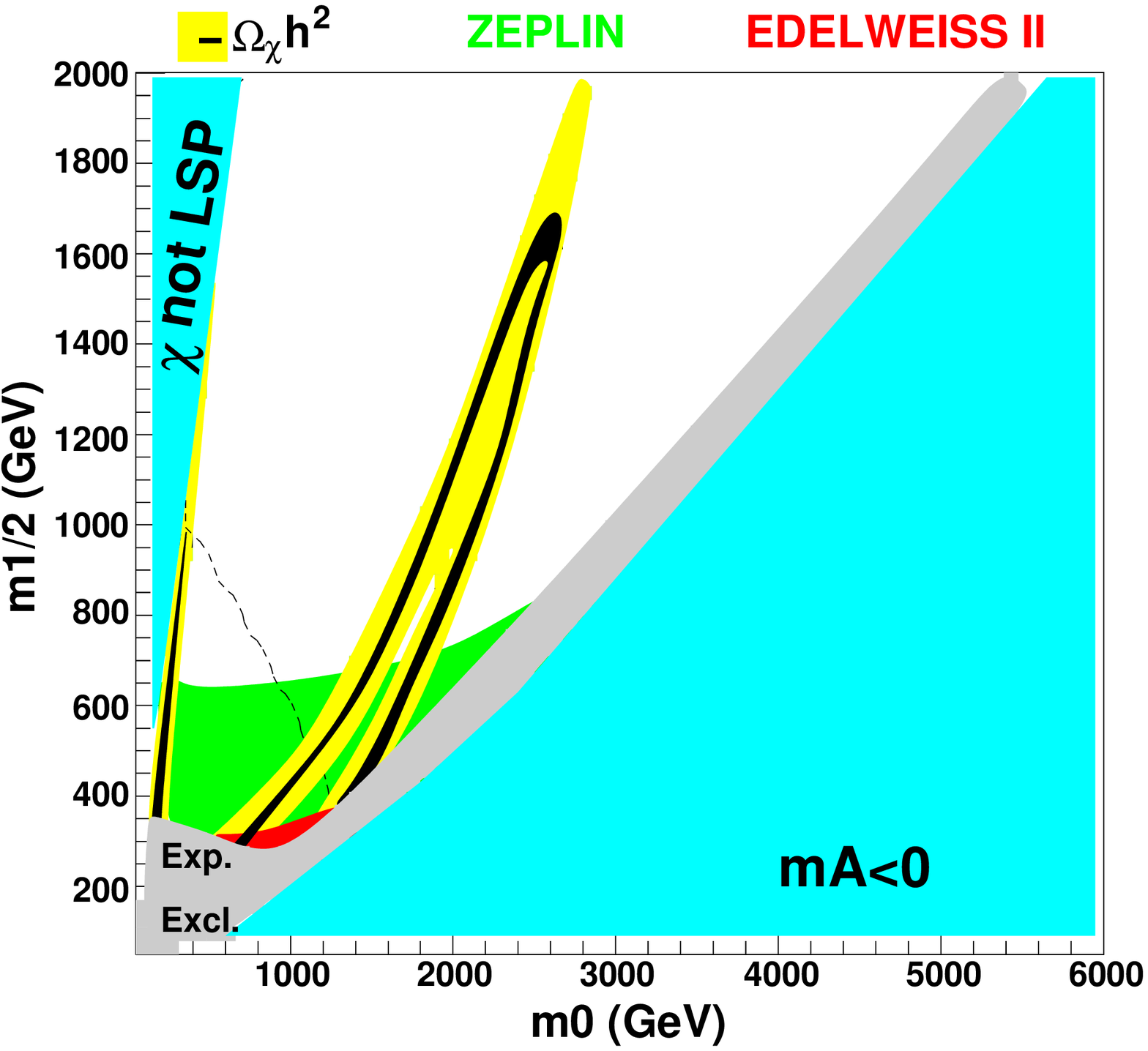}&
\includegraphics[width=0.45\textwidth]{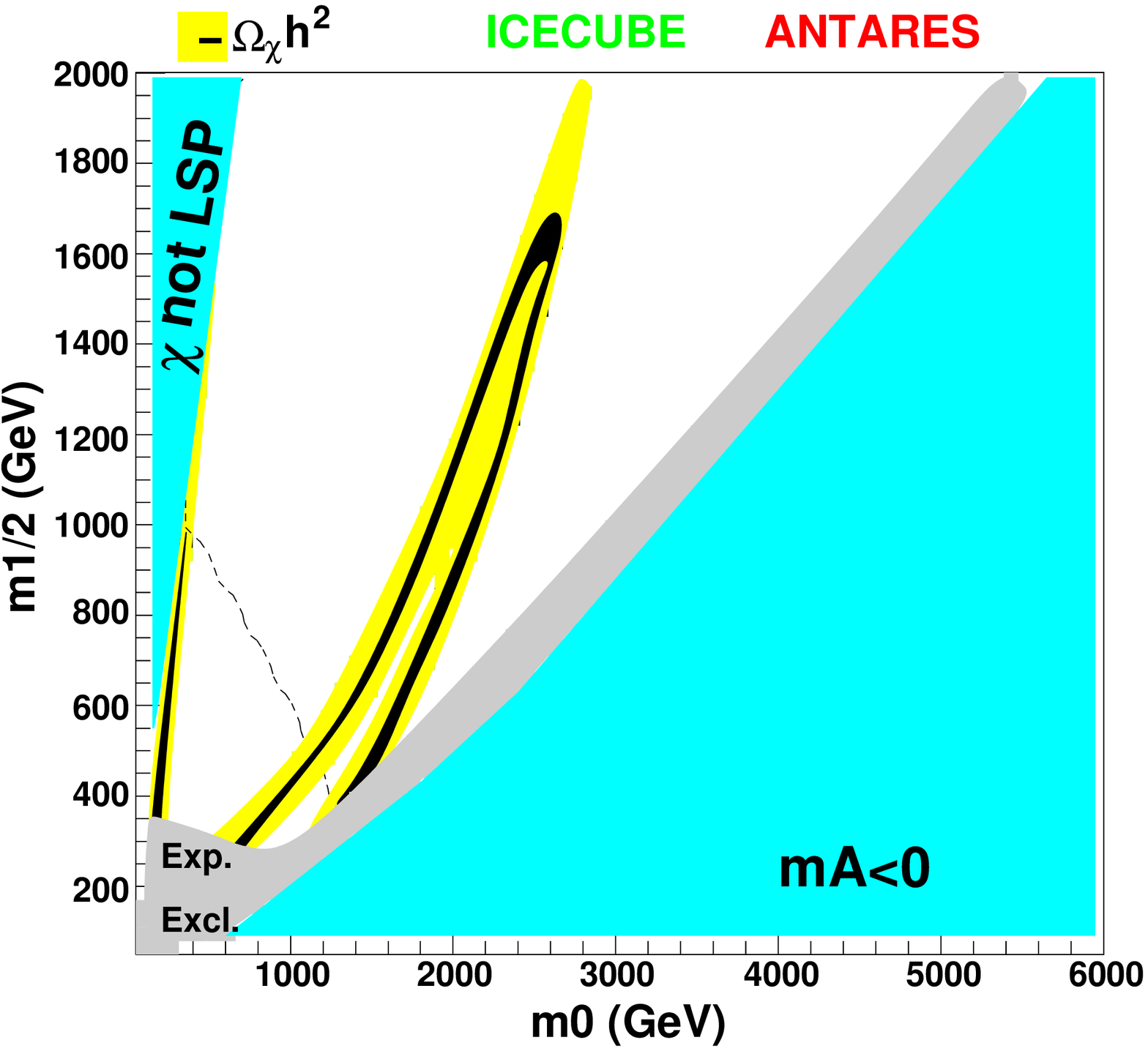}\\
a) Direct Detection & b) $\nu$ Indirect Detection (Sun) \\
\includegraphics[width=0.45\textwidth]{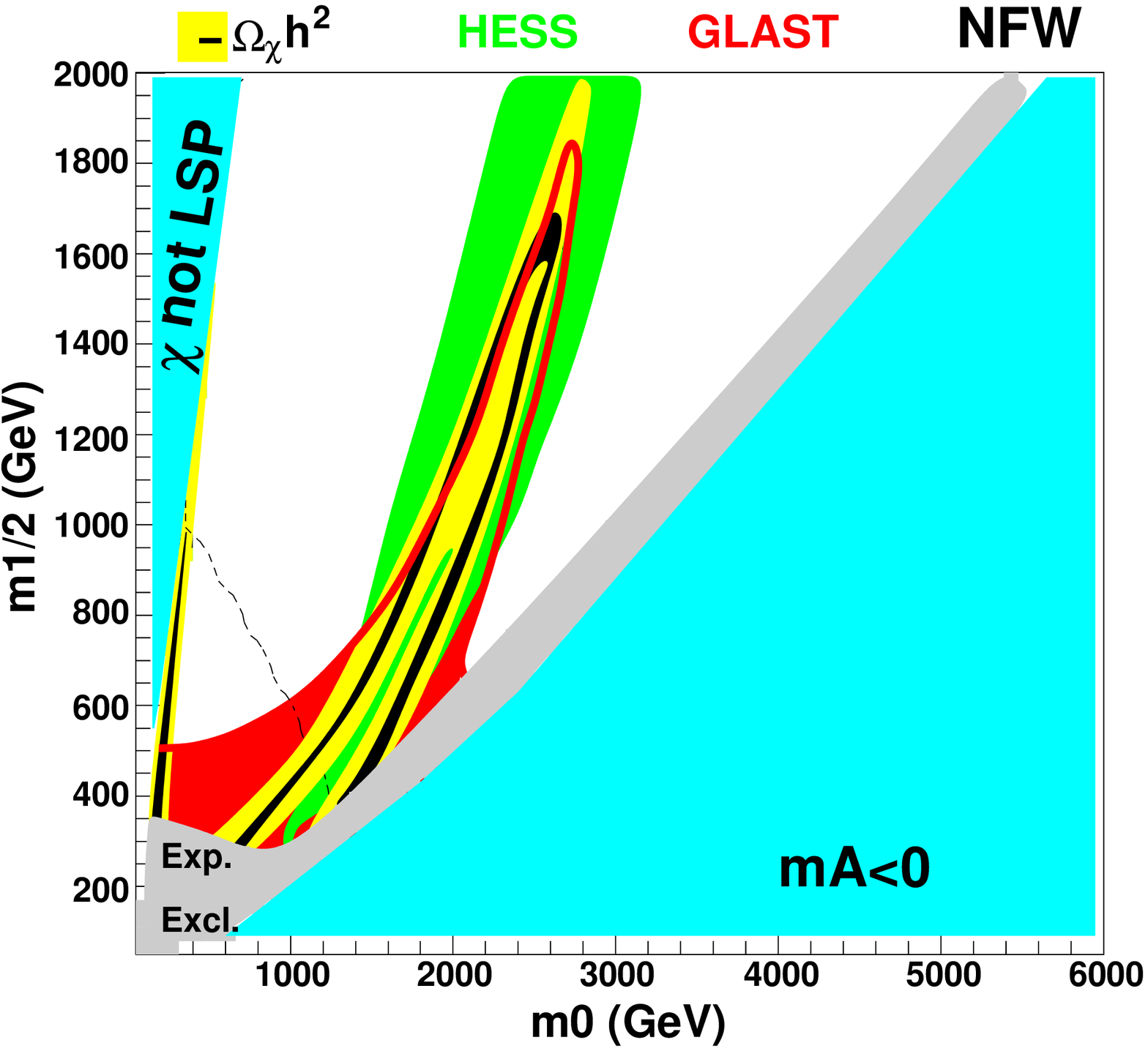}&
\includegraphics[width=0.45\textwidth]{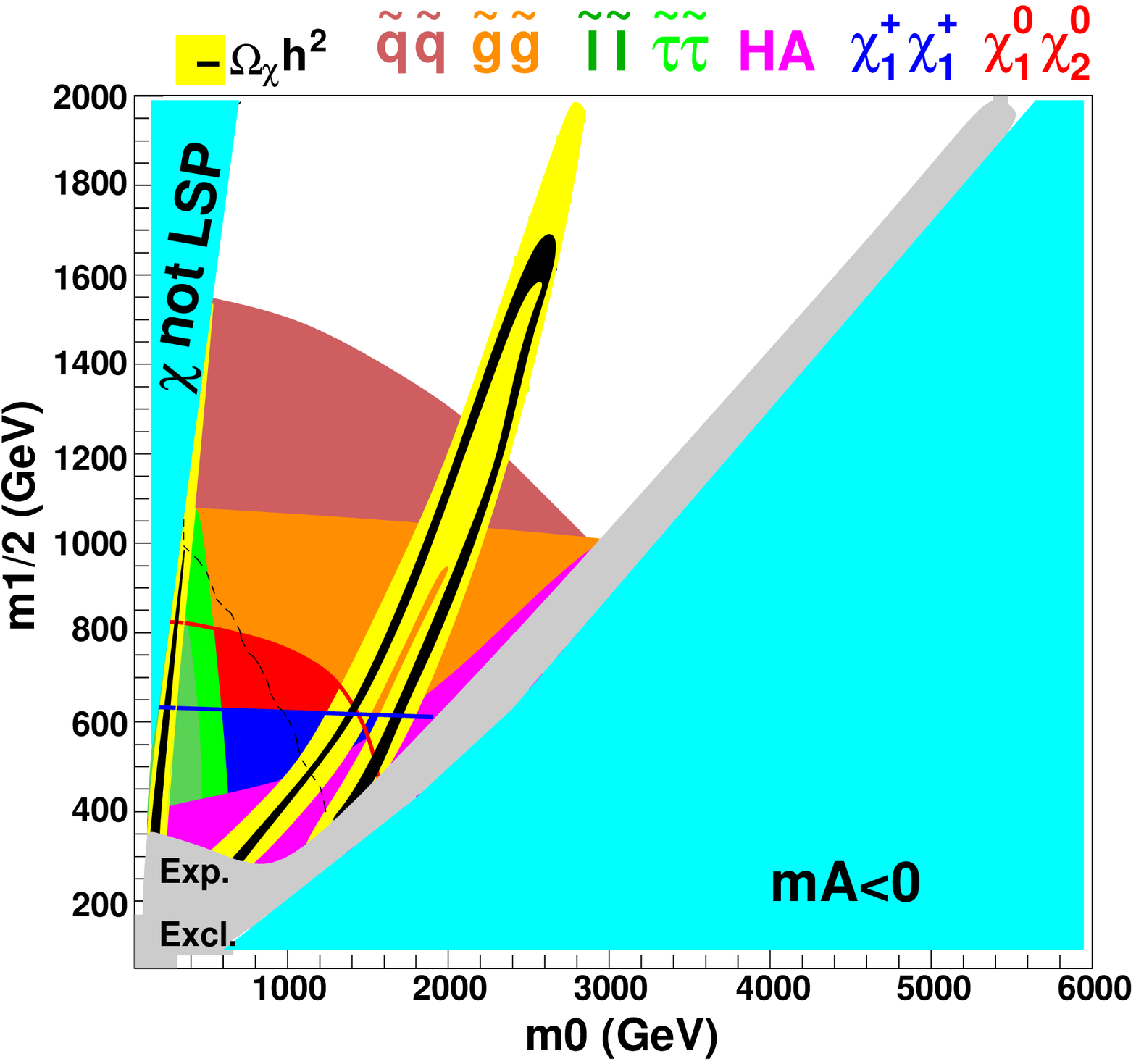}\\
c) $\gamma$ Indirect Detection (GC) & d) Collider production (LHC,ILC)\\
\includegraphics[width=0.45\textwidth]{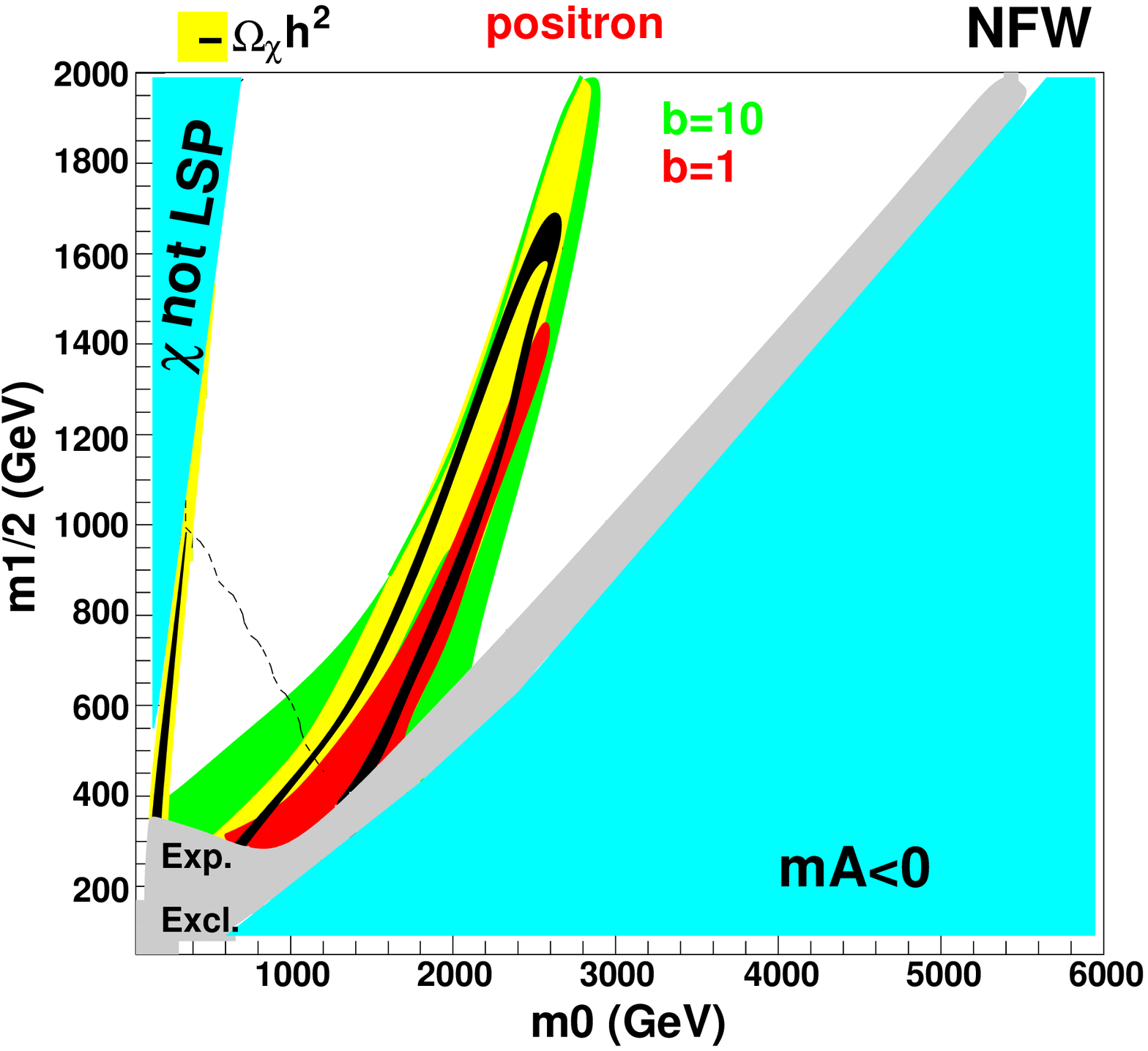}&
\includegraphics[width=0.45\textwidth]{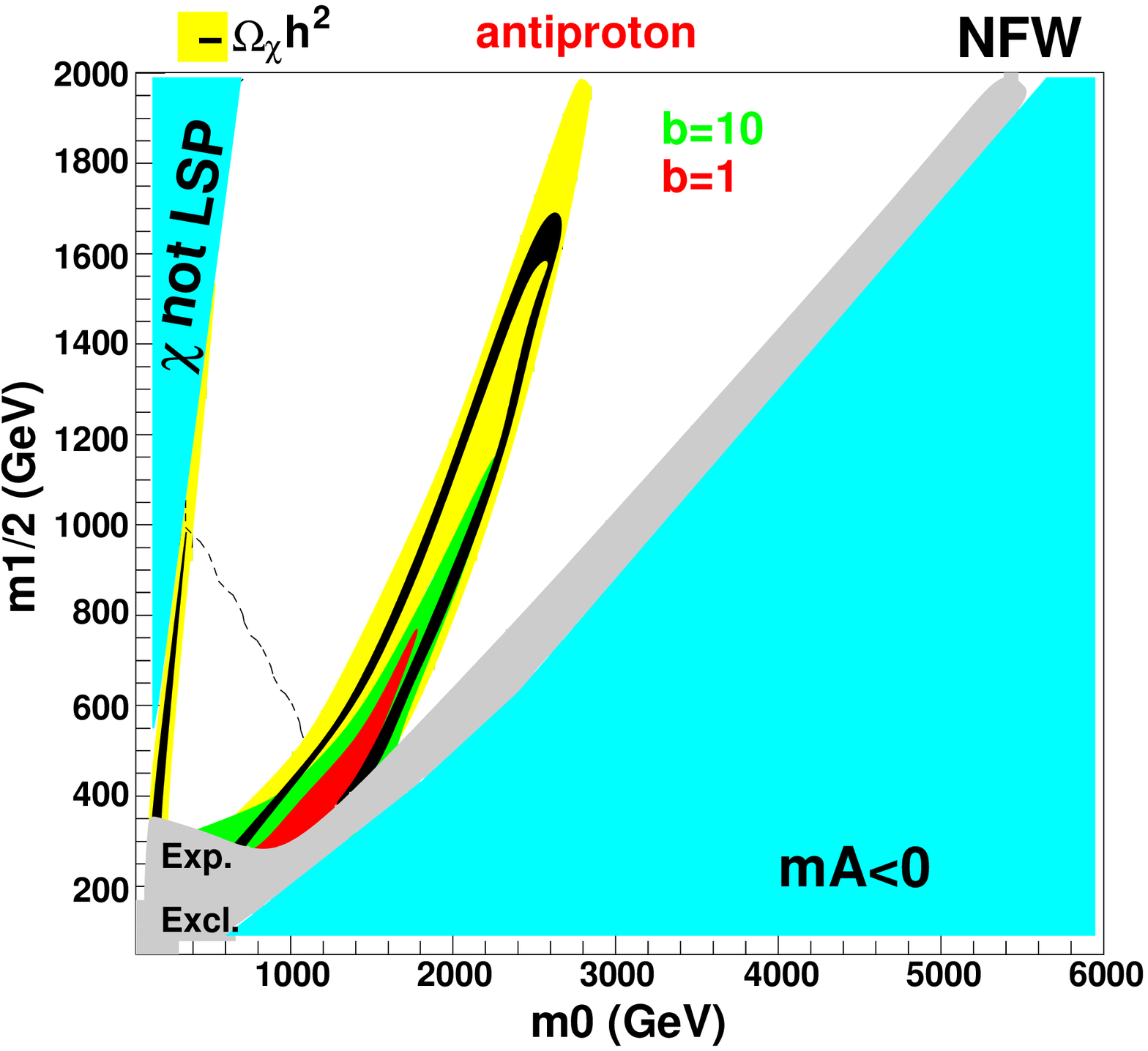}\\
e) $e^+$ Indirect Detection (halo) &f) $\bar{p}$ Indirect Detection (halo)\\
&\\
\end{tabular}
\caption{MSUGRA non Universal $M_{Hd}|_{GUT}=0.4m_{0}$}
\label{fig:MHdtb35}
\end{center}
\end{figure}

Fig. \ref{fig:MHdtb35} illustrates the case of non--universality
$M_{H_d}/m_{0}=0.4$ at GUT scale.
Indeed, a ratio $M_{H_d}/m_{0}<1$ can have interesting phenomenological
consequences concerning the different detection rates.
By decreasing the down-type Higgs mass, $M_{H_d}$ at GUT scale, one essentially
decreases $m_{A,H}$ as can be seen in Eq. \ref{eq:treelevelmumA}. 
The excluded region at high values of $m_0$ results from negative mass of 
the pseudo-scalar in addition to problem of not realizing the EWSB. 
The $A$--pole can be open more easily giving a corridor with interesting  
relic density within the WMAP bounds. In this corridor, neutralino annihilation
 is important, increasing the perspective of discovery through
 $\gamma,e^+$ and $\bar{p}$ indirect detection. The low value of $m_{H_d}$ 
gives also good direct detection rates but we have to keep in mind that
we have nearly bino neutralino in all the remaining region of the
$(m_0,m_{1/2})$ plane such that $\chi\chi H$ coupling is suppressed. For the
same reason of small Higgsino fraction, neutrino telescopes are strongly
disfavored for those kinds of models. The LHC have an equivalent potentiality 
of discovery than in mSUGRA, 
but cover a wider area of the parameter space. Only the Higgs production
$HA$ is enhanced at the Linear Collider compared to the universal case while
gaugino-Higgsino neutralino and chargino productions are suppressed.

\section{Summary-Conclusion}

Dark matter experiments and collider searches will be a major step 
to probe the possibility of low energy supersymmetry and neutralino dark
matter in the Minimal Supersymmetric Standard Model. The possible 
correlations  between (absence of) signals of different kinds of detection 
will bring
many informations on models and scenarios both for supersymmetry and 
astrophysics.

We summarize hereafter the links between the different possibilities to fulfill WMAP
constraint, the parameters involved and the kind of detection concerned.

The $\chi \tilde{\tau}(\tilde{t})$ coannihilation region 
is typically difficult for
dark matter detection. It can be possible for a huge direct detection experiment 
if $m_{\tilde{\tau}}$ and $m_H$ are correlated (low $m_{\tilde{\tau}}$
to favor the coanihilation process and light $H$ to favor direct detection
scattering) as well as for $\gamma$ indirect
detection in the case of favorable galactic profile (with a stronger cusp 
than in the NFW profile). 
The LHC
can be of interest through $\tilde{q}\tilde{q}$ if $m_{\tilde{q}}$ is
correlated to $m_{\tilde{l}}$ and the ILC through $\tilde{\tau}\tilde{\tau}$
to probe the $\tilde \tau$ coannihilation region.

The Higgs funnel region $\chi\chi\xrightarrow{A}b\bar{b}(\tau\bar{\tau})$ is 
favored with respect to the universal case
 for non universal $M_3$ or $M_{H_u}$ or $M_{H_d}$. 
Direct detection is concerned but large $m_A$ value need also a coupling
 enhancement through the Higgsino fraction of the neutralino.
The  $\gamma$ ($e^+,\bar{p}$) indirect detection follow this annihilation
process. However the absolute potentiality depends on astrophysics hypothesis.
The LHC situation depends on $m_{\tilde{q},\tilde{g}}$ with regard to  $m_A$
and for the ILC, $\chi \chi^0_2$ and $HA$ productions are favored up to energy limitation.

The Hyperbolic Branch/Focus Point with mixed bino-Higgsino $\chi$ at
high $m_0$ values and especially non
universal $M_3$ or $M_{H_u}$ parameters is the more interesting region 
concerning the DM
searches. Direct detection can bring
informations on the nature of the neutralino by correlations between
spin dependent/independent experiments as well as through
correlations with neutrino telescope. 
$\gamma(e^+,\bar{p})$ 
Indirect detections of $\gamma$, $e^+$ or $\bar{p}$ are favored
through the enhancement of annihilation processes.
Moreover, this region of mixed neutralino is the only one accessible 
by a neutrino telescope for a signal coming
from the Sun (a ${\rm km^3}$ size telescope will probe up to
$m_{\chi} \sim$ 600 GeV). The chargino production in $e^+e^-$ collider as well
as gluino production for LHC are the relevant processes but become difficult for
very high values of the breaking mass terms, keeping in mind that those regions 
have {\bf $\delta_{\mu}^{\mathrm{susy}}=0$}.

Finally a mixed bino-wino $\chi$ is very sensitive to a non--universal $M_2$
 parameter.
The wino component enhances dark matter rates ($\Omega h^2$ is very sensitive
to the wino fraction). The LHC can be of interest if the gluino mass is
correlated to the wino mass and wino-like neutralino/chargino production in ILC is possible up to $m_{\chi(\chi^+_1)}\simeq$ 500 GeV.

However, even if a part of the supersymmetric spectrum is discovered
at LHC, it will be difficult to measure precisely
the properties of the particles entering 
in the relic density computation. Both types of data (astroparticle
and accelerator physics) will thus be needed to extract more complete
informations about the underlying model.

Whereas the Focus Point (FP) region
 characterized by heavy scalars will be more easily probed by dark matter searches
 projects due to the nature of the neutralino, the region with heavy gaugino and
 light sfermions will be more accessible by collider experiments. Since dark
 matter signals give few informations on the nature of the dark matter and
 since new physics collider signals could not be linked directly to 
dark matter ones, deeper informations on both supersymmetry and astrophysics 
hypothesis can thus be obtained by correlation of the different signals 
or absence of signal.

\section*{Acknowledgments}{E.N. work is supported by the I.I.S.N. and the 
Belgian Federal Science Policy (return grant and IAP 5/27).
Y.M. wants to warmly thank P. Zerwas for sharing his incredible knowledge 
and enthusiasm and the DESY theory group for their
 scientific and financial supports. The authors are grateful to 
A. Djouadi and J.B De Vivie and the referee for their advices, corrections and
 update during the redaction of this work.}


\nocite{}
\bibliography{bmn}

\begin{thebibliography}{99}

\bibitem{Bertone:2004pz}
  G.~Bertone, D.~Hooper and J.~Silk,
  ``Particle dark matter: Evidence, candidates and constraints,''
  Phys.\ Rept.\  {\bf 405}, 279 (2005)
  [arXiv:hep-ph/0404175].


\bibitem{Jungman:1995df}
  G.~Jungman, M.~Kamionkowski and K.~Griest,
  ``Supersymmetric dark matter,''
  Phys.\ Rept.\  {\bf 267}, 195 (1996)
  [arXiv:hep-ph/9506380].


\bibitem{Olive1}
{\rm K.~Olive}
{arXiv:astro-ph/0301505}
{\it Summary of lectures given at the Theoretical Advanced Study Institute in Elementary Particle Physics at the University of Colorado at Boulder - June 2-28, 2002}, and 
references therein.

\bibitem{carlosreview} For a recent review, see C. Mu\~noz, 
`Dark matter detection in the light of recent experimental results', 
{\it Int. J. Mod. Phys.} {\bf A19} (2004) 3093 [arXiv:hep-ph/0309346].


 \bibitem{WMAP}
  C.~L.~Bennett {\it et al.}, 
{\it Astrophys.J.Suppl.} {\rm 148:1,2003}\\
arXiv:astro-ph/0302207;\\
  D.~N.~Spergel {\it et al.}, 
{\it Astrophys.J.Suppl.} {\rm 148:175,2003}
arXiv:astro-ph/0302209.

\bibitem{Planckexp}
J. A. Tauber, ``The PLANCK Mission : Overview and Current Status,''
{\it Astrophys. Lett. Comm.} {\bf 37}, pp. 145-150, 2000.

\bibitem{Freese:1999sh}
K.~Freese, B.~D.~Fields and D.~S.~Graff,
``Death of Stellar Baryonic Dark Matter,''
arXiv:astro-ph/0002058.




\bibitem{Fayet:1976cr}
  P.~Fayet and S.~Ferrara,
  ``Supersymmetry,''
  Phys.\ Rept.\  {\bf 32}, 249 (1977).

\bibitem{Haber:1984rc}
  H.~E.~Haber and G.~L.~Kane,
  ``The Search For Supersymmetry: Probing Physics Beyond The Standard Model,''
  Phys.\ Rept.\  {\bf 117}, 75 (1985).


\bibitem{Barbieri:1987xf}
  R.~Barbieri,
  ``Looking Beyond The Standard Model: The Supersymmetric Option,''
  Riv.\ Nuovo Cim.\  {\bf 11N4}, 1 (1988).


\bibitem{Martin:1997ns}
  S.~P.~Martin,
  ``A supersymmetry primer,''
  arXiv:hep-ph/9709356.

\bibitem{Allanach:2004xn}
B.~C.~Allanach, G.~Belanger, F.~Boudjema and A.~Pukhov,
``Requirements on collider data to match the precision of WMAP on
supersymmetric dark matter,''
JHEP {\bf 0412} (2004) 020
[arXiv:hep-ph/0410091].

\bibitem{Suspect} 
A. Djouadi, J.~L. Kneur and G. Moultaka, 
`SuSpect: a Fortran code for the supersymmetric and Higgs particle 
spectrum in the MSSM', [arXiv:hep-ph/0211331]; 
 
\noindent See also the web page 
 http://www.lpm.univ-montp2.fr:6714/\char126kneur/suspect.html 

\bibitem{Micromegas}
 G. Belanger, F. Boudjema, A. Pukhov and
A. Semenov,
`micrOMEGAs: a program for calculating the relic density in the
MSSM',
{\it Comput. Phys. Commun.} {\bf 149} (2002) 103 [arXiv:hep-ph/0112278];
\noindent G. Belanger, F. Boudjema, A. Pukhov and A.G. Semenov, 
`MicrOMEGAs: Version 1.3',
[arXiv:hep-ph/0405253];
\noindent 
See also the web page http://wwwlapp.in2p3.fr/lapth/micromegas


\bibitem{Darksusy}
  P.~Gondolo, J.~Edsjo, P.~Ullio, L.~Bergstrom, M.~Schelke and E.~A.~Baltz,
  ``DarkSUSY: Computing supersymmetric dark matter properties numerically,''
  JCAP {\bf 0407}, 008 (2004)
  [arXiv:astro-ph/0406204], http://www.physto.se/~edsjo/darksusy/.

\bibitem{Baer:2005bu}
  H.~Baer, A.~Mustafayev, S.~Profumo, A.~Belyaev and X.~Tata,
  ``Direct, Indirect and Collider Detection of Neutralino Dark Matter In SUSY
  Models with Non-universal Higgs Masses,''
  arXiv:hep-ph/0504001.


\bibitem{Baer:2005zc}
  H.~Baer, A.~Mustafayev, E.~K.~Park and S.~Profumo,
  ``Mixed Wino dark matter: Consequences for direct, indirect and collider
  detection,''
  arXiv:hep-ph/0505227.

\bibitem{ellissug1}
  {\rm J.~R.~Ellis, T.~Falk, G.~Ganis, K.~A.~Olive and M.~Srednicki},
  {\it Phys. Lett.} {\bf B510} {\rm (2001) 236}; 
  
\bibitem{ellissug2}
{\rm J.~R.~Ellis, K.~A.~Olive and Y.~Santoso},
  {\it New Jour. Phys.} {\bf 4} {\rm (2002) 32}; 

\bibitem{roszksug}
  {\rm L.~Roszkowski, R.~Ruiz de Austri and T.~Nihei},
  {\it JHEP} {\bf 0108} {\rm (2001) 024}; 

\bibitem{abdelsug}  
{\rm A.~Djouadi, M.~Drees and J.~L.~Kneur},
  {\it JHEP} {\bf 0108} {\rm (2001) 055}; 
  {\rm H.~Baer, C.~Balazs and A.~Belyaev},
  {\it JHEP} {\bf 0203} {\rm (2002) 042}.


  \bibitem{EllisWmap}
  {\rm J.~Ellis, A.~Ferstl, K.~A.~Olive, Y.~Santoso}, 
  {arXiv:hep-ph/0302032}. 

\bibitem{Feng}
  {\rm   J.L.~Feng, K.T.~Matchev, F.~Wilczek},
  {\it Phys.Rev.} {\bf D63} {\rm (2001) 045024},
{arXiv:astro-ph/0008115}.


  \bibitem{Nath1}
  {\rm  U.~Chattopadhyay, A.~Corsetti, P.~Nath}, 
  {arXiv:hep-ph/0303201}. 

  \bibitem{EllisHiggs}  
{\rm J.~Ellis, K.~A.~Olive, Y.~Santoso  and V.~C.~Spanos}, 
  {arXiv:hep-ph/0303043}. 

\bibitem{Bottino1}
  {\rm  V.~Berezinsky, A.~Bottino, J.~Ellis, N.~Fornengo, G.~Mignola,
S.~Scopel}, 
  {\it Astropart.Phys.} {\bf 5} {\rm (1996) 1-26},
{arXiv:hep-ph/9508249}.

\bibitem{Arnowitt1}
  {\rm   P.~Nath, R.~Arnowitt}, 
  {\it Phys.Rev.} {\bf D56} {\rm (1997) 2820-2832},
{arXiv:hep-ph/9701301}. 

\bibitem{Birkedal-Hansen:2001is}
  A.~Birkedal-Hansen and B.~D.~Nelson,
 ``The role of Wino content in neutralino dark matter,''
  Phys.\ Rev.\ D {\bf 64}, 015008 (2001)
  [arXiv:hep-ph/0102075].


\bibitem{Mynonuniv}
  {\rm   V.~Bertin, E.~Nezri, J.~Orloff}, 
  {\it JHEP} {\bf 02} {\rm (2003) 046},
{arXiv:hep-ph/0210034}.

\bibitem{BirkedalnonU}
{\rm A.~Birkedal-Hansen B.~D.~Nelson}
{\it Phys.Rev.} {\bf D67} {\rm (2003) 095006},
{arXiv:hep-ph/0211071}.

\bibitem{Cerdeno:2003yt}
D.~G.~Cerdeno, E.~Gabrielli, M.~E.~Gomez and C.~Munoz,
``Neutralino nucleon cross section and charge and colour breaking
constraints,''
JHEP {\bf 0306} (2003) 030
[arXiv:hep-ph/0304115].

D.~G.~Cerdeno and C.~Munoz,
``Neutralino dark matter in supergravity theories with non-universal scalar and
gaugino masses,''
JHEP {\bf 0410} (2004) 015
[arXiv:hep-ph/0405057].



\bibitem{Binetruy:2003yf}
  P.~Binetruy, Y.~Mambrini and E.~Nezri,
  ``Direct and indirect detection of dark matter in heterotic orbifold
  models,''
  Astropart.\ Phys.\  {\bf 22}, 1 (2004)
  [arXiv:hep-ph/0312155].


\bibitem{Bertone:2004ps}
  G.~Bertone, P.~Binetruy, Y.~Mambrini and E.~Nezri,
  ``Annihilation radiation of dark matter in heterotic orbifold models,''
  arXiv:hep-ph/0406083.

\bibitem{Falkowski:2005ck}
P.~Binetruy, A.~Birkedal-Hansen, Y.~Mambrini and B.~D.~Nelson,
``Phenomenological aspects of heterotic orbifold models at one loop,''
arXiv:hep-ph/0308047;

A.~Falkowski, O.~Lebedev and Y.~Mambrini,
``SUSY Phenomenology of KKLT Flux Compactifications,''
arXiv:hep-ph/0507110.



\bibitem{Nath2}
  {\rm   A.~Corsetti, P.~Nath}, 
  {\it Phys.Rev} {\bf D64} {\rm (2001) 125010},
{arXiv:hep-ph/0003186}.
\bibitem{Profumo:2003em}
S.~Profumo,
 ``Neutralino dark matter, b - tau Yukawa unification and non-universal
sfermion masses,''
Phys.\ Rev.\ D {\bf 68}, 015006 (2003)

\bibitem{Ullio:qe}
P.~Ullio,
 ``Indirect Searches For Neutralino Dark Matter Candidates In Anomaly-Mediated
Supersymmetry Breaking Scenarios,''
Nucl.\ Phys.\ Proc.\ Suppl.\  {\bf 110}, 82 (2002).


\bibitem{Cesarini:2003nr}
A.~Cesarini, F.~Fucito, A.~Lionetto, A.~Morselli and P.~Ullio,
``The galactic center as a dark matter gamma-ray source,''
arXiv:astro-ph/0305075.


\bibitem{Hooper:2003ka}
D.~Hooper and L.~T.~Wang,
``Direct and indirect detection of neutralino dark matter in selected
supersymmetry breaking scenarios,''
Phys.\ Rev.\ D {\bf 69}, 035001 (2004)
[arXiv:hep-ph/0309036].

\bibitem{Cerdeno:2002eg}
D.~G.~Cerdeno and C.~Munoz,
``Phenomenology of heterotic M-theory with five-branes,''
Phys.\ Rev.\ D {\bf 66} (2002) 115007
[arXiv:hep-ph/0206299].


\bibitem{Bottino:2004qi}
A.~Bottino, F.~Donato, N.~Fornengo and S.~Scopel,
 ``Indirect signals from light neutralinos in supersymmetric models without
gaugino mass unification,''
arXiv:hep-ph/0401186.




\bibitem{Donato:2003xg}
  F.~Donato, N.~Fornengo, D.~Maurin and P.~Salati,
  ``Antiprotons in cosmic rays from neutralino annihilation,''
  Phys.\ Rev.\ D {\bf 69}, 063501 (2004)
  [arXiv:astro-ph/0306207].


\bibitem{Lahanas:2003bh}
  A.~B.~Lahanas, N.~E.~Mavromatos and D.~V.~Nanopoulos,
  ``WMAPing the universe: Supersymmetry, dark matter, dark energy, proton
  decay and collider physics,''
  Int.\ J.\ Mod.\ Phys.\ D {\bf 12}, 1529 (2003)
  [arXiv:hep-ph/0308251].


\bibitem{Edsjo:2003us}
  J.~Edsjo, M.~Schelke, P.~Ullio and P.~Gondolo,
  ``Accurate relic densities with neutralino, chargino and sfermion
  coannihilations in mSUGRA,''
  JCAP {\bf 0304}, 001 (2003)
  [arXiv:hep-ph/0301106].

\bibitem{Baltz:1998xv}
  E.~A.~Baltz and J.~Edsjo,
  ``Positron propagation and fluxes from neutralino annihilation in the
  halo,''
  Phys.\ Rev.\ D {\bf 59}, 023511 (1999)
  [arXiv:astro-ph/9808243].



\bibitem{Hooper:2004bq}
  D.~Hooper and J.~Silk,
  ``Searching for dark matter with future cosmic positron experiments,''
  Phys.\ Rev.\ D {\bf 71}, 083503 (2005)
  [arXiv:hep-ph/0409104].


\bibitem{Hooper:2003ad}
  D.~Hooper, J.~E.~Taylor and J.~Silk,
  ``Can supersymmetry naturally explain the positron excess?,''
  Phys.\ Rev.\ D {\bf 69}, 103509 (2004)
  [arXiv:hep-ph/0312076].



\bibitem{Baltz:2001ir}
  E.~A.~Baltz, J.~Edsjo, K.~Freese and P.~Gondolo,
  ``The cosmic ray positron excess and neutralino dark matter,''
  Phys.\ Rev.\ D {\bf 65}, 063511 (2002)
  [arXiv:astro-ph/0109318].


\bibitem{Bergstrom:1999jc}
  L.~Bergstrom, J.~Edsjo and P.~Ullio,
  ``Cosmic antiprotons as a probe for supersymmetric dark matter?,''
APJ {\bf 526} (1999),
  arXiv:astro-ph/9902012.

\bibitem{Ullio:2002pj}
  P.~Ullio, L.~Bergstrom, J.~Edsjo and C.~G.~Lacey,
  ``Cosmological dark matter annihilations into gamma-rays: A closer look,''
  Phys.\ Rev.\ D {\bf 66}, 123502 (2002)
  [arXiv:astro-ph/0207125].



\bibitem{Mambrini:2004ke}
  Y.~Mambrini and C.~Munoz,
  ``Gamma-ray detection from neutralino annihilation in non-universal SUGRA
  scenarios,''
  arXiv:hep-ph/0407158.

\noindent
Y.~Mambrini and C.~Munoz,
``A comparison between direct and indirect dark matter search,''
JCAP {\bf 0410} (2004) 003
[arXiv:hep-ph/0407352].

\bibitem{Baek:2004et}
  S.~Baek, Y.~G.~Kim and P.~Ko,
  ``Neutralino dark matter scattering and B/s $\to$ mu+ mu- in SUSY models,''
  JHEP {\bf 0502}, 067 (2005)
  [arXiv:hep-ph/0406033].

 \bibitem{Mambrini:2005vk}
Y.~Mambrini, C.~Munoz, E.~Nezri and F.~Prada,
``Adiabatic compression and indirect detection of supersymmetric dark matter,''
arXiv:hep-ph/0506204.

\bibitem{deBoer:2004ab}
  W.~de Boer, M.~Herold, C.~Sander, V.~Zhukov, A.~V.~Gladyshev and D.~I.~Kazakov,
  ``Excess of EGRET galactic gamma ray data interpreted as dark matter
  annihilation,''
  arXiv:astro-ph/0408272.




\bibitem{Belanger:2005jk}
  G.~Belanger, S.~Kraml and A.~Pukhov,
  ``Comparison of SUSY spectrum calculations and impact on the relic density
  constraints from WMAP,''
  arXiv:hep-ph/0502079.

\bibitem{Belanger:2004hk}
  G.~Belanger, F.~Boudjema, A.~Cottrant, A.~Pukhov and A.~Semenov,
  ``Relic density of dark matter in mSUGRA and non-universal SUGRA,''
  arXiv:hep-ph/0412309.



\bibitem{Belanger:2004ag}
  G.~Belanger, F.~Boudjema, A.~Cottrant, A.~Pukhov and A.~Semenov,
  ``WMAP constraints on SUGRA models with non-universal gaugino masses and
  prospects for direct detection,''
  Nucl.\ Phys.\ B {\bf 706}, 411 (2005)
  [arXiv:hep-ph/0407218].


\bibitem{Arnowitt:2004xi}
  R.~Arnowitt, B.~Dutta, T.~Kamon and V.~Khotilovich,
  ``Minimal SUGRA model and collider signals,''
  arXiv:hep-ph/0411102.


\bibitem{Stark:2005mp}
  L.~S.~Stark, P.~Hafliger, A.~Biland and F.~Pauss,
  ``New allowed mSUGRA parameter space from variations of the trilinear scalar
  coupling A0,''
  arXiv:hep-ph/0502197.


\bibitem{Baer:2004qq}
  H.~Baer, A.~Belyaev, T.~Krupovnickas and J.~O'Farrill,
  ``Indirect, direct and collider detection of neutralino dark matter,''
  JCAP {\bf 0408}, 005 (2004)
  [arXiv:hep-ph/0405210].






\bibitem{Blumenthal:1985qy}
  G.~R.~Blumenthal, S.~M.~Faber, R.~Flores and J.~R.~Primack,
  ``Contraction Of Dark Matter Galactic Halos Due To Baryonic Infall,''
  Astrophys.\ J.\  {\bf 301}, 27 (1986).

\bibitem{Edsjo:2004pf}
  J.~Edsjo, M.~Schelke and P.~Ullio,
  ``Direct versus indirect detection in mSUGRA with self-consistent halo
  models,''
  arXiv:astro-ph/0405414.

\bibitem{Prada:2004pi}
  F.~Prada, A.~Klypin, J.~Flix, M.~Martinez and E.~Simonneau,
  ``Astrophysical inputs on the SUSY dark matter annihilation detectability,''
Phys. Rev. Lett. 93 241301 (2004).

\bibitem{Gnedin:2004cx}
  O.~Y.~Gnedin, A.~V.~Kravtsov, A.~A.~Klypin and D.~Nagai,
  ``Response of dark matter halos to condensation of baryons: cosmological
  simulations and improved adiabatic contraction model,''
  Astrophys.\ J.\  {\bf 616}, 16 (2004)
  [arXiv:astro-ph/0406247].

\bibitem{silkgondo}
  P.~Gondolo and J.~Silk,
  ``Dark matter annihilation at the galactic center,''
  Phys.\ Rev.\ Lett.\  {\bf 83}, 1719 (1999)
  [arXiv:astro-ph/9906391].
\bibitem{merrit04}
  D.~Merrit,
  ``Dark matter annihilation at the galactic center,''
  Phys.\ Rev.\ Lett.\  {\bf 92}, 201304 (2004)

\bibitem{Goodman} 
M.W. Goodman and E. Witten, Phys. Rev. {\bf D31}, (1985)  3059.

\bibitem{abdeldrees}
A. Djouadi and M. Drees, Phys. Lett. {\bf B484}, (2000)  183.


\bibitem{Sanglard:2005we}
  V.~Sanglard {\it et al.}  [The EDELWEISS Collaboration],
  ``Final results of the EDELWEISS-I dark matter search with cryogenic
  heat-and-ionization Ge detectors,''
  arXiv:astro-ph/0503265.








\bibitem{Cdms} CDMS Collaboration, R. Abusaidi et al., Phys. Rev. Lett. 84 (2000) 5699;
Phys. Rev. D66 (2002) 122003.

\bibitem{Dama} DAMA Collaboration, R. Bernabei at al., Phys. Lett. B480 (2000) 23.

\bibitem{EdelweissII} G.~Chardin. Edelweiss dark matter search, talk
given at the school and workshop on neutrino particle astrophysics, les houches 21 jan -1st feb 2002.


\bibitem{Zeplin} ZEPLIN  Collaboration, R. Luscher et al., talk given
the XXXVIIIth Rencontres de Moriond  ELECTROWEAK INTERACTIONS AND
UNIFIED THEORIES, 15th to March 22nd 2003, Les Arcs France. 


\bibitem{Brink:2005ej}
  P.~L.~Brink {\it et al.},
``Beyond the CDMS-II dark matter search: SuperCDMS,''
  arXiv:astro-ph/0503583.

\bibitem{Turner:1986vr}
M.~S.~Turner,
``Probing The Structure Of The Galactic Halo With Gamma Rays Produced By Wimp
Annihilations,''
Phys.\ Rev.\ D {\bf 34} (1986) 1921.


\bibitem{Bergstrom:1997fj}
  L.~Bergstrom, P.~Ullio and J.~H.~Buckley,``Observability of gamma rays from 
dark matter neutralino annihilations  in the Milky Way halo,''
  Astropart.\ Phys.\  {\bf 9}, 137 (1998)
  [arXiv:astro-ph/9712318].

\bibitem{Navarro:1996he}
J.~F.~Navarro, C.~S.~Frenk and S.~D.~M.~White,
``A Universal density profile from hierarchical clustering,''
Astrophys.\ J.\  {\bf 490} (1997) 493.


\bibitem{Moore:1999gc}
B.~Moore, T.~Quinn, F.~Governato, J.~Stadel and G.~Lake,
Mon.\ Not.\ Roy.\ Astron.\ Soc.\  {\bf 310} (1999) 1147
[arXiv:astro-ph/9903164].


\bibitem{Kravtsov:1997dp}
A.~V.~Kravtsov, A.~A.~Klypin, J.~S.~Bullock and J.~R.~Primack,
``The Cores of Dark Matter Dominated Galaxies: theory vs. observations,''
arXiv:astro-ph/9708176.










\bibitem{Jean:2003ci}
  P.~Jean {\it et al.},
  ``Early SPI/INTEGRAL measurements of galactic 511 keV line emission from
  positron annihilation,''
  Astron.\ Astrophys.\  {\bf 407}, L55 (2003)
  [arXiv:astro-ph/0309484].


\bibitem{EGRET} EGRET Collaboration, 
S.~D. Hunger et al., `EGRET observations of the
diffuse gamma-ray emission from the galactic plane',
{\it Astrophys. J.} {\bf 481} (1997) 205; 
H.~A. Mayer-Hasselwander et al., 
`High-Energy Gamma-Ray Emission from the Galactic Center'
{\it Astron. \& Astrophys.} {\bf 335} (1998) 161.

\bibitem{Kosack:2004ri}
K.~Kosack {\it et al.}  [The VERITAS Collaboration],
``TeV gamma-ray observations of the galactic center,''
Astrophys.\ J.\  {\bf 608} (2004) L97
[arXiv:astro-ph/0403422].



\bibitem{Cangaroo1} CANGAROO-II Collaboration,
K.~Tsuchiya et al.,
`Detection of sub-TeV gamma-rays from the galactic center direction by
CANGAROO-II',
{\it Astrophys.\ J.\ } {\bf 606} (2004) L115
[arXiv:astro-ph/0403592].

\bibitem{Aharonian:2004wa}
  F.~Aharonian {\it et al.}  [The HESS Collaboration],
  arXiv:astro-ph/0408145.

\bibitem{Boehm:2003bt}
C.~Boehm, D.~Hooper, J.~Silk, M.~Casse and J.~Paul,
``MeV dark matter: Has it been detected?,''
Phys.\ Rev.\ Lett.\  {\bf 92} (2004) 101301
[arXiv:astro-ph/0309686].





\bibitem{Casse:2003fh}
  M.~Casse, B.~Cordier, J.~Paul and S.~Schanne,
  ``Hypernovae/GRB in the Galactic Center as possible sources of Galactic
  Positrons,''
  Astrophys.\ J.\  {\bf 602}, L17 (2004)
  [arXiv:astro-ph/0309824].

\bibitem{Bertone:2004ek}
  G.~Bertone, A.~Kusenko, S.~Palomares-Ruiz, S.~Pascoli and D.~Semikoz,
  ``Gamma ray bursts and the origin of galactic positrons,''
  arXiv:astro-ph/0405005.


\bibitem{Hooper:2002ru}
D.~Hooper and B.~L.~Dingus,
``Limits on supersymmetric dark matter from EGRET observations of the galactic
center region,''
Phys.\ Rev.\ D {\bf 70} (2004) 113007
[arXiv:astro-ph/0210617].

 \bibitem{HESS} HESS Collaboration,
 J.~A. Hinton et al.,
 `The status of the HESS project',
 {\it New Astron. Rev.} {\bf 48} (2004) 331 [arXiv:astro-ph/0403052].

\bibitem{GLAST} N. Gehrels, P. Michelson,
`GLAST: the next generation high-energy gamma-ray astronomy mission',
{\it Astropart. Phys.} {\bf 11} (1999) 277;

\noindent 
See also the web page http://www-glast.stanford.edu



\bibitem{Morselli:2002nw}
  A.~Morselli, A.~Lionetto, A.~Cesarini, F.~Fucito and P.~Ullio  [GLAST
                  Collaboration],
  Nucl.\ Phys.\ Proc.\ Suppl.\  {\bf 113}, 213 (2002)
  [arXiv:astro-ph/0211327].


\bibitem{neutrinorate}

  K.~Griest and D.~Seckel,
  Nucl.\ Phys.\ B {\bf 283}, 681 (1987)
  [Erratum-ibid.\ B {\bf 296}, 1034 (1988)].

A.~Gould,
``Resonant Enhancements In Wimp Capture By The Earth,''
Astrophys.\ J.\  {\bf 321} (1987) 571.

\bibitem{Myuniv}
  {\rm   V.~Bertin, E.~Nezri, J.~Orloff}, 
  {\it Eur. Phys. J.} {\bf C26} {\rm (2002) 111-124},
{arXiv:hep-ph/0204135}.


\bibitem{Bertone:2004ag}
  G.~Bertone, E.~Nezri, J.~Orloff and J.~Silk,
  {\it Phys.\ Rev.\ } {\bf D70}, 063503 (2004)
  [arXiv:astro-ph/0403322].









\bibitem{Macro} T.~Montaruli. Search for wimps using upward-going
muons in macro, proceeedings of the 26th icrc in salt lake city 17-25
Aug 1999, 277--280, hep-ex/9905021.
 
\bibitem{Suvorova} O.~Suvorova. Status and perspectives of
indirect search for dark matter, published in tegernsee 1999, beyond
the desert 1999. pages 853--867.
 
\bibitem{SuperK} A.~Habig. An indirect search for wimps with
super-kamiokande, contributed to 27th icrc, hamburg, germany, 7-15 aug 2001.
 
\bibitem{Amanda} http://amanda.uci.edu

\bibitem{Antares} http://antares.in2p3.fr

\bibitem{Ice3} http://icecube.wisc.edu

\bibitem{bailey}
D.~Bailey, 
 Ph.D. thesis 2002. http://antares.in2p3.fr/Publications/thesis/2002/d-bailey-thesis.ps.gz


\bibitem{Ice3Edsjo} J.~Edsjo. Swedish astroparticle physics, talk
given at the conference 'partikeldagarna', uppsala, sweden, march 6 2001.




\bibitem{Moskalenko:2002yx}
  I.~V.~Moskalenko, A.~W.~Strong, S.~G.~Mashnik and J.~F.~Ormes,
  ``Challenging cosmic ray propagation with antiprotons: Evidence for a  'fresh'  
nuclei component?,''
  Astrophys.\ J.\  {\bf 586}, 1050 (2003)
  [arXiv:astro-ph/0210480].


\bibitem{Longair}
M.~S.~.~Longair, ``High-energy astrophysics". Cambridge, University Press 1994.

\bibitem{heat}
S.~W.~Barwick {\it et al.}  [HEAT Collaboration],
``Measurements of the cosmic-ray positron fraction from 1-GeV to 50-GeV,''
Astrophys.\ J.\  {\bf 482} (1997) L191
[arXiv:astro-ph/9703192].

\bibitem{Moskalenko:2001ya}
  I.~V.~Moskalenko, A.~W.~Strong, J.~F.~Ormes and M.~S.~Potgieter,
  ``Secondary antiprotons and propagation of cosmic rays in the galaxy and
  heliosphere,''
  Astrophys.\ J.\  {\bf 565}, 280 (2002)
  [arXiv:astro-ph/0106567].








\bibitem{Maurin:2001sj}
  D.~Maurin, F.~Donato, R.~Taillet and P.~Salati,
  ``Cosmic Rays below Z=30 in a diffusion model: new constraints on propagation  
  Astrophys.\ J.\  {\bf 555}, 585 (2001)
  [arXiv:astro-ph/0101231].


\bibitem{Maurin:2002hw}
  D.~Maurin, R.~Taillet and F.~Donato,
  ``New results on source and diffusion spectral features of Galactic cosmic
  rays: I- B/C ratio,''
  Astron.\ Astrophys.\  {\bf 394}, 1039 (2002)
  [arXiv:astro-ph/0206286].


\bibitem{Lionetto:2005jd}
  A.~M.~Lionetto, A.~Morselli and V.~Zdravkovic,
  ``Uncertainties of cosmic ray spectra and detectability of antiproton mSUGRA
  contributions with PAMELA,''
  arXiv:astro-ph/0502406.

 







\bibitem{charginolimit}
  {\rm ALEPH Collaboration (A. Heister et al.)}
  {\it Phys.Lett.} {\bf B533} {\rm (2002) 223}.



\bibitem{Higgslimit}
 R.~Barate {\it et al.}  [ALEPH Collaboration],
  ``Search for the standard model Higgs boson at LEP,''
  Phys.\ Lett.\ B {\bf 565} (2003) 61
  [arXiv:hep-ex/0306033].

  {\rm LEP Higgs Working Group},
  {\it Searches for the Neutral Higgs Bosons of the MSSM},
  {\rm LHWG Note/2001-04, hep-ex/0107030};

  {\rm LEP Higgs Working Group},
  {\it Search for the Standard Model Higgs Boson at LEP},
  {\rm LHWG Note/2002-01}.


\bibitem{Abazov:2004cs}
V.~M.~Abazov {\it et al.}  [D0 Collaboration],
``A precision measurement of the mass of the top quark,''
Nature {\bf 429} (2004) 638
[arXiv:hep-ex/0406031].


\bibitem{Allanach:2004rh}
  B.~C.~Allanach, A.~Djouadi, J.~L.~Kneur, W.~Porod and P.~Slavich,
  ``Precise determination of the neutral Higgs boson masses in the MSSM,''
  JHEP {\bf 0409} (2004) 044
  [arXiv:hep-ph/0406166].


 
 

\bibitem{Degrassi:2000qf}
S.~Bertolini, F.~Borzumati, A.~Masiero and G.~Ridolfi,
``Effects Of Supergravity Induced Electroweak Breaking On Rare B Decays And
Mixings,''
Nucl.\ Phys.\ B {\bf 353} (1991) 591;

R.~Barbieri and G.~F.~Giudice,
``b $\to$ s gamma decay and supersymmetry,''
Phys.\ Lett.\ B {\bf 309} (1993) 86
[arXiv:hep-ph/9303270];

F.~M.~Borzumati,
``The Decay b $\to$ s gamma in the MSSM revisited,''
Z.\ Phys.\ C {\bf 63} (1994) 291
[arXiv:hep-ph/9310212];

M.~S.~Alam {\it et al.}  [CLEO Collaboration],
``First measurement of the rate for the inclusive radiative penguin decay b
$\to$ s gamma,''
Phys.\ Rev.\ Lett.\  {\bf 74} (1995) 2885;

G.~Degrassi, P.~Gambino and G.~F.~Giudice,
``B $\to$ X/s gamma in supersymmetry: Large contributions beyond the  leading
order,''
JHEP {\bf 0012}, 009 (2000)
[arXiv:hep-ph/0009337].



\bibitem{cleo}  
CLEO Collaboration, S. Chen et al.,  
`Branching fraction and photon energy spectrum for 
$b\to s\gamma$', 
{\it Phys. Rev. Lett.} {\bf 87} (2001) 251807 
[arXiv:hep-ex/0108032]. 
 
\bibitem{belle} 
BELLE Collaboration, H. Tajima, 
`Belle B physics results',  
{\it Int. J. Mod. Phys.} {\bf A17} (2002) 2967 
[arXiv:hep-ex/0111037]. 
 
\bibitem{Heavy}
{\bf Heavy Flavour Averaging Group} Collaboration
{\tt http://www.slac.stanford.edu/xorg/hfag}.


\bibitem{Gambino:2004mv}
  P.~Gambino, U.~Haisch and M.~Misiak,
  ``Determining the sign of the b $\to$ s gamma amplitude,''
  Phys.\ Rev.\ Lett.\  {\bf 94} (2005) 061803
  [arXiv:hep-ph/0410155].
 
\bibitem{Suspect2} 

A.~Djouadi, M.~Drees and J.~L.~Kneur,
``Neutralino dark matter in mSUGRA: Reopening the light Higgs pole window,''
arXiv:hep-ph/0504090.

 


\bibitem{Degrassi:1998es}
J.~R.~Ellis, J.~S.~Hagelin and D.~V.~Nanopoulos,
``Spin 0 Leptons And The Anomalous Magnetic Moment Of The Muon,''
Phys.\ Lett.\ B {\bf 116} (1982) 283;

J.~A.~Grifols and A.~Mendez,
``Constraints On Supersymmetric Particle Masses From (G-2) Mu,''
Phys.\ Rev.\ D {\bf 26} (1982) 1809;

R.~Barbieri and L.~Maiani,
``The Muon Anomalous Magnetic Moment In Broken Supersymmetric Theories,''
Phys.\ Lett.\ B {\bf 117} (1982) 203;

D.~A.~Kosower, L.~M.~Krauss and N.~Sakai,
``Low-Energy Supergravity And The Anomalous Magnetic Moment Of The Muon,''
Phys.\ Lett.\ B {\bf 133} (1983) 305;

G.~Degrassi and G.~F.~Giudice,
``QED logarithms in the electroweak corrections to the muon anomalous  magnetic
moment,''
Phys.\ Rev.\ D {\bf 58} (1998) 053007
[arXiv:hep-ph/9803384].




\bibitem{g-2} Muon g-2 Collaboration, G.~W.~Bennett et al., 
`Measurement of the negative muon anomalous magnetic moment to 
0.7-ppm', {\it Phys. Rev. Lett.} {\bf 92} (2004) 161802 
[arXiv:hep-ex/0401008]. 
   
 
\bibitem{newg2} 
  M.~Davier, S.~Eidelman, A.~Hocker and Z.~Zhang, 
  `Updated estimate of the muon magnetic moment using revised results 
  from $e^+ e^-$ annihilation', 
  {\it Eur. Phys. J. C} {\bf 31} (2003) 503 
  [arXiv:hep-ph/0308213]; 
   
  \noindent 
  K.~Hagiwara, A.~D.~Martin, D.~Nomura and T.~Teubner, 
  `Predictions for $g-2$ of the muon and $\alpha_{QED}(M_Z^2)$', 
{\it Phys. Rev.} {\bf D69} (2004) 093003 
[arXiv:hep-ph/0312250]; 
   
  \noindent 
  J.~F.~de Troconiz and F.~J.~Yndurain, 
  `The hadronic contributions to the anomalous magnetic moment of the 
  muon',  
  [arXiv:hep-ph/0402285]. 
 
 
 
 
 
 
 
%
 
 
 
 
\bibitem{bmumuexp} CDF Collaboration,  
D.~Acosta et al., `Search for $B_s^0\to\mu^+\mu^-$ and $B_d^0\to\mu^+\mu^-$  
decays in $p\bar p$ collisions at $\sqrt{s}=1.96$ TeV', 
{\it Phys.\ Rev.\ Lett.}  {\bf 93} (2004) 032001 
[arXiv:hep-ex/0403032]; 
 
\noindent  
D0 Collaboration, V.~M.~Abazov et al., 
`A search for the flavor-changing neutral current decay  
$B_s^0\to\mu^+\mu^-$ in $p\bar p$ collisions at $\sqrt{s}=1.96$ TeV 
with the D0 detector', 
{\it Phys. Rev. Lett.} {\bf 94} (2005) 071802 [arXiv:hep-ex/0410039]. 
 
 
\bibitem{Babu:1999hn}
K.~S.~Babu and C.~F.~Kolda,
``Higgs-mediated B0 $\to$ mu+ mu- in minimal supersymmetry,''
Phys.\ Rev.\ Lett.\  {\bf 84}, 228 (2000)
[arXiv:hep-ph/9909476].

\noindent
G.~Isidori and A.~Retico,
``Scalar flavour-changing neutral currents in the large-tan(beta) limit,''
JHEP {\bf 0111}, 001 (2001)
[arXiv:hep-ph/0110121].

\noindent
A.~Dedes, H.~K.~Dreiner and U.~Nierste,
``Correlation of B/s $\to$ mu+ mu- and (g-2)(mu) in minimal supergravity,''
Phys.\ Rev.\ Lett.\  {\bf 87}, 251804 (2001)
[arXiv:hep-ph/0108037].

\noindent
G.~Isidori and A.~Retico,
 ``B/s,d $\to$ l+ l- and K(L) $\to$ l+ l- in SUSY models with non-minimal
%

 
\bibitem{ko} 
S. Baek, Y.G. Kim and P. Ko, 
`Neutralino dark matter scattering and $B_s \to \mu^+ \mu^-$ in SUSY 
models',  
{\it J. High Energy Phys.} {\bf 02} (2005) 067 
[arXiv:hep-ph/0406033]. 
 
\bibitem{ko2} 
S. Baek, D.G. Cerde\~no, Y.G. Kim, P. Ko and C. Mu\~noz, 
`Direct detection of neutralino dark matter in supergravity', 
{\it J. High Energy Phys.} {\bf 06} (2005) 017 [arXiv:hep-ph/0505019]. 
`Neutralino dark matter scattering and $B_s \to \mu^+ \mu^-$ in SUSY 
models',  
 


\bibitem{Charles:2001ka}
  F.~Charles,
  ``Higgs and supersymmetry searches at the Large Hadron Collider,''
in {\it Proc. of the 5th International Symposium on Radiative Corrections (RADCOR 2000) } ed. Howard E. Haber,
  arXiv:hep-ph/0105026;

S.~Abdullin and F.~Charles,
``Search for SUSY in (leptons +) jets + E(T)(miss) final states,''
Nucl.\ Phys.\ B {\bf 547} (1999) 60
[arXiv:hep-ph/9811402];

S.~Abdullin {\it et al.}  [CMS Collaboration],
``Discovery potential for supersymmetry in CMS,''
J.\ Phys.\ G {\bf 28} (2002) 469
[arXiv:hep-ph/9806366].


\bibitem{Boehm:1999tr}
  C.~Boehm, A.~Djouadi and Y.~Mambrini,
  ``Decays of the lightest top squark,''
  Phys.\ Rev.\ D {\bf 61} (2000) 095006
  [arXiv:hep-ph/9907428];

  A.~Djouadi and Y.~Mambrini,
  ``Three-body decays of top and bottom squarks,''
  Phys.\ Rev.\ D {\bf 63} (2001) 115005
  [arXiv:hep-ph/0011364].

\bibitem{ILC}
J.~L.~Feng, M.~E.~Peskin, H.~Murayama and X.~Tata,
``Testing supersymmetry at the next linear collider,''
Phys.\ Rev.\ D {\bf 52} (1995) 1418
[arXiv:hep-ph/9502260];

H.~Murayama and M.~E.~Peskin,
``Physics opportunities of e+ e- linear colliders,''
Ann.\ Rev.\ Nucl.\ Part.\ Sci.\  {\bf 46} (1996) 533
[arXiv:hep-ex/9606003];

H.~Baer, R.~Munroe and X.~Tata,
``Supersymmetry studies at future linear e+ e- colliders,''
Phys.\ Rev.\ D {\bf 54} (1996) 6735
[Erratum-ibid.\ D {\bf 56} (1997) 4424]
[arXiv:hep-ph/9606325];

E.~Accomando {\it et al.}  [ECFA/DESY LC Physics Working Group Collaboration],
``Physics with e+ e- linear colliders,''
Phys.\ Rept.\  {\bf 299} (1998) 1
[arXiv:hep-ph/9705442];

S.~Dawson and M.~Oreglia,
``Physics opportunities with a TeV linear collider,''
Ann.\ Rev.\ Nucl.\ Part.\ Sci.\  {\bf 54} (2004) 269
[arXiv:hep-ph/0403015].




\bibitem{Datta:2002mz}
  A.~Datta, A.~Djouadi and M.~Muhlleitner,
  ``Associated production of sfermions and gauginos at high-energy e+ e-
  colliders: The case of selectrons and electronic sneutrinos,''
  Eur.\ Phys.\ J.\ C {\bf 25} (2002) 539
  [arXiv:hep-ph/0204354];

  A.~Datta and A.~Djouadi,
  ``Associated production of sfermions and gauginos at high-energy e+ e-
  colliders,''
  Eur.\ Phys.\ J.\ C {\bf 25} (2002) 523
  [arXiv:hep-ph/0111466].



\bibitem{Djouadi:2001fa}
A.~Djouadi, Y.~Mambrini and M.~Muhlleitner,
``Chargino and neutralino decays revisited,''
Eur.\ Phys.\ J.\ C {\bf 20} (2001) 563
[arXiv:hep-ph/0104115];

A.~Djouadi and Y.~Mambrini,
``Three-body decays of SUSY particles,''
Phys.\ Lett.\ B {\bf 493} (2000) 120
[arXiv:hep-ph/0007174].



\bibitem{Djouadi:2005gi}

 A.~Djouadi, J.~Kalinowski and P.~M.~Zerwas,
  ``Two- and Three-Body Decay Modes of SUSY Higgs Particles,''
  Z.\ Phys.\ C {\bf 70} (1996) 435
  [arXiv:hep-ph/9511342];

 A.~Djouadi, M.~Spira and P.~M.~Zerwas,
  ``QCD Corrections to Hadronic Higgs Decays,''
  Z.\ Phys.\ C {\bf 70} (1996) 427
  [arXiv:hep-ph/9511344];

 A.~Djouadi, J.~Kalinowski and M.~Spira,
  ``HDECAY: A program for Higgs boson decays in the standard model and its
  supersymmetric extension,''
  Comput.\ Phys.\ Commun.\  {\bf 108} (1998) 56
  [arXiv:hep-ph/9704448];

A.~Djouadi,
``The anatomy of electro-weak symmetry breaking. I: The Higgs boson in the
standard model,''
arXiv:hep-ph/0503172;

A.~Djouadi,
``The anatomy of electro-weak symmetry breaking. II: The Higgs bosons in the
minimal supersymmetric model,''
arXiv:hep-ph/0503173.



\bibitem{Djouadi:1996jc}
 A.~Djouadi, J.~Kalinowski and P.~M.~Zerwas,
  ``Exploring the SUSY Higgs sector at e+ e- linear colliders: A Synopsis,''
  Z.\ Phys.\ C {\bf 57} (1993) 569.


A.~Djouadi, P.~Ohmann, P.~M.~Zerwas and J.~Kalinowski,
``Heavy SUSY Higgs bosons at e+ e- linear colliders,''
arXiv:hep-ph/9605437;

A.~Djouadi, J.~Kalinowski, P.~Ohmann and P.~M.~Zerwas,
 ``Heavy SUSY Higgs bosons at e+ e- linear colliders,''
  Z.\ Phys.\ C {\bf 74} (1997) 93
  [arXiv:hep-ph/9605339];



\bibitem{Ellis:2003eg}
{\rm J.~R.~Ellis, A.~Ferstl, K.~A.~Olive and Y.~Santoso},
``Direct detection of dark matter in the MSSM with non-universal Higgs  masses,''
{\it Phys.\ Rev. {\bf D67},  (2003) 123502}
{arXiv:hep-ph/0302032};

{\rm U.~Chattopadhyay and D.~P.~Roy},
``Higgsino dark matter in a SUGRA model with nonuniversal gaugino masses,''
{\it Phys.\ Rev. {\bf D68} (2003) 033010}
[arXiv:hep-ph/0304108];

{\rm H.~Baer, C.~Balazs, A.~Belyaev and J.~O'Farrill},
``Direct detection of dark matter in supersymmetric models,''
arXiv:hep-ph/0305191.


\end{thebibliography}
\bibliographystyle{unsrt}

\end{document}